\documentclass[prb, reprint, 9pt, superscriptaddress,notitlepage, nofootinbib, longbibliography,
floatfix]{revtex4-2}
\usepackage{xcolor}
\usepackage{graphicx}
\usepackage{dcolumn}
\usepackage{bm}
\usepackage{amsmath}
\usepackage{amssymb}
\usepackage{lipsum}
\usepackage{enumitem}
\usepackage{mathtools}
\usepackage{tikz}
\usepackage{soul}
\usepackage{amsmath, braket}
\usepackage[normalem]{ulem}
\usepackage{comment}
\usepackage{bm}
\usepackage{amsthm}

\usepackage[colorlinks]{hyperref}

\newcommand{\be}{\begin{equation}}
\newcommand{\ee}{\end{equation}}

\DeclareMathOperator{\Tr}{Tr}

\newcommand{\beq}{\begin{equation}}
\newcommand{\eeq}{\end{equation}}
\newcommand{\beqn}{\begin{eqnarray}}
\newcommand{\eeqn}{\end{eqnarray}}

\newcommand{\cE}{ {\cal E} }

\begin{document}
\newcommand{\jianhao}[1]{ { \color{violet} \small (\textsf{JHZ}) \textsf{\textsl{#1}} }}
\newcommand{\sgc}[1]{ { \color{teal} \small (\textsf{sg}) \textsf{\textsl{#1}} }}
\newcommand{\sg}[1]{{\color{teal} #1}}
\newcommand{\yy}[1]{ { \color{red} \small (\textsf{YY}) \textsf{\textsl{#1}} }}
\newcommand{\fp}[1]{ { \color{green} \small (\textsf{FP}) \textsf{\textsl{#1}} }}

\newcommand{\pab}[1]{ { \color{olive} \small (\textsf{PS}) \textsf{\textsl{#1}} }}
\newcommand{\olive}[1]{ {\color{olive}#1}}

\newcommand{\peter}[1]{ { \color{blue} \small (\textsf{peter}) \textsf{\textsl{#1}} }}

\definecolor{myblue}{RGB}{0, 0, 140}

\hypersetup{colorlinks,breaklinks,
            urlcolor=[rgb]{0.25,0.41,0.88},
            linkcolor=[rgb]{0.25,0.41,0.88},
            citecolor=[rgb]{0.25,0.41,0.88}}


\newtheorem{theorem}{Theorem}
\newtheorem{prop}{Proposition}
\newtheorem{lemma}[theorem]{Lemma}
\newtheorem{corollary}{Corollary}[theorem]
\newtheorem{thm}{\protect\theoremname}

\newtheorem{rem}[thm]{Remark}
\newtheorem{defn}[thm]{Definition}

\title{Holographic duality between bulk topological order and boundary mixed-state order}

\author{Tsung-Cheng Lu}
\email{tclu@umd.edu}
\affiliation{Joint Center for Quantum Information and Computer Science,
University of Maryland, College Park, Maryland 20742, USA}

\author{Yu-Jie Liu}
\email{yujieliu@mit.edu}

\affiliation{Center for Theoretical Physics - a Leinweber Institute, Massachusetts Institute of Technology, Cambridge, MA 02139, USA}

\author{Sarang Gopalakrishnan}
\email{sgopalakrishnan@princeton.edu}
\affiliation{Department of Electrical and Computer Engineering,
Princeton University, Princeton, NJ 08544, USA}

\author{Yizhi You}
\email{y.you@northeastern.edu}
\affiliation{Department of Physics, Northeastern University, Boston, MA, 02115, USA}

\date{\today}
\begin{abstract}
We introduce a holographic framework for analyzing the steady states of repeated quantum channels with strong symmetries. Using channel–state duality, we show that the steady state of a $d$-dimensional quantum channel is holographically mapped to the boundary reduced density matrix of a $(d+1)$-dimensional wavefunction generated by a sequential unitary circuit. From this perspective, strong-to-weak spontaneous symmetry breaking (SWSSB) in the steady state arises from the anyon condensation on the boundary of a topological order in one higher dimension. The conditional mutual information (CMI) associated with SWSSB is then inherited from the bulk topological entanglement entropy. We make this duality explicit using isometric tensor network states (isoTNS) by identifying the channel’s time evolution with the transfer matrix of a higher-dimensional isoTNS. Built on isoTNS, we further construct continuously tunable quantum channels that exhibit distinct mixed-state phases and transitions in the steady states.

\end{abstract}

\maketitle

\tableofcontents
\section{Introduction}

The classification of many-body states into equivalence classes, called phases of matter, is a central task in many-body physics \cite{wenbook}. Informally, a phase of matter is a set of states that share the same universal long-distance properties: for example, long-range magnetic order, superfluidity, or topological order. The notion of phases has been made precise in the context of ground states of gapped local Hamiltonians: two such states are in the same phase if there exists a finite-time quasilocal evolution that transforms one into the other \cite{hastings2005quasiadiabatic,chen_local_unitary_2010,chen11a,chen11b}.

More generally, for equilibrium ordered phases-namely, those that spontaneously break global symmetries-there is a universal description based on Landau theory. Neither framework generalizes to a family of states that are increasingly relevant to present-day quantum devices: namely, \emph{mixed states} that arise from nonequilibrium processes such as noise, measurements, and error correction. The toric code subject to noise is a canonical example of a model with distinct mixed-state phases: below a certain threshold, the code retains logical information, which can be perfectly retrieved in the thermodynamic limit by error correction; above threshold, the logical information is irretrievable \cite{Dennis_2002,Fan_2024}.
The state of the toric code subject to sub-threshold noise levels does not appear to be a thermal state of any local Hamiltonian; its residual topological order is not captured by any conventional correlation function. Nevertheless, it is a sharply distinct phase from the trivial one. In the past few years, many other examples of mixed-state phases and phase transitions have been found \cite{deGroot2022,Lee2025symmetryprotected,exact_information_lee_2025,css_coherent_lee_2025,MaWangASPT_2023, ma2024topological, Zhang_2023, ma2024symmetry, xue2024tensor, guo2024locally, chen2024separability, chen2024unconventional, chen2023symmetryenforced, lessa2024mixedstate, wang2024anomaly, ellison2024towards,sohal2024noisy,lu2023mixed,Lu_ten_2020,chirame2024stable,albert2014symmetries, zhang2024strong,zhang2024fluctuation,raussendorf2007topological,eckstein2024robust,lee2022decoding,zhu2023nishimori,fang2024probing,coser2019classification,bao2023mixed,sang2023mixedstate,zhang2022strange,brown2016quantum,Intrinsic_wang_2025,Andrea_2023,lu_vijay_2023,you2024intrinsic,Lu_disentanling_2024, lu2025spacetime, Su_2024, lessa2025higher,sang_markov_2025,sahu_2024_symmetry, sang2024approximate_CFT, guo2024new,gu2024spontaneous,Lee_2023_prxq,sala2024spontaneous_published,lessa2025strong,gu2024spontaneous,song2025strong,huang2024hydro,sala2025entanglement,zhang2025strong,liu:2024diss,feng2025hardness}, further motivating the quest for a definition of ``mixed-state phases'' that formally captures these distinctions.

In the context of equilibrium pure-state phases, symmetry plays a key part in the classification. Even phases that were initially described as lying outside a symmetry-based classification, such as a large class of topologically ordered phases, are now understood in terms of higher-form symmetries \cite{McGreevy2022_review}. In mixed states, these symmetries can come in two varieties, ``strong'' or ``weak'': intuitively, strongly symmetric mixed states have a definite eigenvalue of the symmetry charge, while weakly symmetric states can be mixtures (but not superpositions) of charge sectors. (We will define these concepts more precisely in Sec.\ref{sec:basic_holo}.) A state can spontaneously break either a strong or a weak symmetry to nothing: this would be conventional spontaneous symmetry breaking, which can be detected by conventional correlators. However, it can also exhibit the intrinsically mixed-state phenomenon of \emph{spontaneous strong-to-weak symmetry breaking} (SW-SSB) \cite{Lee_2023_prxq,sala2024spontaneous_published,lessa2025strong,gu2024spontaneous,song2025strong,huang2024hydro,ando2024gauge}. An intuitive example of SW-SSB occurs in the maximally mixed state with a definite symmetry charge (for simplicity, consider a simple $U(1)$ charge). This state is akin to a thermal state in the canonical ensemble that has no form of conventional order. However, if one loses information about part of the system, this information cannot be reconstructed locally, since any local reconstruction would not ``know about'' the global charge of the system. Thus, a mixed state with SW-SSB marks a distinct phase of matter from, for example, a single bit string, which can be reconstructed locally. Although this example is simple and classical, SW-SSB is a general framework that can capture a huge variety of distinct mixed-state phases and phase transitions, including error correction transitions \cite{sala2024spontaneous_published,Lee_2023_prxq,lessa2025strong,song2025strong,huang2024hydro,zhang2025strong,kim2024error,wang2024topologically,dai2025steady,aldossari2025tensor,yang2025topological,sang2025mixed}.

\begin{figure}[t]
    \centering
\includegraphics[width=0.4\textwidth]{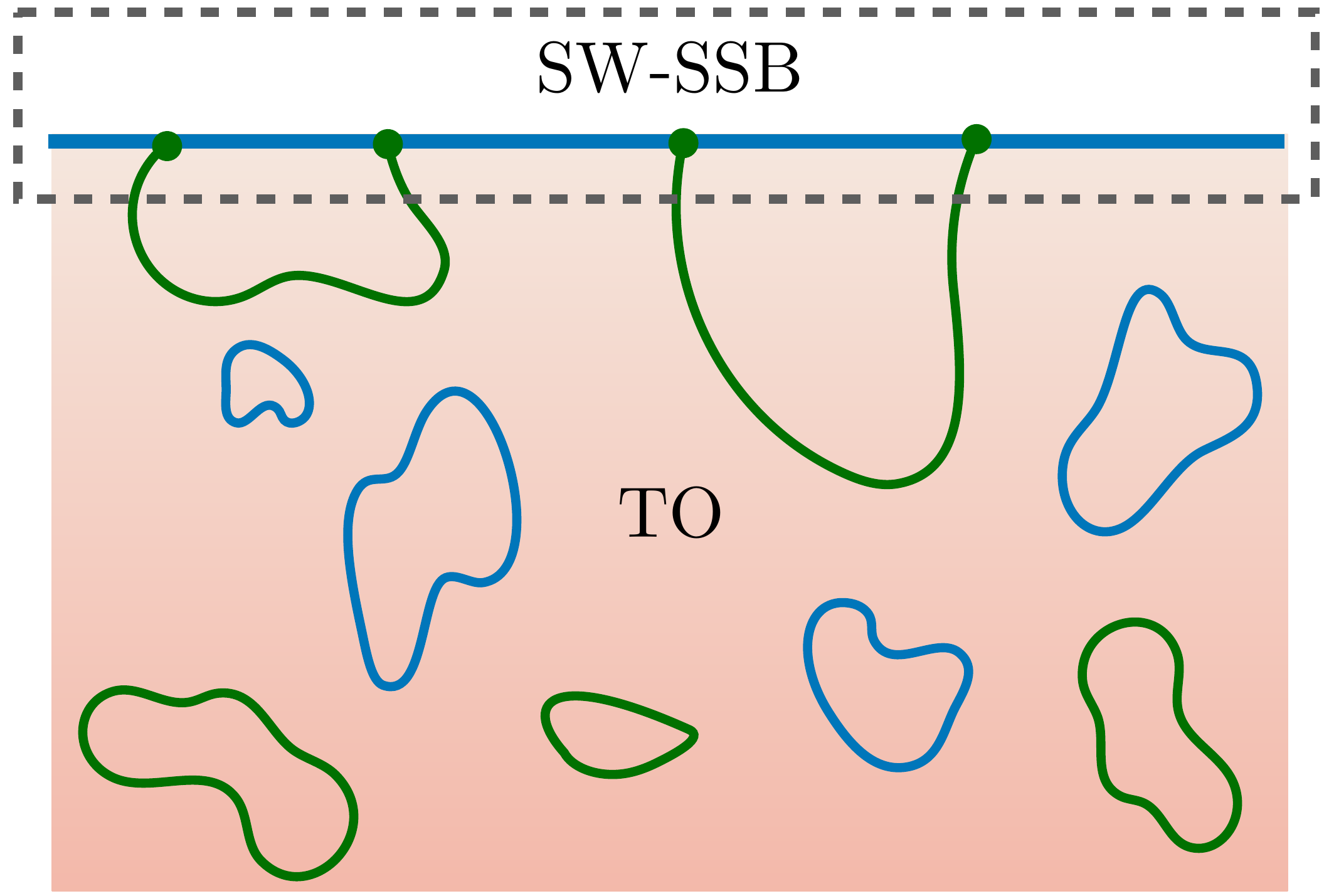}
    \caption{Holographic correspondence between an SW-SSB mixed state in $d$ dimensions and a topological order in $d+1$ dimensions. In a $\mathbb{Z}_2$ topological order, the two loop-like symmetries in the 2d bulk can be pushed to the 1d boundary, where they manifest as a strong symmetry and a weak symmetry. The mutual anomaly between these symmetries signals an SW-SSB in the boundary reduced density matrix.} 
    \label{fig:main_holography}
\end{figure}

So far, much work on mixed-state phases and SW-SSB has focused on transitions that occur in finite-time dynamics, e.g., the toric code subject to noise for a finite time. However, the most natural object to study in a general nonequilibrium process is its late-time \emph{steady state} density matrix \cite{dai2025steady,wang2024topologically,liu2025parent,guo2024new,liu:2024diss,ziereis2025strong}. The classification of steady states of Lindblad master equations was recently explored using principles analogous to those for general mixed states~\cite{rgvk}; whether all mixed states can be realized as steady states of Lindblad master equations, however, remains an open question. Steady states are more structured than generic mixed states because they have a holographic representation in one higher dimension: they correspond to the boundary states of sequential quantum circuits \cite{chen2024sequential,chen2023sequential,lu2025spacetime} (or equivalently, isometric tensor networks \cite{isotns_pollmann_2020}). Physically, this correspondence can be understood as follows. A Markov process consists of a system repeatedly interacting with its environment. This repeated interaction generates a pure state in a spacetime ``bulk'' consisting of the system plus its environments at every time step. From this pure state, the system's steady state (which lives on the top spacetime ``boundary'') can be reached by tracing out the environment qubits that form the spacetime bulk. The question we address in this work is: \emph{what kind of spacetime bulk gives rise to SW-SSB in the steady state}? Our main result can be summarized as follows: we argue (and give explicit constructions demonstrating) that the spacetime bulk could be topologically ordered, and that the SW-SSB boundary arises from boundary condensation of anyons of the bulk topological order; see Fig.\ref{fig:main_holography}. We discuss this correspondence between bulk topological order and boundary mixed-state SW-SSB in a variety of settings. 

The rest of this work is organized as follows. In Sec.\ref{sec:basic_holo}, we review the main concepts of SW-SSB and present the holographic framework that relates a steady state to a higher-dimensional holographic wavefunction via channel/state duality. Sec.\ref{sec:1d} provides a detailed discussion of the holographic correspondence between 1d SW-SSB and 2d topological order through a concrete example. In Sec.\ref{sec:iso}, we reformulate the holographic duality using the isoTNS (isometric tensor network state) formalism, and discuss its applications. Sec.\ref{sec:gen} extends the discussion to various generalized symmetries—including higher-form, subsystem, and fermionic symmetries. In Sec.\ref{sec:relations}, we generalize the holographic framework to non-CPTP (completely positive and trace-preserving) quantum channels and clarify its relation to symTFT (symmetry topological field theory), which is another holographic approach for understanding quantum phases from topological order in one higher dimension. Finally, in Sec.\ref{sec:outlook}, we conclude by highlighting open questions and future directions.

\section{Basic framework and setup}\label{sec:basic_holo}

In this section, we will first briefly review the concept of strong-to-weak spontaneous symmetry breaking (SW-SSB) in many-body mixed states \cite{Lee_2023_prxq,sala2024spontaneous_published,lessa2025strong}, and then show how they can be naturally understood via a holographic perspective.

\subsection{Brief review of SW-SSB} 

In a mixed quantum state $\rho$, a symmetry $U$ can have two different manifestations, i.e., strong or weak. The strong symmetry refers to the property that $U\rho \propto \rho$, in which case all the pure states in any ensemble decomposition carry the same symmetry charges. On the other hand, the weak symmetry only requires $U\rho U^{\dagger} = \rho$, so that pure states in the mixed-state ensemble are allowed to carry distinct charges. Due to these two different notions of symmetry, mixed states may exhibit a new pattern of symmetry breaking, where a strong symmetry can be spontaneously broken down to the weak one. Physically, this indicates the strongly symmetric mixed state does not have long-range order in charged operators, i.e. $\lim_{ |i-j | \to \infty}  \text{Tr}(\rho O_i O_j) \to 0 $, but still, it is possible to decompose the mixed state into an ensemble of long-range ordered state, namely, $\rho =\sum_n p_n  \ket{\psi_n} \bra{\psi_n}$, where $\lim_{ |i-j | \to \infty}  \bra{\psi_n}  O_i O_j  \ket{\psi_n} \neq 0$ \cite{chen2024separability}. In particular, the `averaged' long-range order in the ensemble can be diagnosed by the so-called fidelity correlator \cite{lessa2025strong}, defined as the fidelity between $\rho$ and $O_iO_j\rho O_i^{\dagger}   O_j^{\dagger}$:

\begin{equation}\label{eq:fid}
F_{ij} = \text{Tr} \sqrt{ \sqrt{\rho}     O_iO_j\rho O_i^{\dagger} O_j^{\dagger}     \sqrt{ \rho } } \neq 0
\end{equation}
for $|i-j| \to \infty$. The presence of long-range order in the fidelity correlator further indicates that the resulting mixed state is not locally recoverable. In other words, the steady state features a divergent Markov length and non-vanishing conditional mutual information.

As a simple example, consider a 1d lattice with each qubit per site with a strong $\mathbb{Z}_2$ symmetry generated by $\prod_i X_i$, the maximally mixed state with strong $\mathbb{Z}_2$ symmetry exhibits SW-SSB:
\begin{equation}\label{eq:z2_SWSSB}
    \rho \propto 1+\prod_i X_i.
\end{equation}

It is simple to see that $\text{Tr} (\rho Z_i Z_j   )=0 ~~\forall ~ i\neq j $, so $\rho$ does not have long-range order in local, linear observables. However, the fidelity correlator $F_{ij}=1$, and correspondingly, $\rho$ admits an ensemble of long-range ordered pure states 
\begin{equation}
   \rho \propto \sum_z \ket{\psi_z} \bra{\psi_z}, 
\end{equation}
where $\ket{\psi_z} \propto (1 +\prod_i X_i )  \ket{z}$, i.e., a cat state whose domain wall configuration is fixed by the Pauli-Z basis state $\ket{z}$.

Since SW-SSB only manifests the order in each pure state in the ensemble without showing any order in any linear observables in the density matrix, it is regarded as a novel phenomenon intrinsic to many-body mixed states. As such, there has been a surge of interest in understanding SW-SSB. Below, we will provide one novel perspective, showing that SW-SSB mixed states naturally admit a holographic description in terms of a wave function in one higher dimension.

\subsection{Holographic perspective for quantum channels}\label{sec:holo}
Our holographic description of SW-SSB is based on the duality inspired by Ref.\cite{gopalakrishnan2023push,sun2024holographic}: the output of a repeated quantum channel in $d$ spatial dimensions is exactly the boundary reduced density matrix of a $(d+1)$-dimensional wavefunction generated by sequential unitary circuits. This insight allows us to understand SW-SSB, which generically appears in the steady state of a repeated quantum channel, in terms of a wavefunction in one higher dimension. Here, we will introduce the duality framework for general quantum channels. The special class of channels that preserve strong symmetries and give rise to SW-SSB steady states will be discussed in Sec.\ref{sec:strong_symmetry_channel}.

Given the system qubits in $d$ spatial dimensions, a local quantum channel can be implemented by entangling with the ancilla qubits (in $d$ spatial dimensions as well) via a local unitary circuit $U$, followed by tracing out the ancilla qubits. Specifically, when the ancilla qubits are initialized as the product state $|0\rangle_a\equiv \ket{000...}$, tracing out the ancilla qubits after $U$ implements a completely positive and trace-preserving (CPTP) quantum channel on the system qubits: 
\begin{align}\label{eq:unitarychannel}
\mathcal{E}[\rho_0]
&= \mathrm{Tr}_a\!\left[U\,\bigl(\rho_0 \otimes | 0\rangle\langle 0|_a\bigr)\,U^{\dagger}\right]\nonumber\\
&= \sum_m K_m\, \rho_0\, K_m^{\dagger}.
\end{align} 
where $K_m$ defines the Kraus operator.

\begin{figure}[t]
    \centering
\includegraphics[width=0.48\textwidth]{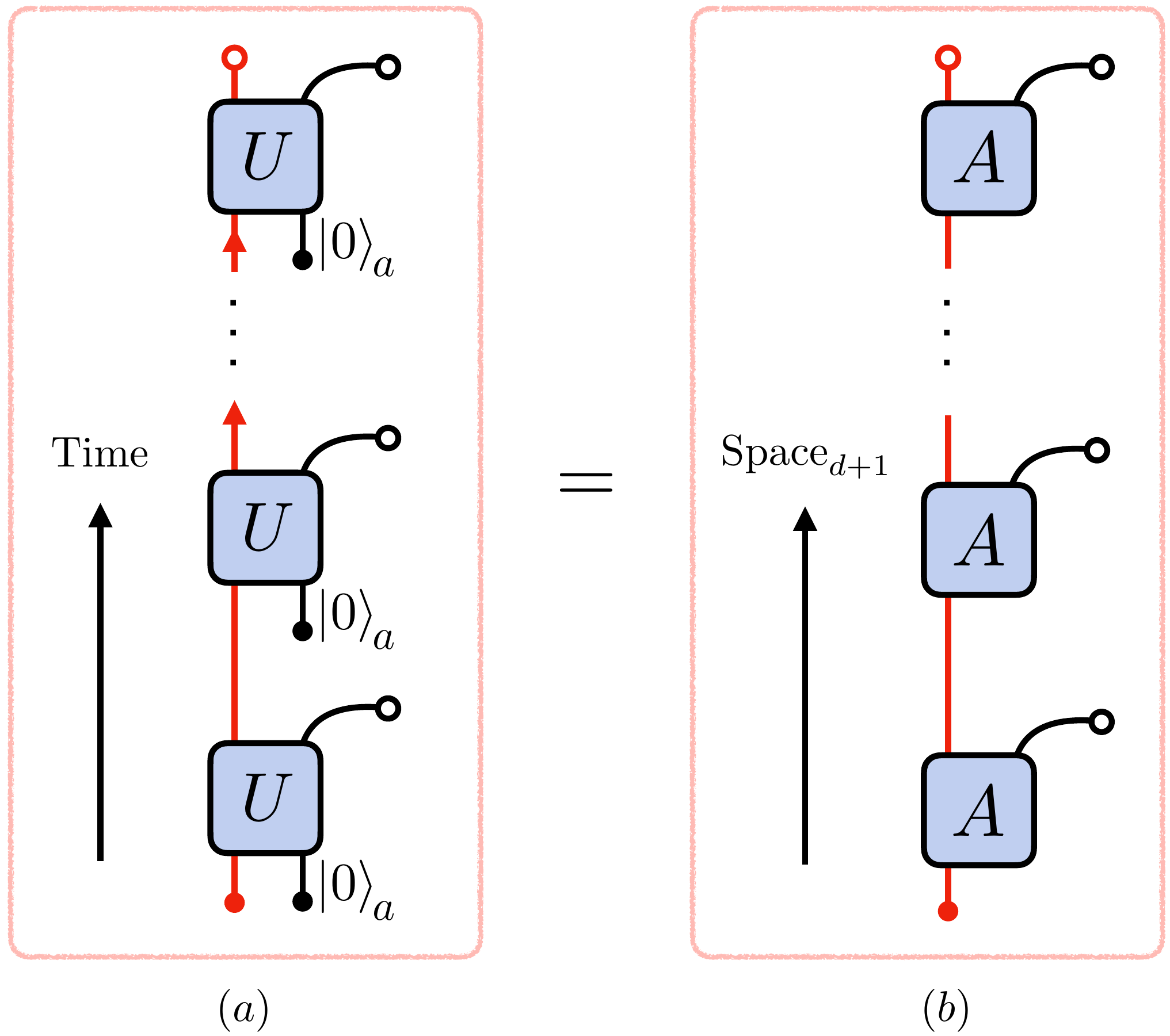}
    \caption{(a) A repeated quantum channel acting on the system (wordline in red) in $d$ spatial dimensions is equivalent to sequentially entangling the system with fresh ancilla qubits in various time slices, which generates a pure-state wavefunction in $(d+1)$ spatial dimensions, as shown in (b). The system output under the repeated channel is the boundary reduced density matrix of this higher-dimensional wave function by tracing out the ancilla qubits.} 
    \label{fig:channel_duality}
\end{figure}

When applying the quantum channel repeatedly, the system qubits are sequentially entangled with fresh $d$-dimensional ancilla qubits at various time slices, followed by tracing out all the ancilla qubits; see Fig.\ref{fig:channel_duality}(a). In particular, the arrangement of these ancilla qubits at various time slices introduces an extra spatial dimension, so that the repeated quantum channel defines a sequential unitary circuit, where the system qubits sequentially interact with ancilla qubits along this new spatial direction. The resulting circuit generates a wavefunction $\ket{\Psi}$ in $(d+1)$ spatial dimensions, where the output of the system qubit lives on the boundary and the bulk is formed by the ancilla qubits; see Fig.\ref{fig:channel_duality}(b). Namely, the reduced density matrix on the boundary

\[
\rho_s = \mathrm{Tr}_{\text{bulk}} \left[\, |\Psi\rangle \langle\Psi| \,\right]
\]
is exactly the output of a repeated quantum channel. This holographic view of a steady state provides a new perspective for SW-SSB, as we will illustrate in the next subsection.

\subsection{Strongly symmetric channel and SW-SSB from holography}\label{sec:strong_symmetry_channel}

To apply the holographic framework to SW-SSB, we consider a quantum channel $\mathcal{E}$ with strong symmetry $S$, meaning that there is no exchange of symmetry charges between the system and ancillae, and hence the charges within the system are conserved. In other words, if the input has the strong symmetry, i.e. $S \rho_0 \propto \rho_0 $, the output of the channel will be strongly symmetric as well: $S\mathcal{E}[\rho_0] \propto \mathcal{E}[\rho_0]$. Mathematically, such a strongly symmetric channel demands that all Kraus operators commute with the symmetry operator $ [K_m,S]=0$, or equivalently, in the dual sequential-circuit picture, the unitary gate $U$ acting on the system and ancillae satisfies $[U, S] = 0 $, where we note that $S$ only acts on the system.

A repeated quantum channel possessing a strong symmetry $S$ will generically lead to a steady state that exhibits SW-SSB. Intuitively, this is because unstructured, repeated interactions with the environment tend to wash out any long-range order, leaving only the information associated with the global symmetry conservation law. In other words, the steady state is expected to be close to a maximally mixed state constrained by the strong symmetry (e.g., $\rho \sim 1 + \prod_i X_i$ in the case of a $\mathbb{Z}_2$ symmetry).

When the steady state of strongly symmetric channels exhibits SW-SSB, we can then adopt the framework in Sec.\ref{sec:holo} to develop a holographic description with a wave function in one higher spatial dimension. This raises a central question: \emph{are the universal features of SW-SSB—such as the strong–weak symmetry anomaly, the non-vanishing conditional mutual information and fidelity correlator—holographically encoded in the higher-dimensional wave function?}

To make the holographic duality explicit, we recast channel dynamics in a unifying isometric tensor-network state (isoTNS) representation \cite{isotns_pollmann_2020}. IsoTNS are an expressive subclass of tensor networks that capture many gapped ground states, including all the 2d string-net topological orders \cite{2020_string_net_isoTNS,liu_stringnet:2022}. Enforcing isometry makes the transfer matrix of a $(d+1)$-dimensional isoTNS naturally dual to the time evolution of local quantum channels in $d$ dimensions, where one of the spatial directions in the isoTNS can be viewed as a temporal direction in the channel dynamics; the dominant eigenvector of this transfer matrix yields the steady state. Thus, the steady state of any repeated quantum channel is the boundary reduced density matrix of a higher-dimensional isoTNS, with boundary degrees of freedom identified with the network’s virtual bonds. The detailed discussion for SW-SSB based on the isoTNS framework will be presented in Sec.\ref{sec:iso}.

\begin{center}
\begin{table*}[htbp]
\renewcommand\arraystretch{1.4}
\begin{tabular}{| l | l |}
\hline
\textbf{1d steady state of a channel} & \textbf{2d topological order} \\ \hline
Strong symmetry & 1-form symmetry generated by m-flux loop \\ \hline
Weak 1-form symmetry &  1-form symmetry generated by e-charge loop  \\ \hline
Strong to weak symmetry breaking & Charge condensation on the boundary\\ \hline
Time evolution of quantum channel & Transfer matrix of isoTNS \\ \hline
Conditional mutual information & Topological entanglement entropy \\ \hline
\end{tabular}
\caption{A summary of the holographic duality discussed in Sec.\ref{sec:1d}}\label{table1}
\end{table*}
\end{center}

As we will elucidate in Sec.\ref{sec:1d} and summarize in Table~I, SW-SSB of a $\mathbb{Z}_2$ symmetry in 1d maps to the boundary reduced density matrix of a 2d $\mathbb{Z}_2$ topological order with a charge-condensed boundary. Notably, the strong symmetry condition maps to a virtual symmetry of the tensors, indicating that the higher-dimensional wavefunction carries a 1-form flux-loop symmetry. Likewise, the fidelity correlator that diagnoses SW-SSB arises from the 1-form charge-loop symmetry, whose open strings can terminate on the boundary~\cite{luo2025topological,qi2025symmetry,schafer2025symtft}. Finally, the nonvanishing conditional mutual information in the boundary reduced density matrix, which signals SW-SSB, is holographically dual to the bulk topological entanglement entropy~\cite{Levin-2006,kitaev-preskill}.

\section{1d SW-SSB as the boundary of 2d topological order}\label{sec:1d}

With the basic formalism established in Sec.\ref{sec:basic_holo}, in this section, we will focus on a particular channel that gives rise to an SW-SSB steady state, and discuss the emergence of topological orders in one higher dimension.

Given a 1d lattice with periodic boundary conditions and one qubit per site, we consider a strong $\mathbb{Z}_2$ symmetry generated by $\prod_i X_i$ and the corresponding SW-SSB mixed state defined in Eq.\ref{eq:z2_SWSSB}:

\begin{equation}\label{eq:1d_Z_2_state}
    \rho \propto 1+\prod_i X_i. 
\end{equation}
To obtain such a mixed state from any pure state $\ket{\psi}$ with the strong $\mathbb{Z}_2$ symmetry, we consider the channel $\cE = \prod_i \cE^z_{i} \cE^x_{i}$, with 

\begin{equation}\label{eq:noise}
\begin{split} 
   &\cE^z_{i}[\rho_0] = (1-p_z) \rho_0 + p_z  Z_i Z_{i+1}\rho_0 Z_i Z_{i+1}, \\
   &  \cE^x_{i}[\rho_0] =  (1-p_x) \rho_0 + p_x X_i \rho_0 X_i,
\end{split}
\end{equation}

The $X$-type noise renders the steady state an incoherent mixture of product states in the $X$ basis, while the $Z$-type noise enforces equal weights among these $X$-basis product states. Therefore, for any $p_x,p_z>0$, the repeated application of these two noise channels drives the system to the SW-SSB steady state (Eq.\ref{eq:1d_Z_2_state}): 
\begin{equation}
\rho= \lim_{N \to \infty}  \cE^{N}  [ \ket{\psi} \bra{\psi}]. 
\end{equation}

\subsection{Holographic description via sequential circuits}\label{sec:1d_channel}

\begin{figure}[t]
    \centering
\includegraphics[width=0.42\textwidth]{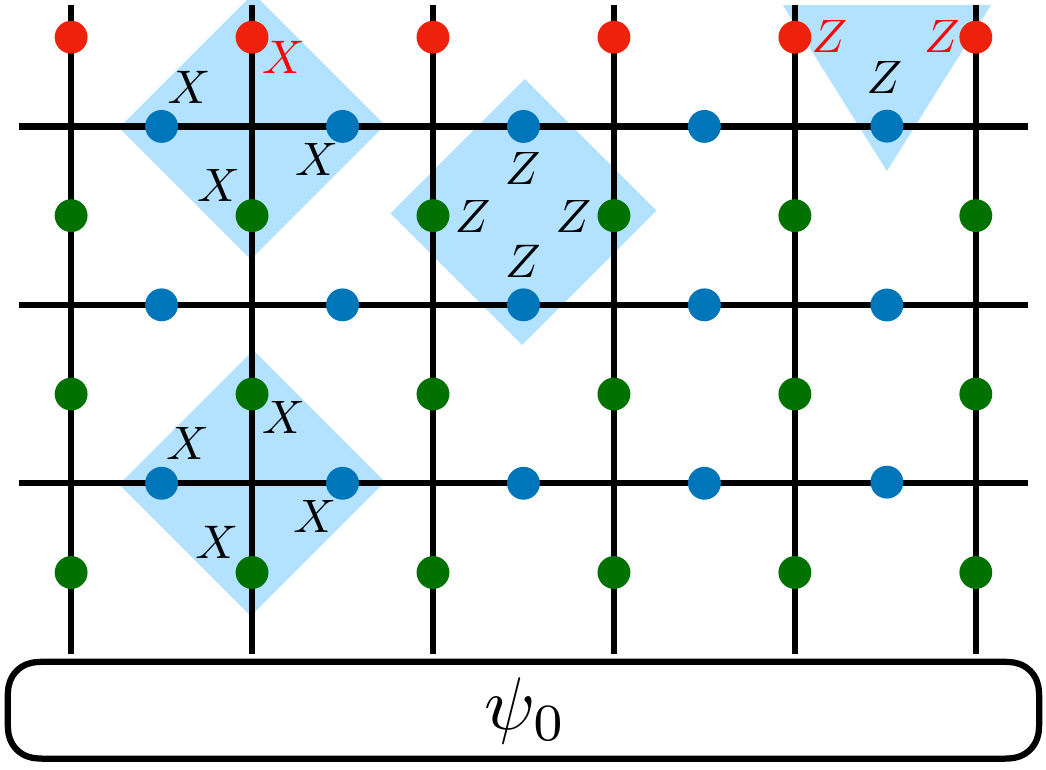}
    \caption{A repeated quantum channel can be viewed as a sequential unitary circuit where a 1d system initialized at $\ket{\psi_0}$ propagates upward and sequentially interacts row by row with ancilla qubits (shown in green and blue). Tracing out the green and blue qubits implements the X and ZZ noise channels, respectively. The reduced density matrix on the 1d top boundary  (colored in red) represents the output of the repeated quantum channel. At $p_z = p_x = \frac{1}{2}$, the bulk has the fixed-point toric-code topological order, and the top boundary is the SW-SSB state, i.e. $\propto (1 + \prod_i X_i) $; see Appendix.\ref{appendix:toric_code_derivation} for derivation.}  
    \label{fig:2d_toric}
\end{figure}

To develop a holographic description, we observe that the 1d noise channel ($ \cE = \prod_i \cE^z_{i} \cE^x_{i}$) can be obtained by entangling system qubits with ancilla qubits on a 1d geometry, followed by tracing out the ancilla qubits. Therefore, the repeated application of the 1d channel, when including both system and ancilla qubits as a whole, can be understood as a sequential unitary circuit where the system qubits propagate and sequentially interact with the ancilla qubits that are arranged on a 2d lattice (Fig.\ref{fig:2d_toric}). In the language of tensor network formalism, the system qubits live in the bond space and the ancilla qubits live in the physical legs of a tensor network. Notably, as we will now show, the sequential circuit that induces the SW-SSB on the system qubits in fact prepares a 2d topological order of those ancilla qubits, thereby establishing an intriguing connection between SW-SSB and the topological order in one higher spatial dimension.

To establish the correspondence, we first discuss the sequential circuit that implements the noise channel in Eq.\ref{eq:noise}. 

\textbf{\textit{ZZ channel}}: to realize the ZZ noise channel acting on two system qubits (say labeled by 1 and 2), we can first initialize an ancilla qubit (labeled by a) in the state $\sqrt{1-p_z} \ket{+}_a + \sqrt{p_z} \ket{-}_a$, so that the system qubits and the ancilla qubit together is described by 

\begin{equation}
\ket{\psi} (\sqrt{1-p_z} \ket{+}_a + \sqrt{p_z} \ket{-}_a),  
\end{equation}
where $\ket{\psi}$ denotes the state of the system qubits. 

Now we apply the controlled gate $u_{zz} = (\ket{+}\bra{+})_a + (\ket{-}\bra{-})_a Z_1Z_2$, resulting in the state 
\begin{equation}
 \ket{\psi_{p_z}} =\sqrt{1-p_z} \ket{\psi}\ket{+}_a + \sqrt{p_z}  Z_1Z_2 \ket{\psi}  \ket{-}_a.   
\end{equation}
Tracing out the ancilla qubit $a$ then realizes the desired Z-type noise channel (Eq.\ref{eq:noise}) on the system qubits, where $Z_1Z_2$ is applied with probability $p_z$. 

We note that when $p_z=\frac{1}{2}$, i.e. maximal decoherence, the ancilla qubit is initialized at $\ket{0}_a$, which is stabilized by $Z_a$, and the subsequent application of $u_{zz}$ generates $Z_1Z_aZ_2$ stabilizer with eigenvalue +1. Away from $p_z=\frac{1}{2}$, the ancilla qubit is initialized at $\sqrt{1-p_z} \ket{+}_a + \sqrt{p_z} \ket{-}_a$, which can also be written as 

\begin{equation}\label{eq:z_deform}
 e^{g_z X_a} \ket{0} \propto  e^{g_z } \ket{+}+  e^{-g_z } \ket{-},
\end{equation}  
with
\begin{equation}\label{eq:g_z}
e^{2g_z } =  \frac{1-p_z}{p_z}.
\end{equation}
Since the imaginary time evolution by $ e^{g_z X_a}$ commutes with the controlled gate $u_{zz}$, the pure state $\ket{\psi_{p_z}   }$ (describing both the system qubits and ancilla qubits) for general $p_z$ can be obtained by applying an imaginary time evolution on the pure state at $p_z = \frac{1}{2}$, in which case the ancilla qubit is initialized at $\ket{0}$:

\begin{equation}
   \ket{\psi_{p_z}} \propto e^{g_z X_a}  \ket{\psi_{p_z=\frac{1}{2}}}  
\end{equation}

This relation will be useful when we investigate the state of the ancilla qubits later.

\textbf{\textit{X channel}}: to realize the X noise channel acting on a system qubit  (labeled by $s$), we can first initialize an ancilla qubit (labeled by a) in the state $\sqrt{1-p_x} \ket{0}_a + \sqrt{p_x} \ket{1}_a$, so that the state is given by 

\begin{equation}
\ket{\psi} (\sqrt{1-p_x} \ket{0}_a + \sqrt{p_x} \ket{1}_a). 
\end{equation}
Now we apply the controlled gate $ (\ket{0}\bra{0})_a + (\ket{1}\bra{1})_a X_s$, resulting in the state 
\begin{equation}
 \ket{\psi_{p_x}} =\sqrt{1-p_x} \ket{\psi}\ket{0}_a + \sqrt{p_x}  X_s \ket{\psi}  \ket{1}_a.   
\end{equation}
Tracing out the ancilla qubit $a$ then realizes the desired X-noise channel on the system qubits, where $X_s$ is applied with probability $p_x$. 

At $p_x=\frac{1}{2}$, i.e. maximal decoherence, the ancilla qubit is initialized at $\ket{+}_a$. Away from $p_x=\frac{1}{2}$, the ancilla qubit is initialized at $\sqrt{1-p_x} \ket{0}_a + \sqrt{p_x} \ket{1}_a$, which can also be written as 

\begin{equation}\label{eq:x_deform}
 e^{g_x Z_a} \ket{+} \propto  e^{g_x } \ket{0}+  e^{-g_x } \ket{1},
\end{equation}  
with
\begin{equation}\label{eq:g_x}
e^{2g_x } =  \frac{1-p_x}{p_x}.
\end{equation}
Therefore, the pure state $\ket{\psi_{p_x}}$ for general $p_x$ can be obtained by an imaginary time evolution acting on the pure state at $p_x= \frac{1}{2}$, in which case, the ancilla qubit is initialized at $\ket{+}$:

\begin{equation}
   \ket{\psi_{p_x}} \propto e^{g_x Z_a} \ket{\psi_{p_x=\frac{1}{2}}}  
\end{equation}

\textbf{Repeated quantum channel:} with the above ingredients, one can consider a 2d lattice with one ancilla qubit per link: the ancilla qubits on the horizontal ($x$) links and vertical ($y$) links are responsible for the ZZ channels and X channels, respectively. The repeated quantum channel can then be understood as a process where the system qubits on 1d move upward to sequentially interact with these ancilla qubits row by row (Fig.\ref{fig:2d_toric}). The output of the system qubits under the repeated quantum channel lives on the top boundary of the 2d lattice, which can be obtained by tracing out the ancilla qubits. Notably,  without tracing out the ancilla qubits, the pure state generated by the sequential circuit is a deformed toric-code wave function (see Appendix.\ref{appendix:toric_code_derivation} for derivation):

\begin{equation}\label{eq:deformed_2d_toric}
|\Psi^{2d}(g_x,g_z)\rangle=\prod_{\ell \in x\,\text{link}}  e^{g_z X_{\ell }} \prod_{\ell \in y\,\text{link}}  e^{g_x Z_{\ell }} \ket{\text{TC}}, 
\end{equation}
where the deformation strengths $g_z$ and $g_x$ are fixed by the noise rates $p_z$ and $p_x$ via Eq.\ref{eq:g_z} and Eq.\ref{eq:g_x}. 

The maximal decoherence, i.e. $p_x=p_z =\frac{1}{2}$, corresponds to vanishing string tensions $g_z=g_x=0$, leading to the fixed-point toric code.  This can be checked explicitly by mapping the stabilizers of the initial ancilla state to the toric-code stabilizers through the sequential circuit that implements the quantum channel, as shown in Appendix.\ref{appendix:toric_code_derivation}. Notably, the bulk toric-code topological order vanishes only when $g_z=\infty$ or $ g_x=\infty$. This can be checked by computing the wavefunction overlap $\braket{\Psi^{2d}(g_x,g_z)|\Psi^{2d}(g_x,g_z)}$, which is only singular when sending $g_z$ or $g_x$ to infinity (see Appendix.\ref{sec:transition}). This guarantees that no phase transition occurs when \(g_z\) or \(g_x\) is tuned to any finite value, and the wavefunction stays in the \(\mathbb{Z}_2\) topologically ordered phase.

This contrasts with the isotropic deformation $e^{g \sum_{\ell} Z_\ell}\ket{\text{TC}}$ \cite{Fradkin_deformed_2007,Chamon_state_2008}, which allows for e-charge condensation at a finite $g$. In that case, the wavefunction overlap maps to the partition function of the 2d Ising model, whose finite-temperature transition corresponds to a finite $g$ transition. In the holographic picture, the robustness of the topological order in our deformed toric-code wavefunction (Eq.\ref{eq:deformed_2d_toric}) implies that the 1d SW-SSB steady state persists for any nonzero \(p_z, p_x\) in the infinite-depth limit of the repeated channel.

We now discuss more details about the holographic wavefunction above. We will focus on the discussion on the fixed-point limit $g_x= g_z = 0$. All universal features will apply equally to the deformed toric code (with finite $g_x, g_z$) since they belong to the same topologically ordered phase. At $p_x= p_z= \frac{1}{2}$, as shown in Eq.\ref{eq:deformed_2d_toric}, the holographic wavefunction in 2d is a fixed-point toric code, with the stabilizers specified in Fig.\ref{fig:2d_toric}. In particular, since the 1d system acted by the repeated quantum channel is defined on a ring (namely, the periodic boundary condition), the toric code has the topology of a cylinder with periodic boundary conditions along the $x$ (horizontal) direction and open boundaries along the $y$ (vertical) direction. The bottom boundary and the top boundary are the input and output of the repeated quantum channel, respectively. In particular, the ZZZ type stabilizer (shown in the upper right of Fig.\ref{fig:2d_toric}), together with the $Z^{\otimes 4}$ stabilizer around any plaquette in the bulk, implies the existence of the following open string stabilizer:

\begin{equation}\label{eq:2d_toric_string}
\includegraphics[width=5.5cm]{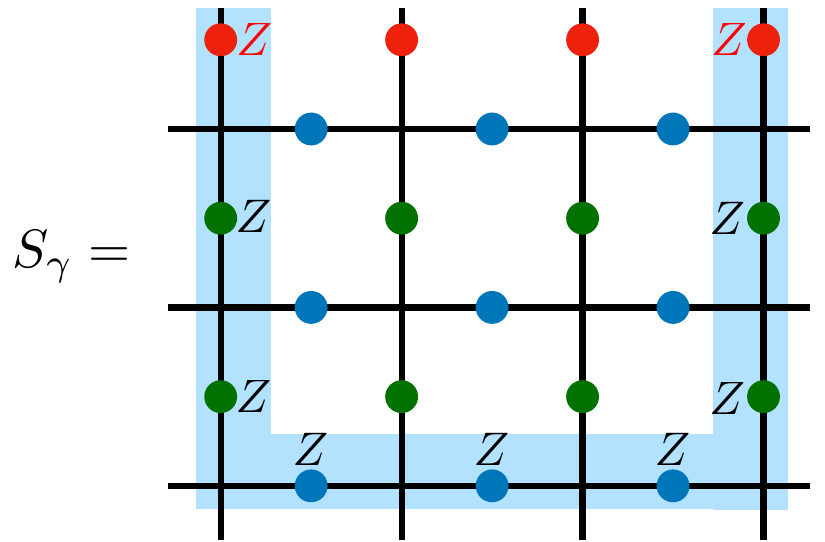},
\end{equation} 
where $\gamma$ can be freely deformable. The physical meaning of this stabilizer can be understood as follows: an open $Z$ string operator in the bulk is responsible for creating two $e$ particles on the two vertices on the endpoints of the string. As such, $S_\gamma$ can be imagined as a process in which a pair of $e$ particles is dragged to the boundary and condenses. In other words, the top 1d edge is specified by the rough boundary condition, interpretable as an $e$-condensed boundary. The condensation of pairs of $e$ charges on the boundary is essential since when tracing out the bulk, the symmetry operator $S_\gamma$ of the 2d pure state becomes a weak symmetry of the reduced density matrix $\rho$ on the top boundary:
\begin{equation}
Z_i Z_j  \rho Z_i Z_j= \rho.
\end{equation}

There is also a stabilizer $\prod_{i \in \text{top edge}} X_i$ on all the red qubits on the top boundary.

\begin{equation}
\prod_{i \in \text{top edge}} X_i |\Psi^{2d}(g_x,g_z)\rangle=  |\Psi^{2d}(g_x,g_z)\rangle
\end{equation}
Note that this equation holds for any finite $g_x, g_z$. The existence of this global symmetry acting on the 1d line is due to the fact that the input state has a fixed $Z_2$ charge, which is preserved throughout the channel dynamics. As such, when tracing out the bulk qubits, leaving on the red qubits, the corresponding reduced density matrix $\rho$ (i.e., the steady state of the quantum channel) has $\prod_i X_i$ as a strong symmetry:

\begin{equation}
\prod_i X_i \rho = \rho.    
\end{equation}

Without tracing out the bulk qubits, this symmetry measures the global flux threading the cylinder and generates the symmetry associated with the $m$-loop, so that the toric-code topological order is defined on a fixed symmetry sector.

Before moving to the next subsection, we make one brief remark here. The holographic correspondence (or more broadly, the bulk-boundary correspondence) is also known in the context of measurement-based quantum computation (MBQC) \cite{one_way_2001_mbqc,mbqc_rbh_2003,mbqc_2009,van2006universal}, in which a spacetime quantum circuit acting on a system in $d$-spatial dimensions is realized by measuring all the bulk qubits from a resource state in $(d+1)$ spatial dimensions. In particular, to implement error correction for a CSS code in $d$-spatial dimensions, one can employ the procedure known as `foliation' to construct a corresponding cluster state in one higher dimension, so that measuring its bulk implements the error correction of the $d$-dimensional CSS code \cite{Foliated_2016}. In that setting, there is a holographic correspondence between the bulk cluster state that exhibits a symmetry-protected-topological (SPT) order and the boundary CSS code that exhibits a topological order, as discussed in great details in Ref.\cite{foliation_anomaly_2024}. We emphasize that such a correspondence is quite different from ours in terms of the actual physical contents. In particular, MBQC focuses on the correspondence between boundary pure-state topological orders and bulk SPT orders through projective measurements, whereas ours reveals the correspondence between boundary mixed-state orders and bulk topological orders through partial trace; see Appendix.\ref{sec:mbqc} for more detailed discussions.

\subsection{Bulk-boundary correspondence from generalized symmetries} \label{sec:genesym}

The concept of generalized symmetries and their anomalies provides a powerful framework for understanding phases of matter (including both pure states and mixed states). In this subsection, we apply such a framework to discuss the correspondence between the 2d bulk topological orders and the boundary SW-SSB mixed-state orders. To begin, we briefly review the notions of generalized symmetry and anomaly to introduce the relevant terminology. We refer readers to Ref.\cite{McGreevy2022_review} for a more comprehensive review.  

In $d$-spatial dimensions, one can define a $p$-form symmetry $S$, which is supported on any $(d-p)$-dimensional closed spatial manifold. One can define another $(d-p)$-form symmetry $G$ that together with $S$ exhibits a `mutual anomaly’. This anomaly manifests in the fact that even though $S$ and $G$ commute, they exhibit a certain non-trivial operator algebra: an $S$ defect obtained by truncating $S$ symmetry must be charged under $G$ symmetry. As a simple example, we can consider a 1d system of $L$ qubits with periodic boundary conditions and define

\begin{equation}\label{eq:1d_anomaly}
\begin{split}
    \mathbb{Z}_2 ~\text{0-form symmetry} ~S: ~~ &\prod_{i=1}^L X_i\\ 
    \mathbb{Z}_2 ~\text{1-form symmetry} ~G:  ~~&Z_i Z_j ~ \forall i,j
\end{split}    
\end{equation}
Obviously, they commute, but yet, the $S$-defect is charged under $G$ symmetry, and the $G$-defect is charged under $S$ symmetry. To see this, one can restrict $S$ to a contiguous interval $\prod_{i \in \mathcal{M}} X_i$, so that it generates two defects on the boundary of $\mathcal{M}$. In particular, this partial symmetry action anti-commutes with the 1-form symmetry $Z_i Z_j$ when $i,j$ lie in and outside of $\mathcal{M}$, respectively. This indicates that the boundary defects created by $\prod_{i \in \mathcal{M}} X_i $ are charged under $G$. Importantly, when considering a pure state $\ket{\psi}$ with strong $G\times S$ symmetry, the mutual anomaly prevents $\ket{\psi}$ from being short-range entangled. Namely, $\ket{\psi}$ cannot be created by any finite-depth local unitary circuit; see e.g. Ref. \cite{yoshida_anomaly_2025} for proof. Indeed, Eq.\ref{eq:1d_anomaly} implies that $\ket{\psi}$ is simply the GHZ state, which exhibits spontaneous symmetry-breaking (SSB) long-range order. Notably, the algebra between $S$ and $G$ is a circuit invariant, so the entire SSB phase results from the same mutual anomaly. In fact, it is widely believed that all SSB gapped phases can be understood as a manifestation of mutual anomaly \cite{pace_2024_ssb}. We also note that the $S\times G$ mutual anomaly implies an obstruction for the symmetry gauging procedure (see Ref.\cite{McGreevy2022_review} and the references therein). 

On the other hand, when considering $S$ as a strong symmetry and $G$ as a weak symmetry of a mixed state $\rho$, the strong-weak anomaly gives rise to the SW-SSB order of $\rho$, where $S$ is spontaneously broken down to the weak one, and $G$ is the corresponding order parameter \cite{sang2025mixed}. For instance, SW-SSB of the state $\rho \propto \frac{1+\prod_i X_i}{2}$ can be understood as the mutual anomaly between a strong $S$ symmetry and a weak $G$ symmetry:

\begin{align}\label{eq:sym}
&S: (\prod_i X_i) \rho =\rho \nonumber\\
&G: (Z_i Z_j) \rho (Z_i Z_j) =\rho.
\end{align}

Moreover, Ref.~\cite{sang2025mixed} proves that this mutual anomaly is preserved in the entire generic SW-SSB mixed-state quantum phase, thereby providing a useful symmetry-based perspective to understand the universal features of SW-SSB. The proof and characterization of the mixed-state phases are beyond the scope of our work, so we refer readers to Ref.\cite{sang2025mixed} for more details. Below, we show that, under our holographic duality, the mutual anomaly of a 1d SW-SSB mixed state can be naturally understood as the mutual anomaly between two strong 1-form symmetries in a topologically ordered bulk when they are pushed to the 1d boundary. For that, we first discuss the $\mathbb{Z}_2$ topological order from the anomaly perspective in the following.

\begin{figure}[t]
    \centering
\includegraphics[width=0.26\textwidth]{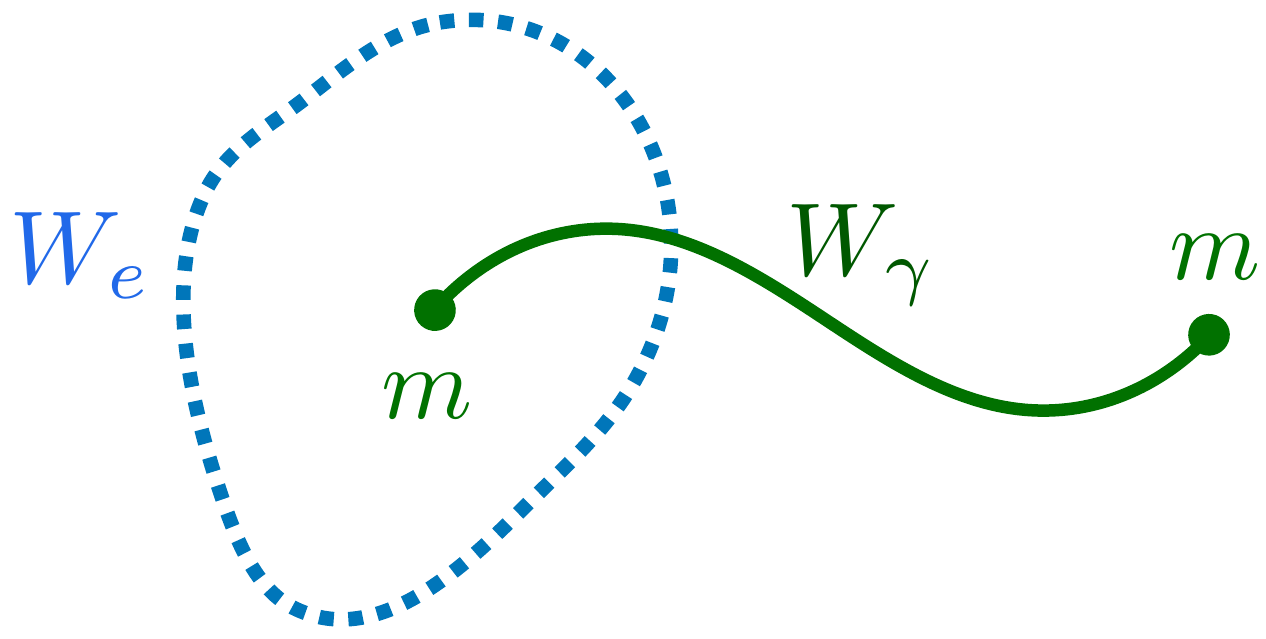}
    \caption{ $W_\gamma$ creates a pairs of $m$ anyons at the endpoints of the string $\gamma$. The $m$ anyon carries a non-trivial charge, which can be detected by $W_e$ 1-form symmetry (marked by the blue dashed line) that encloses it. The resulting $-1$ phase is known as the mutual braiding statistics in the $\mathbb{Z}_2$ topological order.}
    \label{fig:braiding}
\end{figure}

A universal feature of the 2d $\mathbb{Z}_2$ topological order is a $1$-form symmetry $W_m$  (associated with $m$-flux loops) and a $1$-form symmetry $W_e$ (associated with $e$-charge loops), which together carry the mutual anomaly. To see this, we can again follow the general procedure described above: we first truncate $W_m$ to define an open string operator $W_\gamma$, where $\gamma$ denotes an open string; see Fig.\ref{fig:braiding}. The two endpoints of $\gamma$ correspond to the $m$ anyons of the toric code. Importantly, the $m$ anyon is charged under the $W_e$ 1-form symmetry that encloses it. In algebraic terms, one has 

\begin{equation}
W_e W_\gamma W_e^{\dagger} = -1.   
\end{equation}

The mutual anomaly manifests in the $-1$ phase, which is also known as the braiding statistics of $W_m$ and $W_e$ 1-form symmetries. In particular, the braiding statistics is an invariant throughout the $\mathbb{Z}_2$ topologically ordered phase, and can be used to prove that the corresponding state cannot be prepared by finite-depth local unitary circuits \cite{haah_S_matrix_2016,yoshida_anomaly_2025}. Below we show that pushing the $W_m, W_e$ symmetries to the 1d boundary leads to the anomaly between the strong and weak symmetries in Eq.\ref{eq:sym} responsible for the 1d SW-SSB mixed-state order.

When defining the topological order on a cylinder, a contractible $m$-flux loop in the bulk can be deformed into a pair of non-contractible loops $W_{m, \text{bottom}}W_{m, \text{top}}$, with $W_{m, \text{top}}$ winding around the top boundary and $W_{m, \text{bottom}}$ winding around the bottom boundary. In particular, if the 2d topological order is an eigenstate of the $m$-loop operator on a cylinder, then $W_{m, \text{bottom}}$ and $W_{m, \text{top}}$ will individually carry a definite eigenvalue, so the boundary reduced density matrix exhibits the strong $S$ symmetry. 

Meanwhile, a 1-form $e$-charge loop in the bulk can be pushed to the top boundary. Assuming an $e$-condensed boundary, it then becomes an open string operator terminating at two points on the 1d boundary; see Eq.~\ref{eq:2d_toric_string}. This open $e$-string operator remains a symmetry of the full 2d pure state. Physically, it allows a pair of charges to be created in the bulk, moved to the edge, and condensed there.
Equivalently, if the edge qubits are identified with the system degrees of freedom while the bulk qubits constitute the ancilla, this open string operator describes a process in which a pair of charges tunnels from the system into the ancilla and then annihilates. This is precisely the physical mechanism underlying SW-SSB. Notably, after tracing out the bulk degrees of freedom, this open-string symmetry of the 2d pure state reduces to the weak $G$ symmetry of the 1d boundary mixed state. To summarize, given a bulk topological order with a mutual anomaly between two 1-form symmetries, pushing those symmetries to the 1d boundary gives rise to a mutual anomaly between a strong 0-form symmetry and a weak 1-form symmetry, which is responsible for the SW-SSB in the 1d mixed state.  

We conclude this subsection with two remarks:

(1) In Sec.~\ref{sec:1d_channel}, the 2d topological order arises from a 1d quantum channel because this channel can be viewed as a sequential unitary circuit that entangles the 1d system qubits with ancilla qubits in 2d and prepares a specific wavefunction with 2d topological order. By contrast, a generic topologically ordered wavefunction with a finite correlation length is not necessarily exactly preparable by a sequential unitary circuit. Rather, its preparation may require using a sequential \textit{non-unitary} circuit; see Sec.\ref{sec:beyond_channel} for an example. As such, the associated evolution on the 1d system is not trace-preserving and thus does not define a valid quantum channel. In other words, not every higher-dimensional wavefunction can be mapped back to a lower-dimensional quantum channel. Nevertheless, the symmetry-based perspective discussed here does not rely on such precise channel-state duality. Any non-fixed-point $\mathbb{Z}_2$ topological order always possesses two emergent 1-form symmetries, which can acquire a finite dressing width~\cite{hastings2005quasiadiabatic,sal:2023,Liu:2025_1form}. The reduced density matrix on the boundary, whose thickness is set by the dressing length of these emergent 1-form symmetries, is therefore expected to exhibit SW-SSB based on our aforementioned discussion. This observation points to the robustness and generality of the holographic correspondence beyond steady states generated by quantum channels; one such example will be discussed in Sec.~\ref{sec:beyond_channel}.

(2) Based on our discussion of the symmetry anomaly, a topologically ordered bulk implies the boundary SW-SSB mixed-state order. However, we can ask whether the converse also holds. Namely, does a boundary SW-SSB mixed state necessarily imply a bulk topological order? The answer is negative in general, since a given SW-SSB mixed state can arise as the steady state of infinitely many quantum channels, some of which may possess highly fine-tuned structures unrelated to the strong symmetry itself. Nevertheless, in Sec.~\ref{sec:iso}, we argue that for sufficiently generic quantum channels with the strong symmetry of interest, and without additional fine-tuned structure beyond this symmetry, the bulk states obtained through the channel–state duality are topologically ordered. In this way, SW-SSB mixed states can be naturally associated with bulk topological order through their realization as steady states of generic strongly symmetric quantum channels.

\subsection{Conditional mutual information}
\begin{figure}[t]
    \centering
\includegraphics[width=0.4\textwidth]{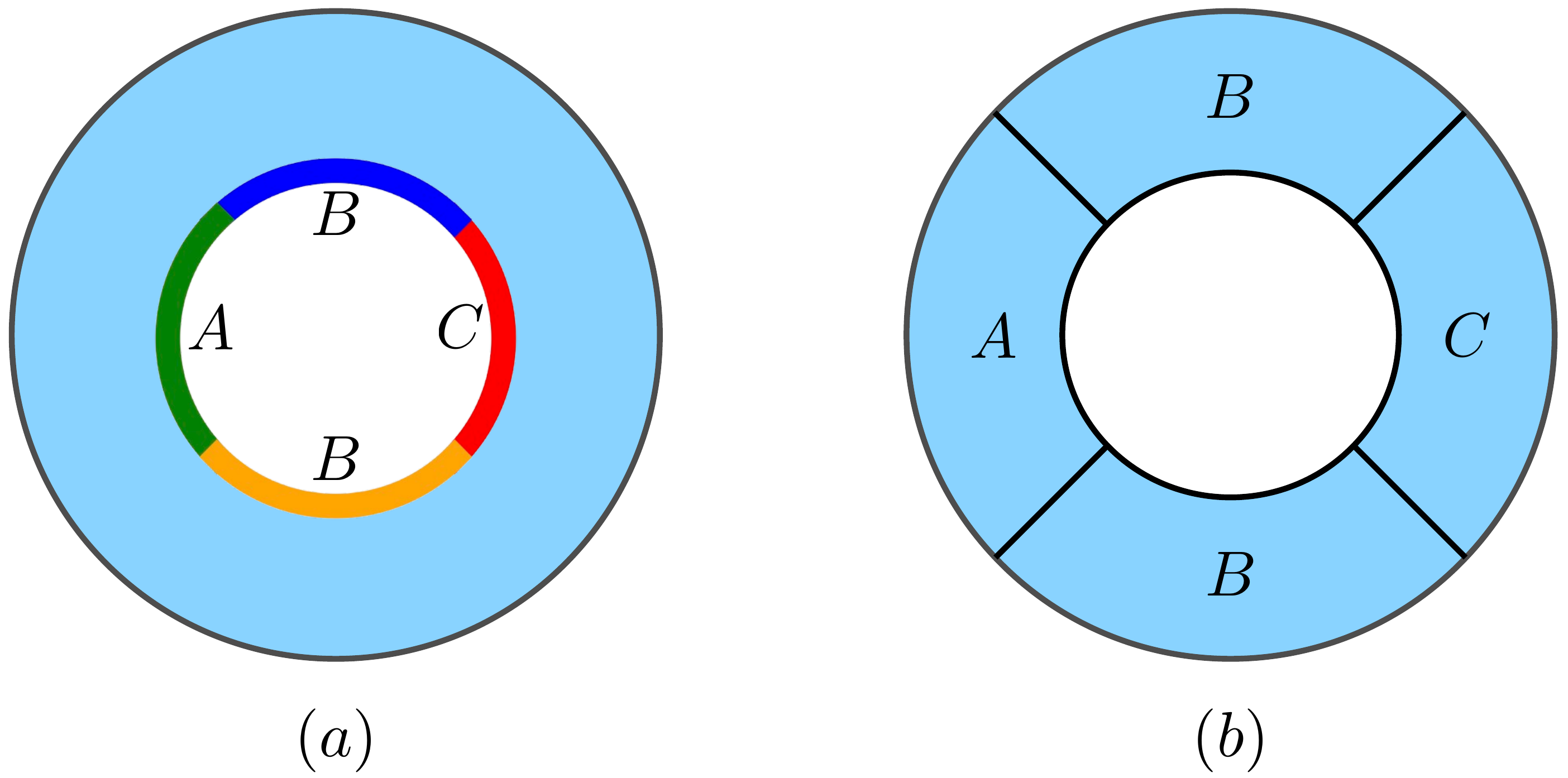}
    \caption{(a) The inner and outer circles represent the top and bottom boundaries of a toric-code state defined on a cylinder. The boundary reduced density matrix supported on the inner circle (corresponding to the 1d steady-state) defines a 1d SW-SSB mixed state with nonzero conditional mutual information $I(A:C|B)$. 
(b) Conventional setup for diagnosing the CMI, or equivalently the topological entanglement entropy, of the 2d topological ordered state. Here the region $ABC$ is chosen as a finite-width buffer strip around the boundary, with width larger than the correlation length. 
The nonzero CMI in both configurations arises from the mutual anomaly between the two emergent 1-form symmetries of the two-dimensional toric-code topological order.}
    \label{cmi}
\end{figure}

The holographic perspective also presents a unified framework for diagnosing SW-SSB and topological order based on certain information-theoretic quantities. Concretely, given an SW-SSB state on 1d, the conditional mutual information (CMI) between the two distant regions \( A \) and \( C \) conditioned on their complement $B$ is non-vanishing \cite{lessa2025strong} (see Fig.\ref{cmi}(a)): 
\begin{equation}
\begin{split}
    &I(A:C|B)\\
    &\equiv  S(AB) + S(BC)-S (B) - S(ABC) >0.     \end{split}
\end{equation}

As a sanity check, one may consider $\rho \propto 1+\prod_i X_i$, and explicitly find $ I(A:C|B) = \log 2$. The non-vanishing CMI implies the mixed state is non-Markovian; when tracing out a subregion, the global correlations in the mixed state cannot is lost and cannot be locally reconstructed.

Interestingly, the non-vanishing CMI in an SW-SSB mixed state can be understood from the aspect of symmetry anomaly discussed in the previous subsection. Specifically, this mutual anomaly between a strong 0-form symmetry and a weak 1-form symmetry in 1d implies a nonzero CMI, as discussed in Ref.~\cite{lessa2025strong,lessa2025higher}. Within our holographic picture, such a mutual anomaly in the 1d system originates from the mutual anomaly between two distinct types of strong 1-form symmetry of the 2d topologically ordered bulk, which implies a non-zero CMI (or equivalently, the topological entanglement entropy) under the partitioning shown in Fig.\ref{cmi}(b) \cite{tee_lower_bound_2023_kim,levin_TEE_2024,lessa2025higher}. Therefore, the phenomenon of the non-vanishing CMI in the 1d SW-SSB and the non-vanishing CMI in 2d topological order, albeit defined via distinct partitioning, shares a common origin based on a symmetry-anomaly perspective, which admits a natural interpretation with our holographic framework. Finally, we note that while the above holographic approach provides a unified physics perspective between the CMIs in the bulk and boundary, we are not able to establish a rigorous relation between the two. For instance, it is known that the topological entanglement entropy can suffer from a spurious contribution \cite{haah16,williamson2019spurious,kato_2020}: it can be non-zero even when a bulk wavefunction is not topologically ordered, in which case, the boundary needs not be in the SW-SSB phase with a non-zero CMI. On the other hand, there can be certain fine-tuned examples with additional symmetries (see e.g. Appendix.\ref{sec:pim}), where the boundary exhibits SW-SSB with a non-zero CMI, but the bulk does not exhibit a topological order with a non-zero CMI. As a result, while the holographic framework provides a physically appealing perspective for connecting certain information-theoretic quantities in the bulk and boundary, having a rigorous connection appears to be difficult.

\color{black}

\subsection{Fidelity correlator} 
The fidelity correlator defined in Eq.\ref{eq:fid} provides a faithful diagnostic of SW-SSB: 
\begin{equation}
F_{ij}=\mathrm{Tr}\left[\sqrt{\sqrt{\rho}, O_i O_j \rho O_i^\dagger O_j^{\dagger} \sqrt{\rho}}\right],
\end{equation}
where $O_i$ is a charged operator localized at point $i$. For instance, when the global strong symmetry is $\prod_i X_i$, one may choose $O_i=Z_i$. Physically, $F_{ij}$ detects the existence of a weak symmetry. This can be manifested by the fixed-point SW-SSB state $\rho \propto 1+ \prod_i X_i$, where the weak symmetry condition $Z_i Z_j \rho Z_i Z_j = \rho$ directly implies $F_{ij} =1$. In particular, a long-range order in $F_{ij}$ signals a divergent Markov length, implying that the mixed state is not locally recoverable and exhibits nonzero conditional mutual information \cite{lessa2025strong}. 

Here we comment that a bulk $1$-form charge symmetry that yields an $e$-condensed boundary provides a lower bound on the fidelity correlator. Suppose the repeated quantum channel admits a dual bulk wavefunction $|\Psi_{2d}\rangle$ whose boundary reduced density matrix is $\rho$. If the bulk has a $1$-form symmetry with an $e$-condensed edge, there exists an open Wilson string operator $W$ (e.g. the string operator in Eq.\ref{eq:2d_toric_string}) in the bulk that extends to and terminates at boundary points $i,j$, with a nonvanishing expectation value $\langle\Psi_{2d}|W|\Psi_{2d}\rangle = O(1)$. This Wilson string terminates on the edge at $i,j$, and its boundary endpoints are represented by operators $O_i,O_j$ that—while not fixed microscopically—must be charged under the $X$ $1$-form symmetry.

Since $|\Psi_{2d}\rangle$ is a purification of the boundary density matrix $\rho$, the state $W|\Psi_{2d}\rangle$ is a purification of $O_i O_j \rho O_i^\dagger O_j^\dagger$. By Uhlmann’s theorem~\cite{uhlmann1976transition}, the fidelity between two density matrices equals the maximum overlap over all of their purifications. Hence,
\begin{align}
|\langle\Psi_{2d}|W|\Psi_{2d}\rangle| \le \mathrm{Tr}\left[\sqrt{\sqrt{\rho}, O_i O_j \rho O_i^\dagger O_j^\dagger \sqrt{\rho}}\right],
\end{align}
Therefore, the bulk Wilson-line expectation value $|\langle \Psi_{2d}|W|\Psi_{2d}\rangle|$ provides a lower bound for the fidelity correlator.

Finally, we note that in the above discussion, we implicitly assume the Wilson-line operator $W$ factorized as $W=  O_i O_j W_{\text{bulk}}$, with $W_{\text{bulk}}$ being an operator supported in the bulk. This condition is essential so that $W\ket{\Psi_{2d}}$ is a valid purification of $O_i O_j  \rho O_i O_j$. It remains unclear whether one can lower bound the fidelity correlator on the 1d top boundary when the above condition is relaxed.

\subsection{Necessity of infinite-depth for SW-SSB in 1d}\label{sec:nece}
In the previous discussion, we have focused mainly on the steady state of the channel defined in Eq.\ref{eq:noise}, namely the mixed state $\rho\propto 1+ \prod_i X_i$ that exhibits SW-SSB. An important question is: what is the depth required for the output of the repeated quantum channel to exhibit SW-SSB with a non-zero fidelity correlator? Consider the symmetric input state $\ket{+...}$ as an example, under a \textit{single} application of the $ZZ$ channel, the output exhibits SW-SSB only when $p_z=\frac{1}{2}$, in which case the output is simply $\rho\propto 1+ \prod_i X_i$. For any $p_z< \frac{1}{2}$, in order to develop a non-vanishing fidelity correlator, the depth of the corresponding quantum channel needs to be infinite. While this has been known in the literature \cite{lessa2025strong,sala2024spontaneous_published} via a mapping to 1d classical statistical models, below we provide a \textit{heuristic} argument for this phenomenon from a holographic perspective.

In our holographic duality framework, the depth of the repeated quantum channel is the size of the vertical (y) direction of the deformed toric-code wavefunction (Eq.\ref{eq:deformed_2d_toric}):

\begin{equation}
|\Psi^{2d}(g_x,g_z)\rangle=\prod_{\ell \in x\,\text{link}}  e^{g_z X_{\ell }} \prod_{\ell \in y\,\text{link}}  e^{g_x Z_{\ell }} \ket{\text{TC}}. 
\end{equation}

\begin{figure}[t]
    \centering
\includegraphics[width=0.32\textwidth]{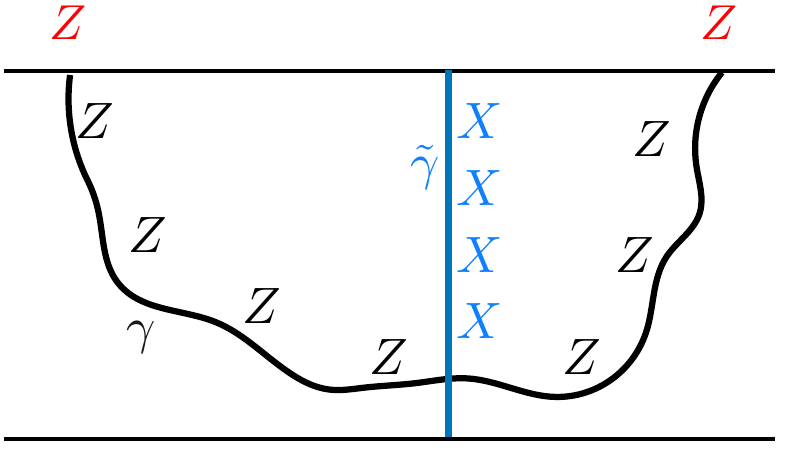}
    \caption{The formation of the Pauli X string extending from the top boundary to the bottom boundary will modify the expectation value of $S_\gamma$ string. Only when $L_y\to \infty$, the probability of such an X string will vanish, leading to the robustness of the $S_\gamma$ string, and therefore, the emergence of SW-SSB on the top boundary.}  
    \label{fig:2d_toric_intersection}
\end{figure}

At $p_z=\frac{1}{2}$, the holographic wavefunction is the fixed-point toric code, which has the deformable open $Z$ string operator $S_\gamma$ along any path $\gamma$ (Eq.\ref{eq:2d_toric_string}; see also Fig.\ref{fig:2d_toric_intersection} for the schematic), with two Pauli-Zs attached to the 1d top boundary. The symmetry operator $S_\gamma$ for the pure-state wavefunction gives the weak symmetry for the 1d reduced density matrix, and leads to the non-decaying fidelity correlator.

For the deformed toric code, by expanding the exponential $\prod_{\ell \in x\,\text{link}}  e^{g_z X_{\ell }}$, one sees that it creates various Pauli-X insertions on horizontal ($x$) links of the toric code state. Since the string operator $S_\gamma$  (Eq.\ref{eq:2d_toric_string}) can be deformed freely, its expectation will be modified only when the Pauli-X insertion form a string operator $W_{\tilde{\gamma}}$ that extends from the top boundary to the bottom boundary, in which case the intersection between $\gamma$ and $\tilde{\gamma}$ cannot be avoided (shown in Fig.\ref{fig:2d_toric_intersection}). What is the probability of forming such an extended string $\tilde{\gamma}$? To provide a heuristic estimation, we can first write $ e^{g_z X_{\ell }} \propto  1 + X_\ell\tanh g_z$, so the probability amplitude for a single-X insertion is $\tanh g_z$. It follows that the weight for forming a string extending from the top to the bottom goes as $(\tanh g_z)^{L_y}$. On the other hand, the endpoint of the vertical string on the top boundary can occur at any position between $i$ and $j$. Taking into account this multiplicity gives the probability weight $\sim   |i-j| (\tanh g_z)^{L_y} =   |i-j| (1-2p_z)^{L_y}$. For a large and fixed $|i-j|$, for any $0<p_z<\frac{1}{2}$, the probability of the extended string will vanish only at $L_y \to \infty$. This provides a heuristic argument for the fact that SW-SSB at the top boundary requires an infinite depth for $0<p_z<\frac{1}{2}$. On the other hand, the local channel can induce SW-SSB in 2d in finite depth since an ``error" membrane can be suppressed for a large but finite depth (see the example in Sec.\ref{sec:2d} and Appendix.\ref{appendix:finite_depth} for discussions). 

\section{Holographic duality from isoTNS}\label{sec:iso}

In this section, we elaborate on the holographic duality using the framework of isometric tensor-network states (isoTNS). This framework provides a generic tool to relate SW-SSB steady states and topological order in one higher dimension, offering the possibility to go beyond the simple example considered in the previous section.

We begin by first recalling the mapping between 2d isoTNS and repeated application of a quantum channel on a 1d system. The relation generalizes to higher dimensions straightforwardly. The isoTNS on a square lattice can be described by the local tensor with two physical legs and four virtual legs $T^{\sigma\gamma}_{ijmn}$, with the graphical notation
\begin{equation}
    \includegraphics[scale = 1]{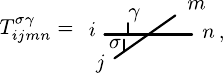}
\end{equation}
where the physical leg labels $\sigma,\gamma$ enumerate the local Hilbert space dimension, and the dimension of the virtual legs is referred to as the bond dimension of the tensor networks. By contracting the virtual legs of the adjacent local tensors, we obtain a wavefunction with physical sites on the edges of a square lattice (we adopt this convention from Refs.\cite{liu_iso:2024,boesl_iso:2025,Boesl:2025_stringnetiso}).
The local tensors of the isoTNS additionally satisfy the following isometry condition
\begin{equation}\label{eq:iso_condition}
    \sum_{\sigma,\gamma, m,n}\left(T^{\sigma\gamma}_{ijmn}\right)^*T^{\sigma\gamma}_{i'j'mn} = \delta_{ii'}\delta_{jj'},
\end{equation}
where $\delta$ denotes the Kronecker delta function. With the isometry condition, we can view the tensor $T$ as the Kraus operators acting on the virtual legs and mapping the legs $i,j$ to the legs $m,n$. This defines a quantum channel acting on the two-site density matrix defined on the virtual bond space: 
\begin{equation}\label{eq:channel}
\Lambda[\rho] = \sum_{\sigma,\gamma}T^{\sigma\gamma} \rho (T^{\sigma\gamma})^\dag.
\end{equation}
It can be graphically visualized as a double-stacked tensor $T$ acting on the virtual density matrix $\rho$:
\begin{equation}
    \includegraphics[scale = 1]{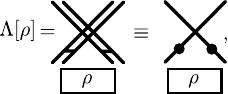}
\end{equation}
where we simplify the graphical notation at the last step. In particular, the trace-preserving condition of the channel $\Lambda$ follows from the isometry condition
\begin{equation}
\sum_{\sigma,\gamma} (T^{\sigma\gamma})^\dag T^{\sigma\gamma} = I,
\end{equation}
where $I$ denotes the identity operator. This shows that the contraction of the bulk of an isoTNS can be understood as a repeated application of the quantum channel $\Lambda$ on a density matrix defined on the virtual bond space. Conversely, the isoTNS is a particular purification of the repeated quantum channels (Stinespring dilation theorem). Schematically, this channel-state duality in 2d systems can be visualized as
\begin{equation}
    \includegraphics[scale = 0.5]{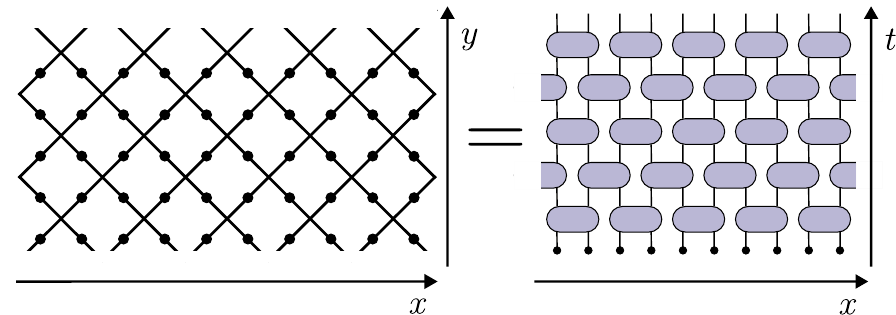}\nonumber.
\end{equation}
The left-hand side shows the 2d isoTNS, and the right-hand side shows the repeated quantum channel acting on a 1d system, where each purple oval denotes a local quantum channel defined by the isoTNS local tensor $T$. 

It is worth noting that repeated quantum channels are dual to isoTNS, which always represent ground states of frustration-free local Hamiltonians. Physically, the dual isoTNS ground state can be viewed as a superposition of the space–time trajectories generated by the dynamics of the corresponding quantum channel. More precisely, denote the repeated quantum channel induced by the transfer operator of the isoTNS as $\mathcal{E}[\rho] = \sum_i M^i\rho (M^{i})^\dag$, where for the isoTNS, $i$ denote the physical degrees of freedom (we group $\sigma,\gamma$ in Eq.\eqref{eq:channel} along each row into $i$). Suppose the dynamics is initialized with a pure state $\ket{\psi_0}$, then the isoTNS takes the form $\sum_{\{i\},v} c(\{i\},v) |\bra{v}\cdots M^{i_3}M^{i_2}M^{i_1} \ket{\psi_0}|\ket{i_1,i_2,i_3,\cdots, v}$, where $v$ label the virtual degrees of freedom and $c(\{i\},v)$ is some complex phase. Here, we focus only on the unraveling along the temporal direction and ignore the internal structure within each row; those rows encode spatial information, which can be unraveled in an analogous way. In this sense, the isoTNS wavefunction can be viewed as a superposition of all unraveled trajectories of the quantum channel, each weighted by the (square root of) corresponding trajectory probability and an associated complex phase. This duality thus establishes an explicit connection between the properties of the quantum channel and the ground-state ordering in the many-body system.

\subsection{Forced nontrivialness of isoTNS from channel-state duality}\label{sec:forced_degeneracy}

A fundamental implication of the channel–state duality is that the strong symmetries of a repeated quantum channel correspond to the virtual symmetries of the isoTNS, which act exclusively on the virtual Hilbert space, spanned by the virtual degrees of freedom of the tensor network. This duality highlights the pivotal role of strong and virtual symmetries in determining, respectively, the emergence of SW-SSB in the channel picture and bulk topological order in the tensor-network picture.

Here, using the channel–state duality, we provide a physical argument that an isoTNS with a nontrivial global virtual symmetry necessarily has a parent Hamiltonian with nontrivial ground-state degeneracy on a cylinder. While such degeneracy alone is insufficient to guarantee topological order, we expect this to hold for sufficiently generic isoTNSs that are not fine-tuned and do not possess any additional accidental virtual symmetries. We will make the notion mathematically precise using the so-called \textit{$G$-injectivity condition}, which will be discussed in the next subsection (Sec.\ref{sec:main_g_injectivity}).

Our argument proceeds in two steps. First, we show that the coexistence of the local isometry condition and a nontrivial virtual symmetry guarantees that the transfer operator of the isoTNS has degenerate fixed points. Second, we argue that this fixed-point degeneracy implies ground-state degeneracy of the parent Hamiltonian.

We make use of the channel–state duality by first considering the effect of strong symmetries in quantum channels. The existence of a strong symmetry in a quantum channel necessarily implies a degenerate steady-state manifold: each symmetry sector of the strong symmetry must contain at least one steady state~\footnote{This follows directly from trace preservation~\cite{albert2014symmetries}. Given a quantum channel $\mathcal{E}$, viewed as a linear map acting on operators in the Hilbert space, trace preservation ensures that the identity operator $I$ is always a left eigenvector of $\mathcal{E}$ with eigenvalue 1 ($\mathcal{E}^{\dag}[I] = I)$. Let $P_\lambda$ denote the projector onto the eigensubspace of the strong symmetry $U$ with distinct eigenvalue $\lambda$, satisfying $\sum_s P_s = I$. Since each $P_\lambda$ commutes with $\mathcal{E}^{\dag}$, the operators $P_\lambda \cdot I$ form linearly independent left eigenvectors associated with the same eigenvalue 1. Consequently, there must exist the same number of right eigenvectors, i.e., steady states, with eigenvalue 1.}. Through the channel–state duality, this degeneracy maps to the isoTNS transfer operator, whose fixed-point degeneracy arises from a nontrivial virtual symmetry. In particular, since the virtual symmetry acts solely on the virtual legs while leaving the physical degrees of freedom unaffected, these degenerate fixed points correspond to bulk states with identical energy with respect to the frustration-free local parent Hamiltonian constructed directly from the tensor-network representation (e.g., as described in~\cite{schuch2010peps}). Since distinct fixed points indicate that the associated isoTNS wavefunctions are distinct, the parent Hamiltonian of the isoTNS, when defined on a cylinder, possesses degenerate ground states. The argument extends to isoTNS in higher spatial dimensions as well.

The forced nontrivialness directly leads to intriguing physical consequences. For example, if a nontrivial virtual symmetry is preserved, the parent Hamiltonian of the isoTNS cannot possess a unique trivial ground state. Similarly, in the dual channel picture, a strongly symmetric quantum channel cannot have a unique paramagnetic steady state~\cite{ziereis2025strong,gu2024spontaneous}. Importantly, both the isometry condition and the presence of a nontrivial virtual symmetry must coexist; violating either one immediately evades this constraint.

\subsection{Holographic correspondence from $G$-injectivity}\label{sec:main_g_injectivity}

As discussed in the previous section, we expect a generic isoTNS with virtual $G$ symmetry to exhibit topological order. This parallels the expectation that generic strongly $G$-symmetric quantum channels possess SW-SSB steady states. The notion of such “genericity” can be made precise using concepts from the tensor-network literature, particularly Refs.~\cite{schuch2010peps,schuch2011classifying}.

A central concept is $G$-injectivity. Mathematically, this condition requires that, when the local tensor is viewed as a map from the virtual legs to the physical leg, the map is injective within the $G$-symmetric subspace of the virtual Hilbert space. Physically, this means that the virtual $G$ symmetry is essentially the only symmetry present, excluding additional accidental virtual symmetries. The condition is therefore not particularly restrictive, and is expected to hold generically once no extra fine-tuned structure beyond the $G$ symmetry is imposed. In this sense, $G$-injectivity provides a precise formulation of the notion of “genericity” introduced above. A more detailed review of this condition is given in Appendix~\ref{sec:ginjective}.

Ref.~\cite{schuch2010peps} showed that the parent Hamiltonian of a $G$-injective TNS exhibits a $K$-fold ground-state degeneracy on the torus, where $K$ equals the number of anyon types in the quantum double model of $G$. While ground-state degeneracy alone does not fully establish $G$-topological order, the situation becomes stronger when the inverse map on the $G$-symmetric virtual subspace is given by the conjugate transpose. In this case, the corresponding tensor network is proven to be a fixed-point wavefunction of the $G$ quantum double topological phase.

Within our holographic framework, we show in Appendix~\ref{sec:ginjective} that such fixed-point wavefunctions are $G$-injective isoTNS, and therefore correspond to valid $G$-symmetric fixed-point quantum channels, namely completely positive and trace-preserving maps. Here, “fixed-point channel” means that, after finitely many iterations, the channel projects any initial state onto its steady-state manifold.
More generally, we expect generic $G$-injective isoTNS, with provable topological ground-state degeneracy, to be smoothly connected to these fixed-point representatives, and therefore to belong to the same topological phase.
Consequently, both the fixed-point and generic $G$-injective isoTNS provide a mathematically controlled realization of our holographic framework. From the dynamical perspective, they correspond to strongly $G$-symmetric quantum channels without additional fine-tuned structures beyond the symmetry itself, and therefore naturally realize SW-SSB steady states.

It is also instructive to mention a counterexample of holographic duality associated with the violation of the $G$ injectivity, in which case additional symmetries appear. This example is given by the channel considered in Sec.\ref{sec:1d} when choosing $p_x=0$ in the X-noise channel, so that only the ZZ-noise channel is applied. In this case, the number of steady states grows exponentially with system size; the boundary steady state needs not exhibit SW-SSB and the bulk need not exhibit topological order.  (see Appendix~\ref{sec:pim} for details).

To summarize, the channel-state duality offers a powerful tool that translates the tensor-network theoretic results to classify distinct quantum channels and their steady-state structures. In the next subsection, we will further employ such duality to construct a continuously-tuned quantum channel that exhibits a phase transition in the steady states.

\subsection{Continuously-tuned quantum channels and $U(1)$ SW-SSB critical point}

The isoTNS framework provides a useful tool to extend the holographic correspondence beyond the fixed-point states with zero correlation length, enabling the exploration of continuous deformations and phase transitions between distinct quantum phases and steady states. In particular, recent works have demonstrated that isoTNS with finite bond dimension can represent families of continuously varying ground states across a topological quantum phase transition~\cite{liu_iso:2024,boesl_iso:2025,Boesl:2025_stringnetiso}. Through the channel–state duality of isoTNS, such a topological quantum phase transition directly maps onto a continuously tuned transition in the corresponding repeated quantum channels and their steady-state manifolds.

To explicitly construct such a transition path, we employ the plumbing method \cite{liu_iso:2024}. In this approach, the local tensor $T$ is required to have the following form
\begin{equation}
    T^{\sigma\gamma}_{ijmn} = \sum_{i'j'} \delta^\gamma_{ii'}\delta^{\sigma}_{jj'} W_{i'j'mn},
\end{equation}
or pictorially, 

\begin{equation}\label{eq:plumbing}
    \includegraphics[scale = 0.35]{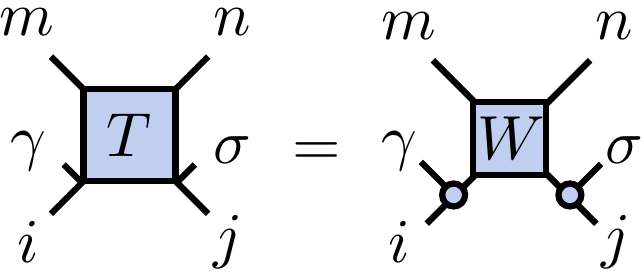}\nonumber,
\end{equation}

where $\delta$ =  \includegraphics[height=1em]{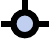} denotes the Kronecker delta such that it takes the value one when all indices agree, and zero otherwise. With this form, the tensor-network state is entirely specified by the $W$-tensor. The isometry condition (Eq.\ref{eq:iso_condition}) becomes a normalization condition on the $W$-tensor as $\sum_{mn}|W_{ijmn}|^2 = 1$ for all $i,j$. Due to the plumbing $\delta$, the virtual system coincides with the physical system. We introduce a parametrized $W$-tensor defined for $g\in [-1,1]$
\begin{equation}\label{eq:w_tensor}
W(g) \propto\left[ \begin{matrix}
1 & f(g) & f(g) & |g| \\
f(g) & \sqrt{\frac{1+g^2}{2}} &  \sqrt{\frac{1+g^2}{2}} & f(g) \\
f(g) & \sqrt{\frac{1+g^2}{2}} &  \sqrt{\frac{1+g^2}{2}} & f(g) \\
|g| & f(g) & f(g) & 1 
\end{matrix}\right],
\end{equation}
where the indices $i,j$ and $m,n$ are grouped into the row and column of a matrix, respectively. The function $f(g) = 0$ for $g\geq 0$ and $f(g) = |g|$ for $g<0$. By the channel-state duality, this family of isoTNS is dual to a family of continuously-tuned repeated quantum channels.

\begin{figure}
    \centering
    \includegraphics[width=0.8\linewidth]{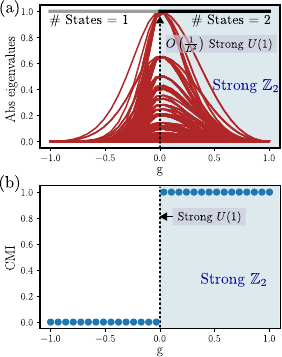}
    \caption{Tuning a transition in the quantum channel defined from the W tensor in Eq.\ref{eq:w_tensor}. This is equivalent to the transition in the isoTNS under the channel-state duality. The isoTNS is in the toric-code topologically-ordered phase for $g>0$ and in the trivial paramagnetic phase for $g<0$. The figures present the numerical results based on the exact diagonalization of the transfer matrix that defines the quantum channel for $L = 10$ with a periodic boundary condition. A strong symmetry $\mathbb{Z}_2$ is preserved for $g>0$. (a) The absolute value of the eigenvalues of the transfer matrix spectrum. The red lines represent the eigenvalues with magnitude less than 1. The steady-state degeneracy is 2 and 1 for $g>0$ and $g<0$, respectively. At $g = 0$, the channel preserves the strong $U(1)$ symmetry, leading to $L+1$ steady-state degeneracy; the gap closes as $O(1/L^2)$, by mapping the transfer operator to that of the classical six-vertex model~\cite{Baxter:1982zz}. (b) The CMI, in units of $\log 2$, of the steady states computed according to the partition in Fig.\ref{cmi}(a), with $|A| = |C| = 3$ and $|B| = 4$ (two sites per subregion of $B$). For $g>0$, the steady states are degenerate; we compute the CMI of one of the strongly symmetric steady states.}
    \label{fig:isotns_transition}
\end{figure}

At $g = 1$, the isoTNS represents the toric-code ground state, as can be seen from the $W$-tensor:

\begin{equation}
 W_{ijmn} = 
\begin{cases}
&1, \text{ for }  i+j+m+n=0  ~\text{  mod }2 \\  
&0, \text{ for }  i+j+m+n=1 ~\text{  mod }2 \\  
\end{cases}
\end{equation}
with $i,j,m,n =  0,1$ denoting the configuration in Pauli-Z basis. As $g$ is tuned away from this limit, the bulk wavefunction has a finite correlation length and exhibits the same topological order until $g = 0$, where the gap of the transfer operator closes. Importantly, for $g\in[0,1]$, $W_{ijmn}$ remains vanishing for $i+j+m+n=1$ mod 2, which translates to a virtual $\mathbb{Z}_2$ symmetry in the isoTNS:

\begin{equation}\label{eq:virtual}
    \includegraphics[scale = 0.35]{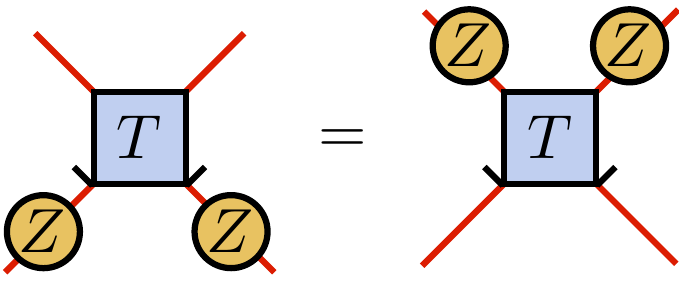}\nonumber.
\end{equation}
Namely, it is a symmetry acting solely on the virtual (red) legs. Such a local virtual symmetry on a tensor also implies that the isoTNS defines a strongly symmetric quantum channel; given any 1d input pure state with the global symmetry $\prod_i Z_i$, the output state of the channel will have the strong symmetry $\prod_i Z_i$ as well~\footnote{Note that although the $\mathbb{Z}_2$ symmetry here is expressed in a different basis from the one used in Sec.~\ref{sec:1d}, generated by $\prod_i X_i$, the two are related by a virtual Hadamard rotation $H$ satisfying $H X H = Z$ and $H^2 = I$. This basis change cancels out when the virtual legs are contracted and therefore does not alter the bulk isoTNS wavefunction. In other words, the basis change is merely a ``gauge'' transformation of the 2d tensor-network state.}. 

For $g<0$, such a virtual symmetry no longer exists. The isoTNS describes a trivial paramagnetic phase and become the product state $\ket{++\cdots }$ at $g = -1$ \footnote{At $g=-1$, $W_{ijmn}=1$ for any $i,j,m,n$, so the corresponding 2d tensor-network state is a superposition of all Z-basis product state, or equivalently, $\ket{++\cdots}$.}. A transition thus happens at $g = 0$, where the transfer matrix defining the quantum channel is exactly that of the critical six-vertex model \cite{Baxter:1982zz}, and the gap closes as $O(1/L^2)$. In particular, there is a $U(1)$ virtual symmetry, corresponding to particle-number conservation. More concretely, a virtual bond can take the value 0 or 1, which can be interpreted as the absence or presence of a particle. Using the explicit form of the $W$ tensor at $g=0$, one can see that the total particle number in the lower two virtual legs is equal to that in the upper two legs. Correspondingly, the bulk isoTNS wavefunction can support an anisotropic power-law correlation along the vertical $y$ direction due to the diffusive dynamics of these $U(1)$ charges, while developing a long-range order along the horizontal direction, with details depending on the choice of the boundary condition~\cite{liu_iso:2024}. Relatedly, the virtual $U(1)$ symmetry also implies the $L+1$ exact degeneracy of the transfer matrix, due to the $L+1$ possible total U(1) charges on a 1d line, which we numerically verify. 

Under the channel-state duality, the tunable isoTNS corresponds to a continuously tuned quantum channel with a transition, as shown in Fig.\ref{fig:isotns_transition}. At $g=1$, the isoTNS is the fixed-point toric code, whose top boundary is a steady state (of the corresponding channel) that exhibits $\mathbb{Z}_2$ SW-SSB. The SW-SSB persists in the range $g\in (0,1]$, which is evident in Fig.\ref{fig:isotns_transition}, e.g. indicated by the non-vanishing CMI (conditional mutual information). For $g<0$, the isoTNS and the corresponding steady state on the top boundary are in a paramagnetic phase, with CMI $=0$. This confirms the validity of the holographic picture beyond the fixed-point limit of the topological wavefunctions. Notably, at $g=0$, the virtual $Z_2$ symmetry is further enhanced to $U(1)$ symmetry, and the channel describes the diffusion of $U(1)$ charges. Correspondingly, given an input pure state with a fixed total $U(1)$ charges, the steady state is a classical mixture of all possible $U(1)$ charge configurations with a fixed total charge, hence exhibiting a $U(1)$ SW-SSB. 

Importantly, this transition does not contradict the fact that strongly symmetric quantum channels necessarily possess degenerate steady states, as discussed in the previous section. In particular, as long as the symmetry is preserved, one cannot tune a transition between a quantum channel with degenerate steady states, including the SW-SSB state, and a quantum channel with a unique steady state. A similar observation was made in Ref.\cite{ziereis2025strong}, which emphasized that the completely positive and trace-preserving (CPTP) condition forbids a transition from SW-SSB to a paramagnetic phase across different steady states—whether under repeated quantum channels or long-time Lindbladian evolution. Such a transition could only occur by introducing post-selection into the non-unitary process, which necessarily violates the CPTP condition. The present example we discussed circumvents this restriction by preserving a strong $\mathbb{Z}_2$ symmetry of the quantum channel for $g>0$, while removing the symmetry for $g<0$. 

As a final remark, the isoTNS framework can be employed to interpolate between the toric code and double-semion ground states, and more generally, between a wide class of string-net states \cite{Boesl:2025_stringnetiso}. The presence of distinct bulk topological orders implies that the corresponding dual quantum channels possess distinct SW-SSB steady states. For example, the virtual boundary of the double-semion isoTNS (see Appendix G of~\cite{liu_iso:2024}, and also~\cite{Boesl:2025_stringnetiso}) will lead to a 1d steady state of the form
\begin{equation}
    \rho \propto I+ CZX,
\end{equation}
where the $CZX$ symmetry is defined as $CZX = \left(\prod_i X_i\right)\left(\prod_i CZ_{i,i+1}\right)$ with $CZ$ being the control-$Z$ operator such that $CZ\ket{11} = -\ket{11}$ and it acts trivially otherwise. The intriguing feature is that the anomaly in $CZX$ symmetry implies the steady state $\rho$ is long-range tri-partite entangled: it cannot be created from any tripartite-separable state $\rho_A \otimes \rho_B \otimes \rho_C$ using finite-depth local channels \cite{lessa2024mixedstate}. This is in stark contrast to the SW-SSB steady state ($\propto 1+ \prod_i X_i$) arising from the fixed-point toric code, which is fully separable (i.e., not entangled). As a notable application, the isoTNS framework provides a concrete protocol to prepare exotic many-body mixed states. For example, one can input any pure state with $CZX$ symmetry (e.g., a GHZ state); applying a depth-1 channel defined by the isoTNS tensor of the double semion will immediately prepare the $ I+ CZX$ mixed state. This gives a concrete state-preparation protocol, which would otherwise be difficult to design without the insight of isoTNS. More generally, the continuous tuning of the isoTNS provides an analytically tractable framework for exploring quantum channels with distinct SW-SSB steady states and transitions between them.

\color{blue}

\color{black}

\section{Examples for SW-SSB under holographic duality}\label{sec:gen}
In this section, we present a variety of examples of SW-SSB, including higher-dimensional generalizations and cases involving various generalized symmetries—such as subsystem and higher-form symmetries—and discuss the corresponding emergent topological orders through holographic duality. The results are summarized in Table \ref{table_II}.

\begin{table*}
\centering
\renewcommand\arraystretch{1.4}
\begin{tabular}{| l | l |}
\hline
\textbf{Steady state in $d$-dim} & \textbf{Topological order in $d{+}1$-dim} \\ \hline
$Z_2$ 0-form SW-SSB in 1d & 2d toric code with $e$-charge-condensed boundary \\ \hline
$Z_2$ 0-form  SW-SSB in 2d & 3d toric code with $e$-charge-condensed boundary \\ \hline
$Z_2$ 1-form SW-SSB in 2d & 3d toric code with $m$-loop–condensed boundary \\ \hline
$Z_2$ subsystem SW-SSB in 2d & 3d fracton order with charge-condensed boundary \\ \hline
$Z_2$ fermionic 0-form SW-SSB in 1d  & 2d toric code with boundary condensation of $f$-charges and Majorana fermions. \\ \hline
$Z_2$ fermionic 1-form SW-SSB in 2d & 3d fermionic toric code with flux-loop–condensed surface \\ \hline
\end{tabular}
\caption{Various examples that illustrate the holographic duality between SW-SSB and topological orders.}\label{table_II}
\end{table*}

\subsection{2d 0-form SW-SSB}\label{sec:2d}

Consider a 2d lattice with one qubit per site, any initial state with the global strong $\mathbb{Z}_2$ symmetry $\prod_i X_i$ will converge to the SW-SSB state $ \propto  1+\prod_i X_i$ under the repeated application of the noise channel $\cE =  \prod_{ \langle ij \rangle}  \cE^z_{ij}  \prod_i  \cE^x_{i}$: 

\begin{equation}\label{eq:2d_noise}
\begin{split} 
   &\cE^z_{ij}[\rho_0] = (1-p_z) \rho_0 + p_z  Z_i Z_{j}\rho_0 Z_i Z_{j}, \\
   &  \cE^x_{i}[\rho_0] =  (1-p_x) \rho_0 + p_x X_i \rho_0 X_i.
\end{split}
\end{equation}
 $\cE^z_{ij}$ is a local channel acting on two neighboring qubits and $ \cE^x_{i}$ is a single-site channel. Similar to the discussion in Sec.\ref{sec:1d}, the repeated application of these channels can be understood as a sequential unitary circuit, where the 2d system qubits sequentially interact with the ancilla qubits in the 3d bulk, layer by layer. Tracing out all ancilla qubits then effectively implements the channel acting on the system qubits.

Notably, at $p_x= p_z =  \frac{1}{2}$, the 3d bulk has the stabilizers:

\begin{equation}\label{}
\includegraphics[width=6.4cm]{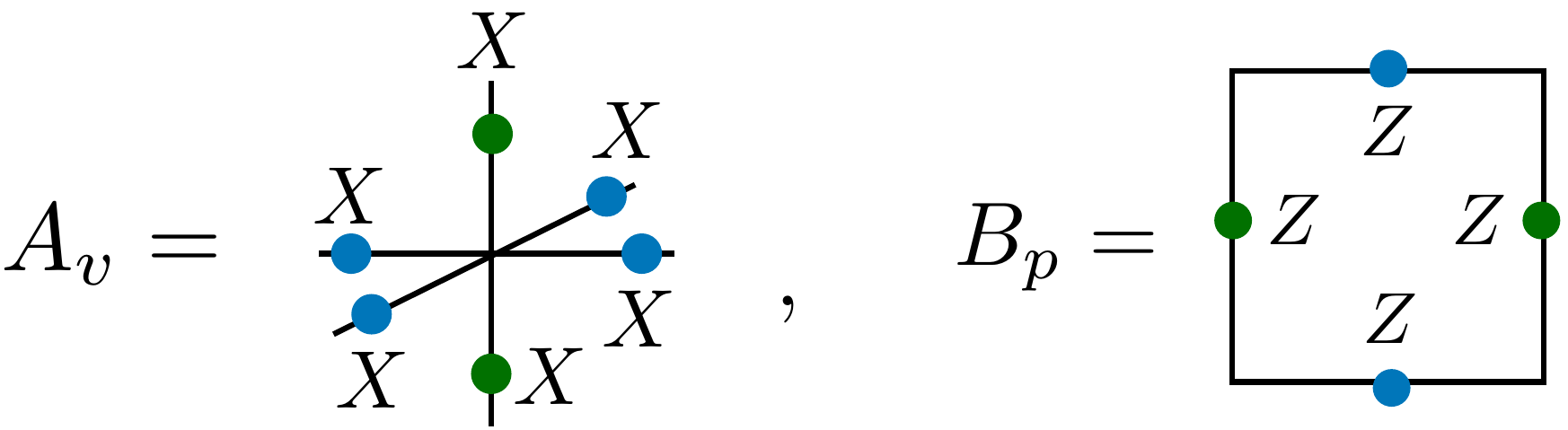}
\end{equation}  
The vertex stabilizer $A_v$ is a product of 6 Pauli-Zs around a vertex, and the plaquette stabilizer $B_p$ is a product of 4 Pauli-Xs around a plaquette. These stabilizers indicate the 3d toric code, which exhibits a $\mathbb{Z}_2$ topological order. In addition, the top boundary is specified by the stabilizers

\begin{equation}\label{3d_toric_top}
\includegraphics[width=6.4cm]{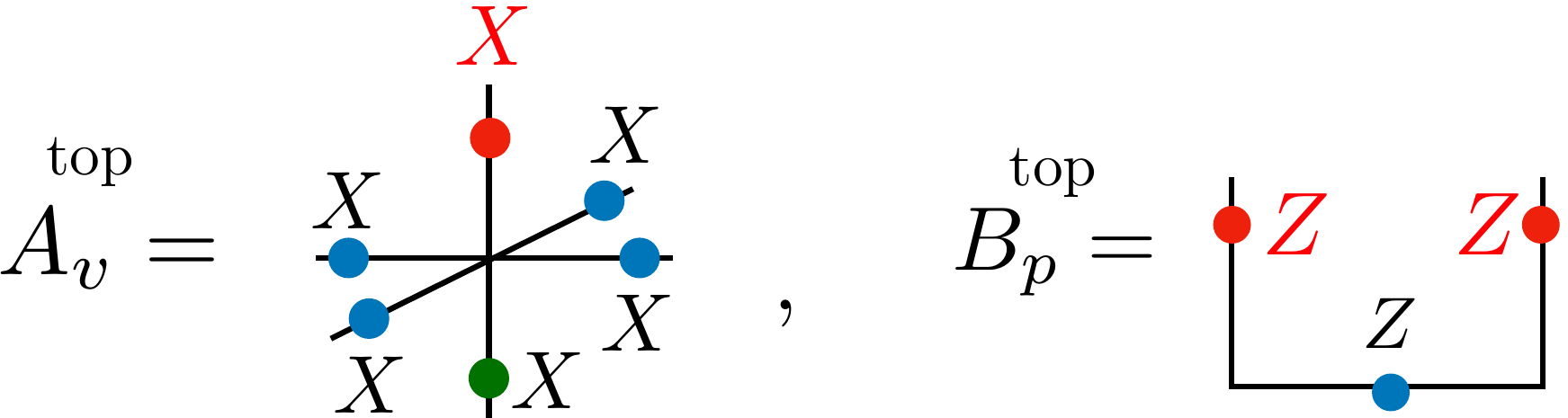}
\end{equation}   
where the system qubits (i.e., the output state of the repeated quantum channel) are colored in red.

Away from this maximal decoherence limit, i.e. $p_x, p_z \neq \frac{1}{2}$ , one instead obtains a deformed toric code

\begin{equation}\label{eq:deformed_3d_toric}
\ket{\text{TC}}(g_x,g_z)=\prod_{\ell \in x-y \,\text{plane}}  e^{g_z X_{\ell }} \prod_{\ell \in z\,\text{link}}  e^{g_x Z_{\ell }} \ket{\text{TC}}, 
\end{equation}
with 
\begin{equation}
e^{2g_z } =  \frac{1-p_z}{p_z}~, ~~~e^{2g_x } =  \frac{1-p_x}{p_x}. 
\end{equation}
Here the imaginary evolution $ e^{g_z X_{\ell }}$ occurs on all links on the $x-y$ plane, and $e^{g_x Z_{\ell }}$ occurs on the $z$-link. The $\mathbb{Z}_2$ topological order will be robust under these deformations, unless $g_z \to \infty$ or $g_x \to \infty$, which corresponds to the zero-noise limit $p_z = 0$ or $p_x = 0$. This wavefunction is defined on a 3d lattice with periodic boundary conditions along $x, y$ directions, and with open boundaries along $z$. The reduced density matrix on the top surface (say $z=L_z$) is the output steady state of the repeated applications of the channel.

At the fixed point limit ($p_x=p_z =\frac{1}{2}$), a string operator of Pauli-Zs generates the two $e$-charges on the two endpoints of the string. Notably, the Z-type boundary stabilizer in Eq.\ref{3d_toric_top} indicates that the top surface realizes an $e$-condensed boundary, where a pair of $e$-charges can condense on the boundary. Such $e$-condensation also gives rise to the weak symmetry for the SW-SSB state $\rho$ on the boundary: $Z_iZ_j  \rho Z_iZ_j = \rho$, where $i, j$ can be any two points on the top surface.

On the other hand, one can define a non-contractible closed-membrane operator on the top surface:
\[
V_{\text{surface}}=\prod_{i\in \text{surface}} X_i,
\]
which measures the global $m$-flux and generates the associated $\mathbb{Z}_2$ $1$-form symmetry. Since the wavefunction in Eq.\ref{eq:deformed_3d_toric} is an eigenstate of $V_{\text{surface}}$ for all finite $g_x,g_z$, the reduced density matrix on the top surface has a strong $\mathbb{Z}_2$ symmetry. Thus, under the holographic duality, the strong symmetry of the $2d$ boundary reduced state descends from the $1$-form symmetry in the 3d bulk.

Unlike the 1d case discussed in Sec.\ref{sec:1d}, which requires an infinite depth to realize SW-SSB with non-maximal decoherence, the channel acting on the 2d lattice can drive a transition to the SW-SSB phase with a finite constant depth, even with non-maximal decoherence \cite{lessa2025strong,sala2024spontaneous_published}. Heuristically, this is because the $Z$ string operator with two endpoints on the top boundary can emerge at finite depth. We provide a more detailed discussion about this perspective in Appendix.\ref{appendix:finite_depth}.

\subsection{2d 1-form SW-SSB}\label{sec:sw-SSB_1-form}
Consider a 2d lattice with qubits defined on edges, one can define the $\mathbb{Z}_2$ 1-form symmetry generator $B_p= \prod_{\ell \in \partial p } X_\ell$, which is the product of four Xs along a plaquette $p$. These $B_p$ operators generate the 1-form symmetry: $\prod_{\ell \in \mathcal{C}} X_\ell$, which is the product of Xs along any closed (contractible) loop $\mathcal{C}$. A canonical example of the mixed state, where the strong 1-form symmetry is spontaneously broken down to the weak one, is the maximally mixed state with this strong 1-form symmetry \cite{sang2025mixed,zhang2025strong}:

\begin{equation}\label{eq:1-form_SWSSB}
\rho    \propto \prod_p (1+B_p).
\end{equation} 
An essential feature of this mixed state is the existence of the weak 1-form symmetry along loops in the dual lattice, which is generated by 
\begin{equation}\label{eq:star}
A_v = \prod_{\ell \in v } Z_{\ell}, 
\end{equation}
namely, the product of four Zs around a vertex $v$. The universal feature of the SW-SSB can then be understood as the mutual anomaly between the strong $X$ loops and the weak Z loops.

Again, to develop a holographic description, we will consider a repeated quantum channel that (i) preserves a strong higher-form symmetry and (ii) allows for a weak higher-form symmetry to emerge in the steady state. Concretely, we define the following quantum channel $\cE = \prod_v \cE_v^z \prod_\ell \cE_\ell^x$, with
\begin{align}\label{eq:2dhigher}
   \cE^x_{\ell}[\rho] =  (1-p_x) \rho+ p_x X_\ell\rho X_\ell . \nonumber\\
     \cE^z_{v}[\rho] = (1- p_z)  \rho   + p_z     A_v \rho A_v    
\end{align}
$\cE^x_{\ell}$ is an $X$-dephasing channel, and $\cE^z_{v}$ is the channel that applies the star operator $A_v$ defined in Eq.\ref{eq:star} with probability $p_z$. It is straightforward to check that the decoherence channel $\cE$ preserves the strong X 1-form symmetry. In particular, starting with any pure state with the X 1-form symmetry, it will converge to the SW-SSB mixed state defined in Eq.\ref{eq:1-form_SWSSB}. 

Building on our holographic framework, where the repeated quantum channel is viewed as a sequential unitary circuit that generates a wavefunction in one higher dimension, we find that the bulk ancilla qubits form a 3d $\mathbb{Z}_2$ topological order (see Appendix.\ref{appendix:1-form_swssb} for details). In particular, at  $p_x=p_z = \frac{1}{2}$ corresponds to the fixed-point limit, described by the toric-code stabilizer in the bulk

\begin{equation}\label{}
\includegraphics[width=6.5cm]{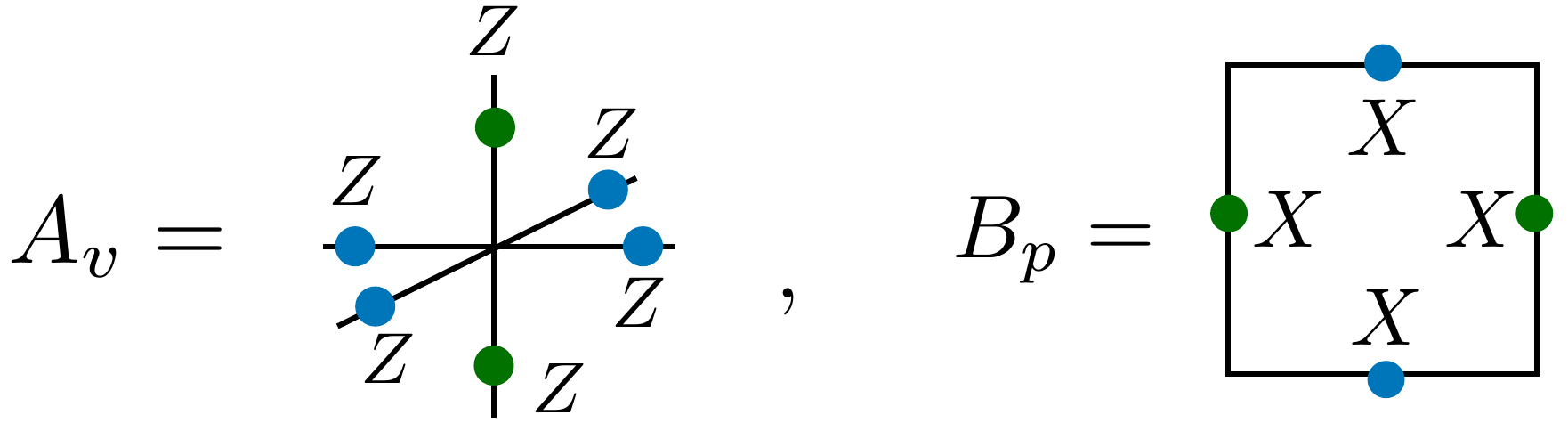}
\end{equation}

\begin{figure}[h]
    \centering
\includegraphics[width=0.48\textwidth]{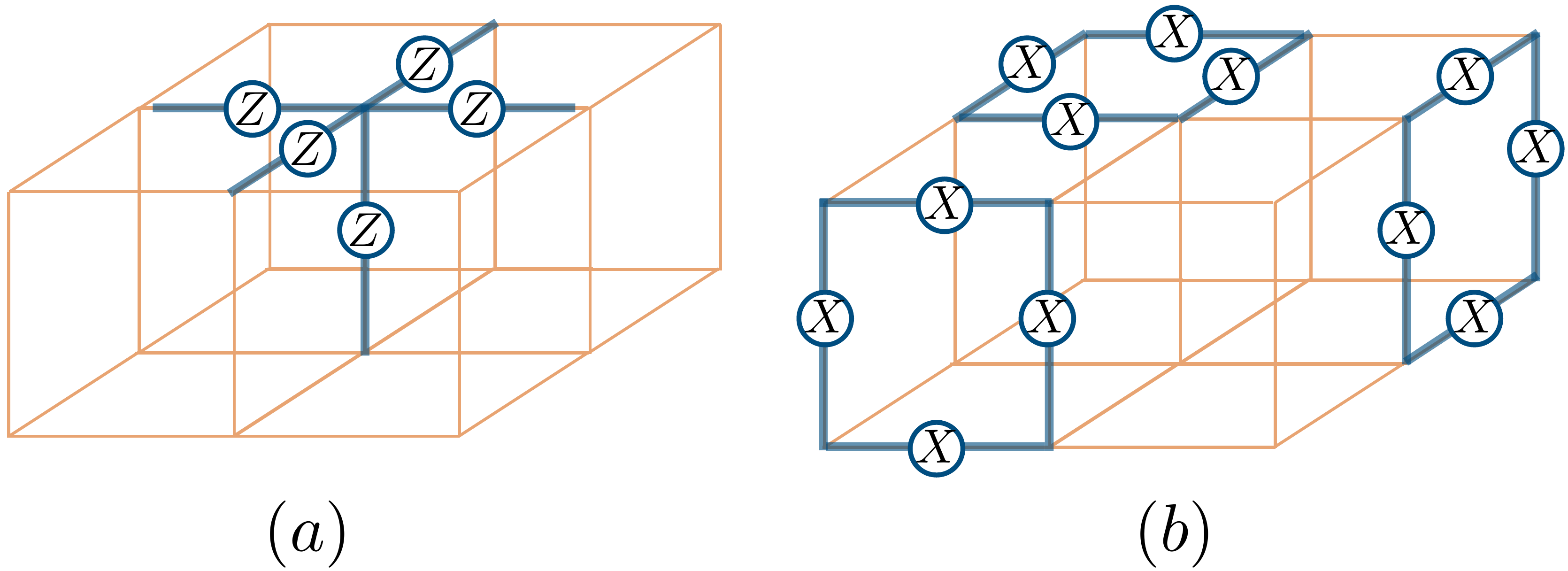}
    \caption{The surface stabilizers for the 3d toric code emerging from the repeated quantum channel (Eq.\ref{eq:2dhigher}). (a) The Z-type stabilizer indicates the condensation of the magnetic-flux (m) excitation on the top surface. Tracing out the bulk leads to the weak-1-form symmetry of the top-surface reduced density matrix. (b) The top surface has the $X^{\otimes 4}$ stabilizer, leading to the strong 1-form symmetry of the top-surface reduced density matrix.}  
    \label{fig:3dtcsurface}
\end{figure}

The steady state (Eq.\ref{eq:1-form_SWSSB}) of the repeated quantum channel in Eq.\ref{eq:2dhigher} is the reduced density matrix on the top boundary of the $3d$ toric code with a smooth ($m$-loop–condensed) surface, as shown in Fig.\ref{fig:3dtcsurface}. In this holographic picture, the strong symmetry of the steady state descends from a bulk 2-form symmetry (i.e., the product of Pauli-Xs along any 1d closed loops). On the other hand, the weak symmetry originates from a bulk 1-form symmetry whose membrane operators pierce the bulk and intersect the surface as loops. Consequently, the mixed anomaly between the strong and weak 1-form symmetries in the steady state is inherited from the topological order of the 3d bulk.

\subsection{2d subsystem SW-SSB}
We here discuss a novel SW-SSB state in 2d with strong subsystem symmetries, and show that it is holographically dual to a topological fracton model in 3d. 

To begin, we consider a 2d lattice with one qubit per site, and define the following subsystem symmetries \cite{xu2004strong,you2018majorana,you2018subsystem,foliation_anomaly_2024}:

\begin{equation}\label{eq:subsystem}
U_x = \prod_{y=1}^{L_y} X_{x,y}, ~~~~U_y = \prod_{x=1}^{L_x} X_{x,y}.  
\end{equation} 
$U_x$ and $U_y$ act on the 1d lines along the vertical (y) direction and the horizontal (x) direction, respectively. 

The maximally mixed state with these subsystem symmetries is 

\begin{equation}
  \rho_{\text{sub}}  \propto  \prod_x (1 +U_x)    \prod_y (1 +U_y)  
\end{equation}
In particular,  $\rho_{\text{sub}}$ is weakly symmetric under $W_{x_1,x_2,y_1,y_2}$ defined as

\begin{equation}
W_{x_1,x_2,y_1,y_2} = Z_{x_1,y_1}  Z_{x_2,y_1} Z_{x_2,y_2} Z_{x_1,y_2},
\end{equation}
which is the product of four Pauli-Zs on the corners of a rectangle. The SW-SSB can be understood as the mutual anomaly between the strong subsystem symmetries $U_x, U_y$ and the weak symmetry $W_{x_1,x_2,y_1,y_2}$.

This mixed state can be obtained by the repeated action of the channel $\cE =  \prod_P \cE^z_{P} \prod_i \cE^x_{i}$, with 

\begin{equation}\label{eq:subsystem_noise}
\begin{split} 
   &\cE^z_{P}[\rho_0] = (1-p_z) \rho_0 + p_z (\prod_{i\in P }Z_i) \rho_0 (\prod_{i\in P }Z_i) . \\
   &  \cE^x_{i}[\rho_0] =  (1-p_x) \rho_0 + p_x X_i \rho_0 X_i,
\end{split}
\end{equation}
where $\prod_{i\in P }Z_i$ denotes the product of four Pauli-Zs on the corners of a plaquette $P$.

This channel can be realized as a sequential circuit where the 2d system qubits sequentially interact with the ancilla qubits on the plaquette on the x-y plane (responsible for $\cE^z_{P}$ channel) and the ancilla qubits on the z-direction link (responsible for $\cE^x_{i}$ channel). 

At $p_z = p_x = \frac{1}{2}$, the 3d bulk theory is described by two types of 6-body stabilizers:

\begin{equation}\label{}
\includegraphics[width=8.5cm]{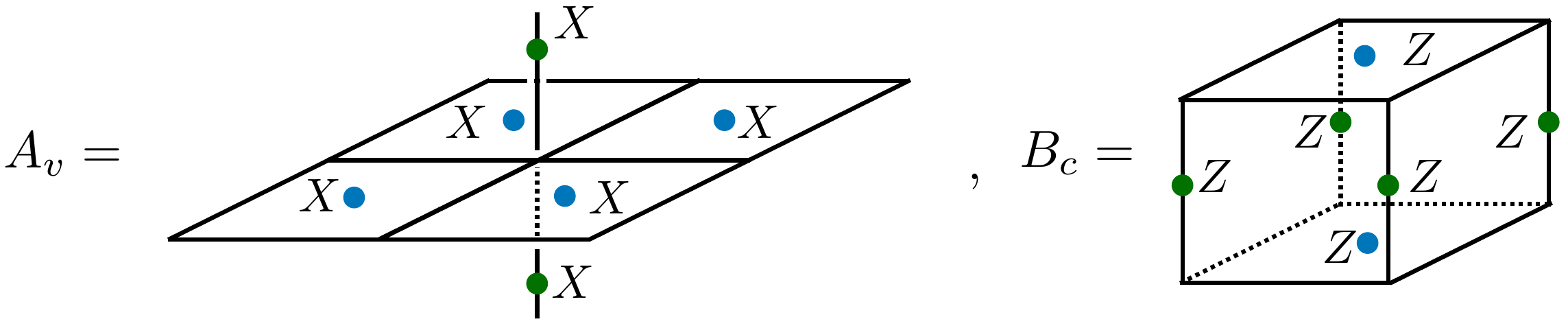}
\end{equation} 

$A_v$ is the product of the two Xs on the z-directional link and the four Xs on the x-y plane plaquettes neighboring the vertex $v$. $B_c$ is the product of the four Zs on the z-directional links and the two Zs on the x-y plane plaquettes in the cube $c$. This is an anisotropic fracton model with lineons and planons, first discovered in Ref.\cite{fracton_chen_2019}. It represents a foliated fracton phase with 2 foliations composed of toric code layers along the x-z and y-z planes. 

The top boundary is specified by the following stabilizers: \begin{equation}\label{}
\includegraphics[width=8.5cm]{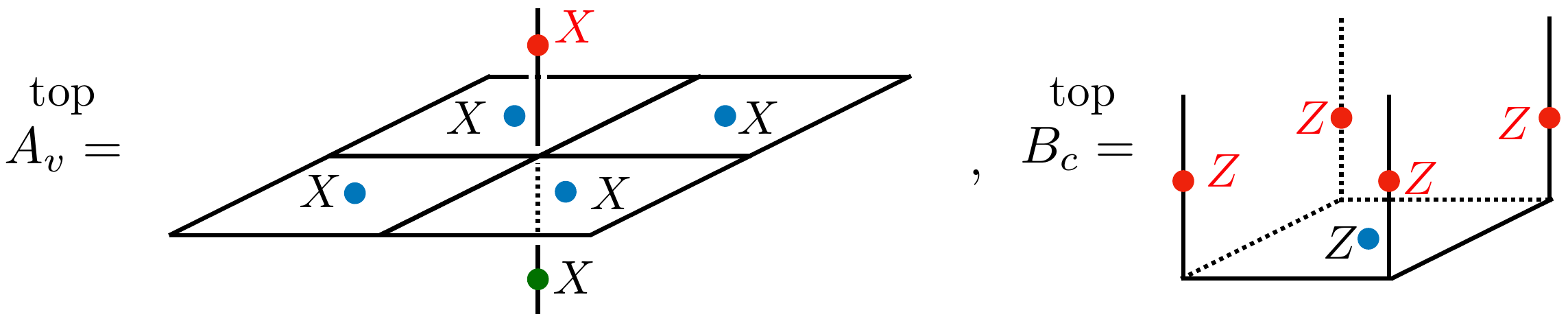}
\end{equation} 
where the red qubits are the output of the repeated quantum channel. Those red qubits have the strong subsystem symmetries $U_x, U_y$ in Eq.\ref{eq:subsystem}, and the weak symmetry comes from tracing out the bulk qubits given the $B_c^{\text{top}}$ stabilizer. In particular, the boundary condition specified by $B_c^{\text{top}}$ indicates that four electric lineons condense on the top boundary (see Ref.\cite{fracton_chen_2019,you2018majorana} for details on the lineon excitations).

\subsection{1d fermion parity SW-SSB}
Here, we apply our holographic framework to study SW-SSB in a fermionic system. Consider a one-dimensional fermionic chain with two Majorana modes per site, $\eta_i$ and $\eta'_i$. The system possesses a global $\mathbb{Z}_2^f$ symmetry generated by $\prod_i P_i$, where $P_i = -i\eta_i\eta'_i$ denotes the onsite fermion parity. We focus on an SW-SSB state $\rho_f$, in which the strong $\mathbb{Z}_2^f$ symmetry is spontaneously broken down to the weak $\mathbb{Z}_2^f$:

\begin{equation}
  \rho_f  \propto 1+ \prod_i  P_i.   
\end{equation}
The essential properties of $\rho_f$ are (i) the strong $Z_2^f$ symmetry $\prod_i P_i $ and (ii) the weak 1-form symmetry $\eta_i' \eta_j\rho_f \eta_j \eta_i' = \rho_f$. The SW-SSB follows from the mutual anomaly between the strong and weak symmetries. Below, we will show that it is holographically dual to a toric-code order in 2d, with a key feature that the condensed boson on the 1d boundary is a composite of a physical Majorana and a fermionic excitation of the toric code.

\begin{figure}[t]
    \centering
\includegraphics[width=0.4\textwidth]{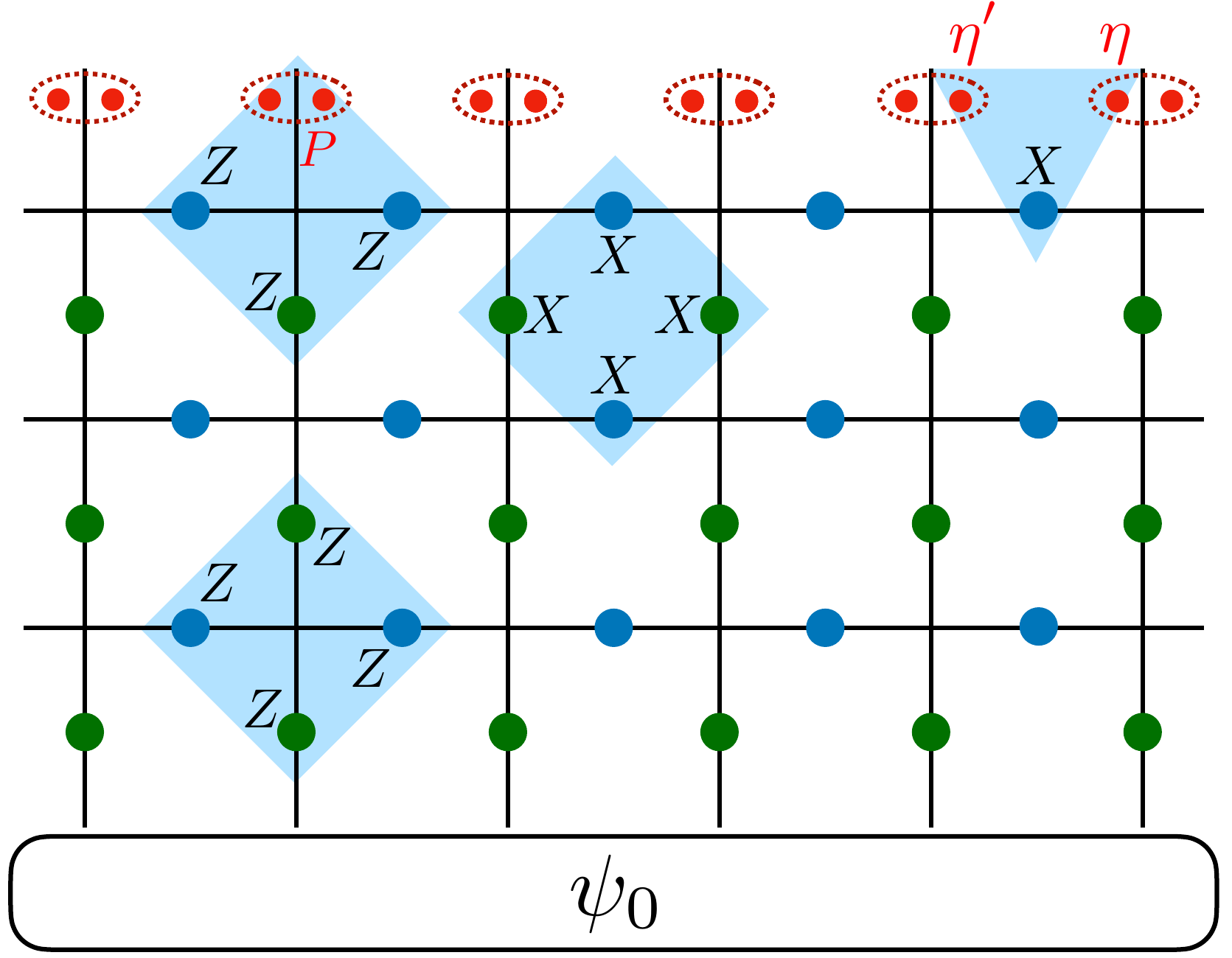}
    \caption{Holographic duality between the $\mathbb{Z}_2^f$ SW-SSB on the top boundary and the bulk $\mathbb{Z}_2$ topological order. At $p_x=p_z= \frac{1}{2}$, the bulk is the fixed-point toric code with the star stabilizer $A_v = Z^{\otimes 4}$ and plaquette stabilizer $B_p = X^{\otimes 4}$.} 
    \label{fig:fermion_main}
\end{figure}

To obtain a holographic description, we consider the repeated action of the channel: 
\begin{equation}\label{eq:fermion}
\begin{split} 
   &\cE^a_{i}[\rho_0] = (1-p_z) \rho_0 + p_z  \eta_i' \eta_{i+1}\rho_0  \eta_{i+1} \eta_i' \\
   &  \cE^b_{i}[\rho_0] =  (1-p_x) \rho_0 + p_x P_i \rho_0 P_i.
\end{split}
\end{equation}
This quantum channel can be viewed as the Jordan–Wigner transformation of the Ising quantum channel in Eq.\ref{eq:noise}. Given this duality, the channel admits a sequential-circuit realization: ancilla qubits on the $x$ and $y$ links form a toric-code wavefunction in the bulk, while the Majorana degrees of freedom on the boundary vertices couple to the ancilla qubits in the bulk. (see Fig.\ref{fig:fermion_main}).

At $p_x=p_z= \frac{1}{2}$, the bulk qubits form the fixed-point toric code state, and the boundary coupling between the Majorana fermions and qubits is indicated by the $\eta' X\eta$ type stabilizer (the phase is omitted) and $PZZZ$ type stabilizer. Notably, the combination of these two types of stabilizers implies the long-string stabilizer, where the two boundary Majoranas are connected by a fermion string - a composite object of the X string and Z string -  through the bulk, shown as follows:

\begin{equation}\label{}
\includegraphics[width=4cm]{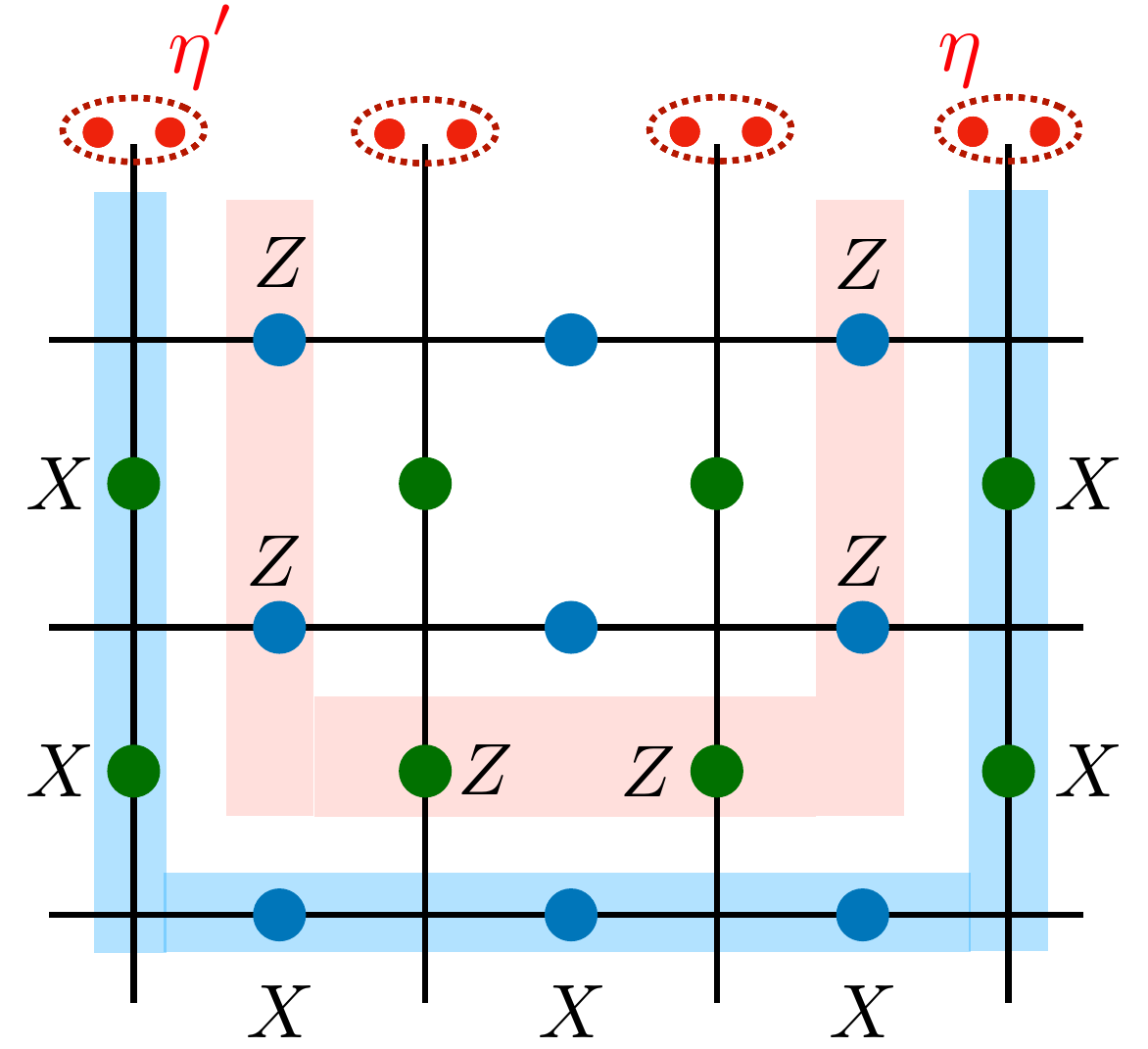}
\end{equation}

In other words, the fermion charges from the toric code bind with a physical Majorana fermion to form a composite bosonic particle that condenses on the 1d top boundary. When tracing out the bulk to obtain the boundary density matrix $\rho_f$, such a strong fermion-line symmetry becomes a weak fermionic symmetry: $\eta_i' \eta_j\rho_f \eta_j \eta_i' = \rho_f$, which together with the strong symmetry $\prod_iP_i$ indicates the 1d SW-SSB.

\subsection{2d fermionic 1-form SW-SSB}\label{sec:fermion_1_form}
As our final example, we consider a quantum channel whose steady state realizes an intrinsically mixed-state topological order \cite{ellison2024towards,sohal2024noisy}, namely the fermion-decohered toric code, defined as \cite{Intrinsic_wang_2025}

\begin{equation}
\rho_f \propto \prod_v (1+A_v B_{p(v)}).
\end{equation}
$A_v = Z^{\otimes 4}$ is the product of four Pauli-Zs around a vertex, and $B_p = X^{\otimes 4}$ is the product of four Pauli-Xs around a plaquette. In particular, $B_{p(v)}$ is the bottom-right plaquette adjacent to the vertex $v$, so the stabilizer $A_vB_{p(v)}$ takes the form

\begin{equation}\label{eq:fermion_figure}
\includegraphics[width=2.4cm]{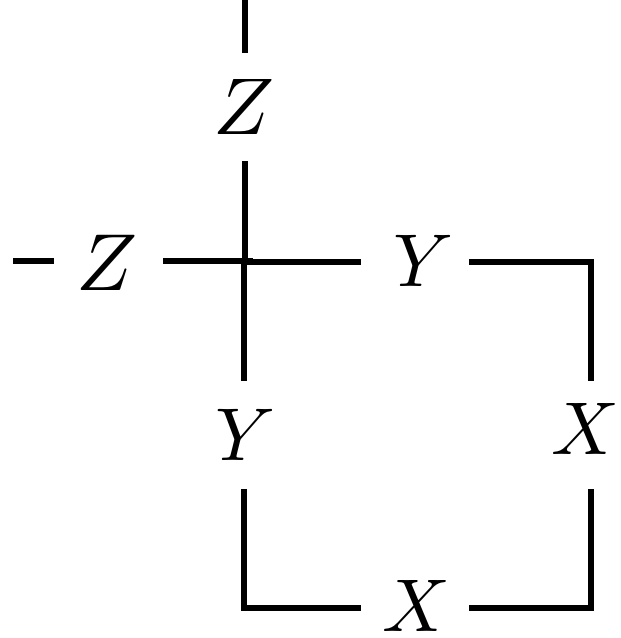} 
\end{equation} 
These stabilizers define the standard fermionic 1-form symmetry in the 2d toric code: $\prod_{\ell \in \mathcal{C}} Z_\ell X_{\ell + \frac{\hat{x}}{2} - \frac{\hat{y}}{2} }$, with $\mathcal{C}$ being any closed loops. When cut open, it creates the $f$ excitation (the $e$–$m$ bound state) on the two endpoints of the string.

To generate $\rho_f$ starting from any pure state with fermionic 1-form symmetry, we consider the repeated application of the channel $ \cE = \prod_v \cE_v  \prod_\ell  \cE_\ell $ with

\begin{equation}
\begin{split}
   &\cE_{\ell}[\rho] = p_\ell \rho +  (1-p_\ell) X_\ell Z_{\ell +\frac{\hat{x}}{2}-\frac{\hat{y}}{2}}\rho    X_\ell Z_{\ell +\frac{\hat{x}}{2}-\frac{\hat{y}}{2}}\\
   & \cE_v[\rho] = p_v \rho + (1-p_v) A_v \rho A_v 
    \end{split}
\end{equation}\label{eq:2dhigherfer}
When $p_{\ell}, p_v \neq 0$, $\rho_f$ is the unique steady state in the strong fermionic 1-form symmetry section.

Notably, at $p_{\ell} = p_v = \frac{1}{2}$, the corresponding 3d holographic wavefunction is described by the following stabilizers, akin to the 3d fermionic toric code discussed in Ref. \cite{fermion_levin_2003,Walker_wang_2015,Bosonization_chen_2015,chen_supercohomology_2021,zhou2025finite}:

\begin{equation*}
\includegraphics[width=8cm]{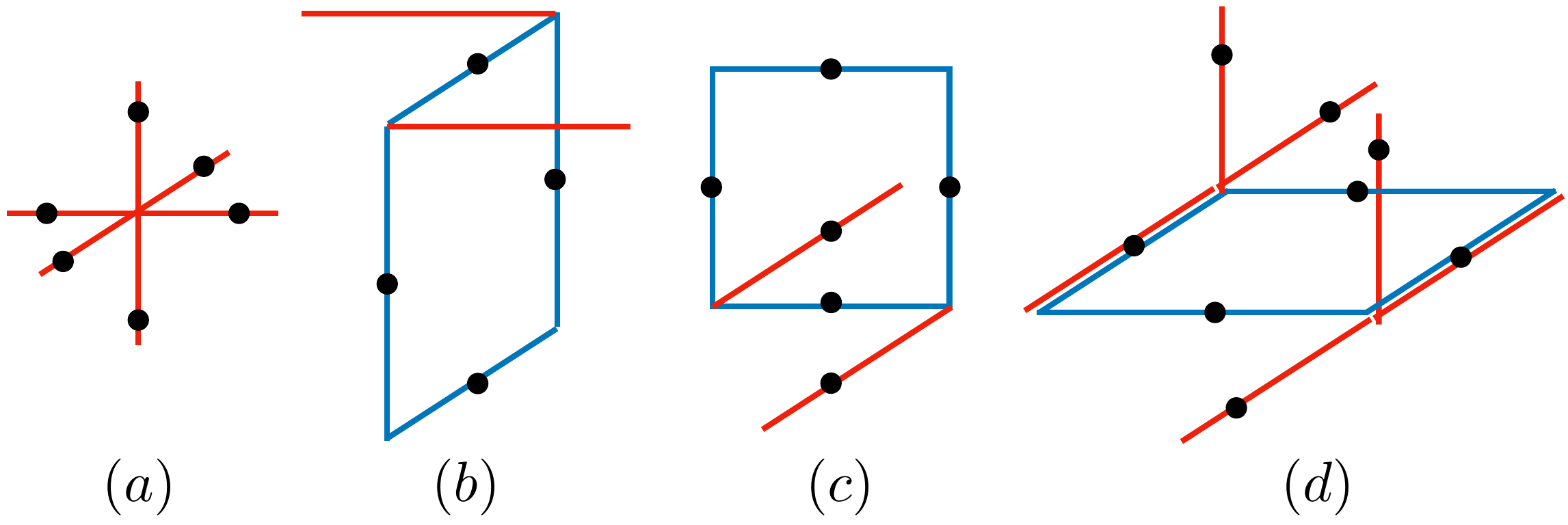} 
\end{equation*} 

The qubits defined on the x-y plane are the ancilla qubits responsible for $\cE_\ell$ noise, and the qubits defined on the z-direction (vertical) links are the ancilla qubits responsible for $\cE_v$ noise. We use the red line and blue line to label the location of the Pauli-Z and Pauli-X. For example, the stabilizer (a) is a vertex term of the form $Z^{\otimes 6}$. In stabilizer (d), the simultaneous presence of blue and red lines indicates the coexistence of Pauli-X and Pauli-Z, or equivalently, the Pauli-Y. One can check that all the stabilizers commute with each other. 

\begin{figure}[t]
    \centering
\includegraphics[width=0.46\textwidth]{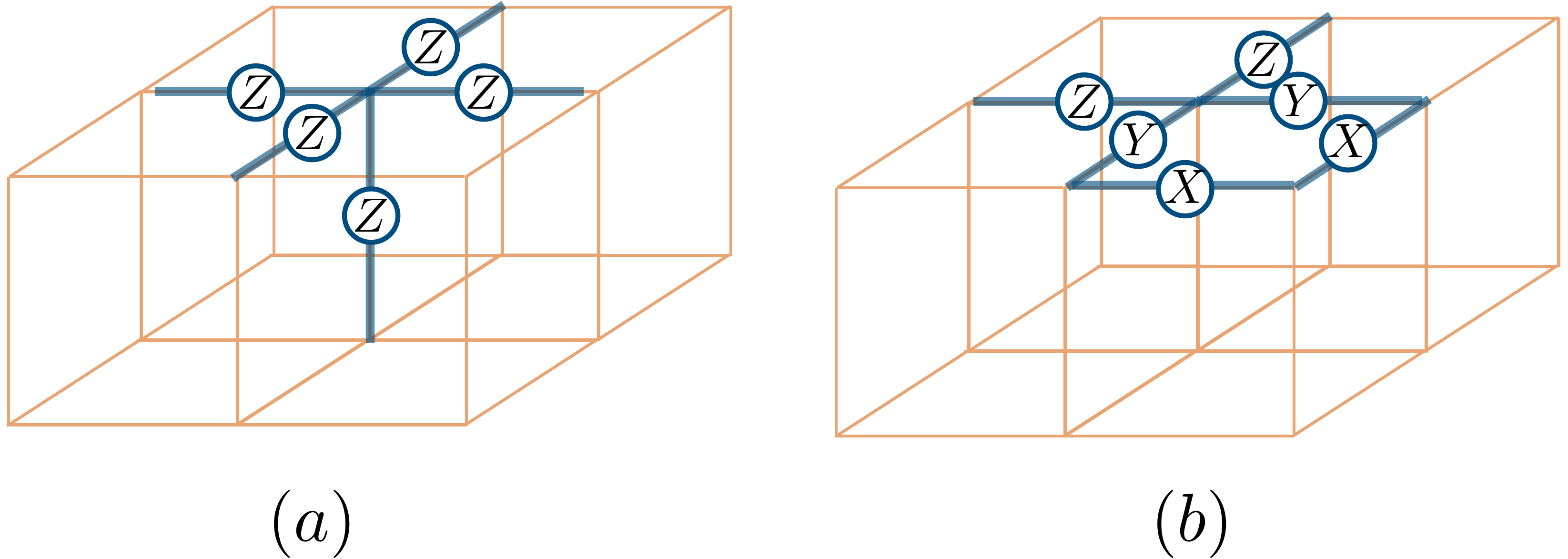}
    \caption{Stabilizers on the top boundary of the fermionic toric code. The stabilizer in (a) indicates the condensation of m-flux-loop excitations, which gives rise to the weak 1-form symmetry upon tracing out the bulk qubits. The stabilizer in (b) is exactly the one in Eq.\ref{eq:fermion_figure}, which gives the strong fermionic 1-form symmetry on the 2d surface.}  
    \label{fig:3dfermionsurface}
\end{figure}

The top boundary, which corresponds to the steady state in 2d, is specified by the stabilizers shown in Fig.\ref{fig:3dfermionsurface}. The truncated vertex-type operator implies the $m$-flux condensation, and it gives rise to the weak symmetry $A_v \rho_f A_v = \rho_f$ of the 2d boundary reduced density matrix; see Fig.\ref{fig:3dfermionsurface}(a). On the other hand, the stabilizer in Fig.\ref{fig:3dfermionsurface}(b) acts purely on the top surface, corresponding to the strong fermionic 1-form symmetry.  

\section{General discussions and comments}\label{sec:relations}

\subsection{SW-SSB transitions in non-CPTP maps}\label{sec:beyond_channel}

In this work, we primarily focus on the steady states arising in strongly symmetric quantum channels and demonstrate that such mixed states can be holographically dual to higher-dimensional topological (fracton) orders. Within this holographic framework, the time evolution of a $d$-dimensional quantum channel is dual to the transfer matrix of an isoTNS in one higher dimension.

Our holographic perspective can be further extended to scenarios beyond quantum channels defined as CPTP (completely positive and trace-preserving) maps. For instance, consider a non-unitary process where the system interacts with ancilla qubits, some of which are subsequently post-selected. Such dynamics are not CPTP \cite{hauser2024information,wang2025decoherence,negari2024spacetime}, yet if this process is repeated infinitely many times, the system can still relax to a steady state that exhibits strong-to-weak symmetry breaking (SW-SSB). Remarkably, even in these non-CPTP scenarios, the resulting time evolution can still be encoded as the transfer matrix of a higher-dimensional tensor network wavefunction, although it needs not be of the isoTNS form. Hence, our duality framework establishes a unified viewpoint: many mixed states arising in open quantum systems, regardless of whether the underlying evolution is CPTP, can be viewed either as reduced density matrices (RDM) of higher-dimensional wavefunctions. The strong symmetry conditions imposed on the quantum channel translate into constraints on the tensor symmetries in the higher-dimensional wavefunction, such as emergent 1-form symmetries. Conversely, starting from a nontrivial (topologically ordered) wavefunction, one can always examine its reduced density matrix as a lower-dimensional mixed state \cite{li2025subdimensional}; in such cases, the mutual anomaly associated with the bulk topological order guarantees the presence of SW-SSB in the RDM.

To illustrate the above perspective explicitly, we consider the boundary RDM of the toric code state with homogeneous string tension $\ket{\psi (\beta)}  \propto \prod_i e^{\beta Z_i}\ket{\text{TC}}$ \cite{Fradkin_deformed_2007,Chamon_state_2008}. While this deformed state admits an exact tensor-network representation of finite-bond dimension, it is not an isoTNS. This is in contrast to the toric code with inhomogeneous deformation in Sec.~\ref{sec:1d_channel}, which admits an isoTNS representation. Hence, the corresponding transfer matrix defines a non-CPTP channel, which can be implemented by entangling the system qubits with ancilla qubits with unitary gates, followed by post-selecting on certain ancilla qubits; see Appendix~\ref{sec:non_cptp_appendix} for the construction. We also note that each local tensor of $\ket{\psi (\beta)}$ has a Z-type $\mathbb{Z}_2$ virtual symmetry, which corresponds to the fact that the 1d non-CPTP channel preserves the 1d global $\mathbb{Z}_2$ symmetry generated by $\prod_i Z_i$. It is known that as \(\beta\) is varied, \(|\psi(\beta)\rangle\) undergoes a transition from a \(\mathbb{Z}_2\) topologically ordered phase to a topologically trivial phase at a critical $\beta_c$. This follows from the Kramers-Wannier (KW) duality transformation, under which $\ket{\psi(\beta)}$ is mapped to $e^{\beta \sum_{\langle v,v' \rangle} Z_v Z_{v'}} \ket{+}^{N_v}$ (see e.g. Ref.\cite{sahay2025_gapless}). The latter exhibits a 2d Ising order-disorder transition, which can be diagnosed by the two-point correlation function $\langle Z_v Z_{v'} \rangle$.

\begin{figure}[h]
    \centering
    \includegraphics[width=0.8\linewidth]{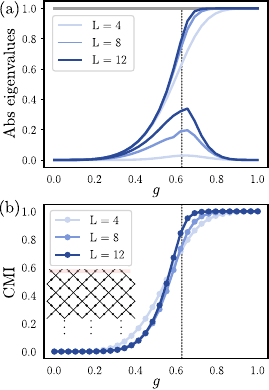}
    \caption{Topological phase transition and boundary CMI of the deformed toric code with string tension $\prod_i e^{\beta Z_i}\ket{\text{TC}}$ on an $L\times \infty$ cylinder, where $\beta(g) = -\log g$. A transition from $\mathbb{Z}_2$ topological order to the trivial phase happens at $g_c\approx 0.63$ (dashed line), solved by a mapping to the classical 2D Ising model. The results are obtained from the exact diagonalization of the transfer operator (constructed with the same orientation convention as in Sec.\ref{sec:iso}, see also the inset panel) of the deformed toric code for different circumference $L$. The transfer operator is a channel with a strong $\mathbb{Z}_2$ symmetry; it is, however, not trace-preserving and therefore not a quantum channel. (a) The absolute value of the three eigenvalues closest to 1 (including the eigenvalue 1 itself, marked as a gray line). The gap tends to zero with increasing system size at the critical point. (b) The CMI (in units of $\log 2$) of the boundary subsystem computed according to the partition in Fig.\ref{cmi}.}
    \label{fig:deformtc}
\end{figure}

We numerically study how this bulk transition is reflected in the conditional mutual information (CMI) of the boundary RDM; see Fig.~\ref{fig:deformtc}. With increasing system size, the boundary CMI develops a sharpening jump across the transition, changing from the trivial value \(0\) to \(\log 2\). Furthermore, the disorder operator $\prod_{i\in M} Z_i$ defined on a contiguous interval $M$ in the 1D boundary mixed state can be mapped to the two-point function $Z_v Z_{v'}$ with $v$, $v'$ on the two endpoints of $M$ in the dual Ising description via a KW duality. This implies that the disorder parameter $\prod_{i\in M} Z_i$ takes a non-vanishing $O(1)$ value only when $\beta> \beta_c$, consistent with a trivial strongly symmetric mixed state without SW-SSB.

To summarize, the steady state can undergo a phase transition from an SW-SSB phase to a trivial, strongly symmetric mixed state by tuning a non-CPTP channel, which can be implemented with post-selection. This is consistent with earlier observations that, while SW-SSB is generic for strongly symmetric CPTP channels, the inclusion of post-selection can stabilize a symmetric paramagnetic steady state \cite{ziereis2025strong}.

\subsection{Relation to SymTFT}

The bulk-edge correspondence developed here bears a close resemblance to the framework of symmetry topological field theory (symTFT)\cite{JiWen2020,GaiottoKulp2021,Apruzzi2023,FreedMooreTeleman2024}, or topological holography, where a \(d\)-dimensional quantum phase is encoded in a \((d+1)\)-dimensional topological order defined on an open cylinder, together with appropriate anyon-condensation conditions at its two boundaries. In the simplest \(1+1\)d \( \mathbb Z_2 \) setting, the bulk symTFT is the \(2+1\)d toric code. Choosing the lower boundary to be \(e\)-condensed identifies the horizontal \(m\)-loop as the generator of the global \( \mathbb Z_2 \) symmetry acting on the 1d boundary. Distinct boundary phases are then selected by different condensation patterns at the upper boundary. In particular, an \(e\)-condensed upper boundary realizes the Ising symmetry-broken phase, with the vertical \(e\)-string playing the role of the order parameter. In this way, the lower boundary specifies the symmetry and holonomy data, while the upper boundary captures the dynamics and determines the boundary phase. Recent works have further proposed extending this symTFT construction to mixed states by expressing the density matrix in the Choi-double representation, which naturally promotes the bulk topological order to a bilayer topological order\cite{sun2024holographic,QiSohal2025,SchaferNameki2025}.

Although the symTFT framework and our holographic correspondence share the common goal of characterizing lower-dimensional phases via higher-dimensional topological orders, they differ in several essential aspects:

(1) In symTFT, it is important to consider a thin slab, so that the topological operators in the topological order can be directly identified as the operators acting on the lower-dimensional phase. In contrast, we do not make this identification, and one needs to explicitly trace out the bulk degrees of freedom, thereby reducing the bulk topological operator to a weak symmetry that describes the mixed-state phases in lower dimensions. From this perspective, our holographic framework naturally involves a thick bulk region generated by repeated applications of quantum channels.

(2) symTFT can describe general pure-state quantum phases, and has also been proposed to extend to general mixed-state phases through the Choi-double construction \cite{sun2024holographic,QiSohal2025,SchaferNameki2025}. By contrast, our holographic framework is restricted to steady states of strongly symmetric quantum channels described by CPTP maps. This imposes nontrivial constraints on the allowed holographic bulk wavefunctions and the allowed mixed-state phases of the steady state. For example, for the strongly \(Z_2\)-symmetric quantum channel discussed in Sec.~\ref{sec:1d_channel}, the strong symmetry is broken in the steady state, making it impossible to realize a paramagnetic steady state with a nonvanishing disorder operator for the strong symmetry. This is in contrast to symTFT, where both the ferromagnetic and paramagnetic phases in \(1d\) can be naturally realized by tuning the boundary conditions (\(e\)- or \(m\)-condensed boundary) in the \(2d\) toric code.

\section{Outlook}\label{sec:outlook}
To summarize, we propose a holographic correspondence between steady states with SW-SSB and topologically ordered states in one higher dimension. Our results show that a mixed state under repeated noise can be viewed as the reduced density matrix of a higher-dimensional wavefunction represented by an isoTNS. Theoretically, this holographic duality allows us to infer salient features of SW-SSB mixed states, including long-range CMI and nonlocal recoverability, from the corresponding topological order. Conversely, since any quantum channel admits an isoTNS dual, the duality provides a route to classify and analyze steady-state transitions by manipulating the virtual bonds of the isoTNS.

In this work, we have mostly focused on SW-SSB of discrete symmetries; extending the framework to various continuous symmetries \cite{huang2024hydro,gu2024spontaneous,su2025spin} remains open. In particular, Refs.\cite{sahu_2024_symmetry,salo_steady_state_strong_sym_2025} reports that a strongly SU(2) symmetric 1d quantum channel can exhibit steady states whose bipartite mixed-state entanglement (e.g. entanglement of formation \cite{eof_1996} and logarithmic negativity \cite{peres1996separability,negativity_1996_Horodecki,Eisert_1999,Vidal_nega_2002}) scales logarithmically with size. It would be illuminating to identify the associated bulk wavefunction and infer steady-state entanglement from the holographic dual. Likewise, quantum channels with strong continuous symmetry can host hydrodynamic modes \cite{gu2024spontaneous,huang2024hydro} with gapless transfer-matrix spectra and long mixing times; clarifying how these features manifest in the holographic bulk is an essential next step. Meanwhile, recent developments have shown that there is no chiral steady state in CPTP evolution, a result that can be proved via channel–circuit duality \cite{Fan2025toappear}. It would be illuminating to see how these constraints shed light on isoTNS states.

As discussed in Sec.\ref{sec:iso},  the framework of isoTNS provides a powerful tool to explore steady-state mixed-state phases and their transitions. For instance, Refs.\cite{liu_iso:2024,Boesl:2025_stringnetiso} constructed an isoTNS path that interpolates between the $\mathbb{Z}_2$ toric-code topological order and the  $\mathbb{Z}_2$ double-semion topological order, with a continuous transition between them. Viewing the isoTNS as a repeated quantum channel immediately leads to the identification of two distinct 1d mixed-state phases separated by a mixed-state transition—an insight that would otherwise be difficult to obtain. One immediate open question is to develop a complete theory for this intriguing mixed-state quantum criticality. More broadly, the isoTNS framework may provide a unifying platform for uncovering and classifying a wide range of mixed-state phenomena\cite{li2025subdimensional,Fan2025toappear,Lee2025toappear}.

Finally, in this work we have mostly focused on the cases where a strongly symmetric quantum channel gives rise to an SW-SSB steady state and argue that the corresponding holographic bulk is topologically ordered. An important open question we did not address is: what are the minimal requirements for a quantum channel to exhibit SW-SSB in its steady state? While SW-SSB steady states are expected in general, there are counterexamples where the steady states can exhibit strong-to-nothing SSB (such as GHZ state) or symmetric product state (such as $\ket{+++ \cdots}$), although these examples require certain additional fine-tuned structure, and are not stable under symmetric perturbation to quantum channels; see e.g. Ref. \cite{gu2024spontaneous,shu2026universal}. It would be interesting to explore whether the holographic approach, combined with tensor-network theory, helps identify the mathematical conditions underlying SW-SSB steady states.

 \acknowledgments 
 We thank Michael Knap for helpful discussions.
This research was completed in part by grant NSF PHY-2309135 to the Kavli Institute for Theoretical Physics (KITP).  TCL acknowledges the support of the RQS postdoctoral fellowship through the National Science Foundation (QLCI grant OMA-2120757). YJL acknowledges support from the MIT Center for Theoretical Physics - a Leinweber Institute, NSF CIQC Award No. 10434 and NSF Early Career DMR-2237244. YY
acknowledges support from NSF under award number
DMR-2439118.

\appendix

\section{Derivation of the deformed toric code}\label{appendix:toric_code_derivation}

Given the quantum channel $\cE = \prod_i \cE^z_{i} \cE^x_{i}$ acting on a 1d system, with 

\begin{equation}\label{}
\begin{split} 
   &\cE^z_{i}[\rho_0] = (1-p_z) \rho_0 + p_z  Z_i Z_{i+1}\rho_0 Z_i Z_{i+1}. \\
   &  \cE^x_{i}[\rho_0] =  (1-p_x) \rho_0 + p_x X_i \rho_0 X_i, 
\end{split}
\end{equation}
we show that the repeated application of this channel gives rise to a deformed toric-code wave function in two spatial dimensions. 

For simplicity, we first consider the channel at $p_x = p_z = \frac{1}{2}$ to illustrate the emergence of a fixed-point toric code state. The derivation for the general $p_x$ and $p_z$ will be discussed later.

As we have discussed in the main text, a repeated quantum channel acting on the 1d system is a process where this 1d system propagates upward to sequentially interact with ancilla qubits row-by-row, as shown in the following figure:

\begin{equation*}
\includegraphics[width=6cm]{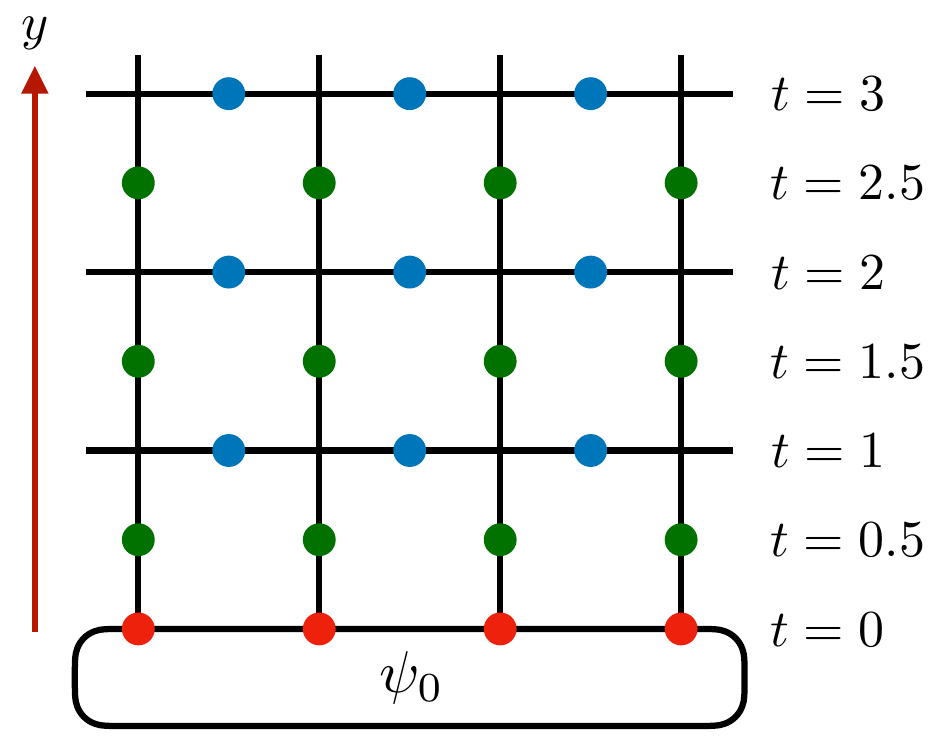}
\end{equation*}

In other words, system (red) qubits are initialized in the state $\ket{\psi_0}$ at time $t=0$, and they travel along the vertical $y$ direction to entangle with the green and blue qubits, which are responsible for X noise and ZZ noise, at various time slices. Note that the spatial ($y$) direction can also be identified as the temporal direction, so we may use $y$ or $t$ interchangeably. For instance, in the figure above, we use $t$ to label the ancilla qubits at various rows; $t=0.5, 1.5, 2.5 ~... $ label the time slices when the system (red) qubits interact with the ancilla qubits responsible for the $X$ noise, and  $t=1, 2, 3~... $ label the time slices when the system (red) qubits interact with the ancilla qubits responsible for the $ZZ$ noise. 

\begin{figure}[h]
\centering
\includegraphics[width=\columnwidth]{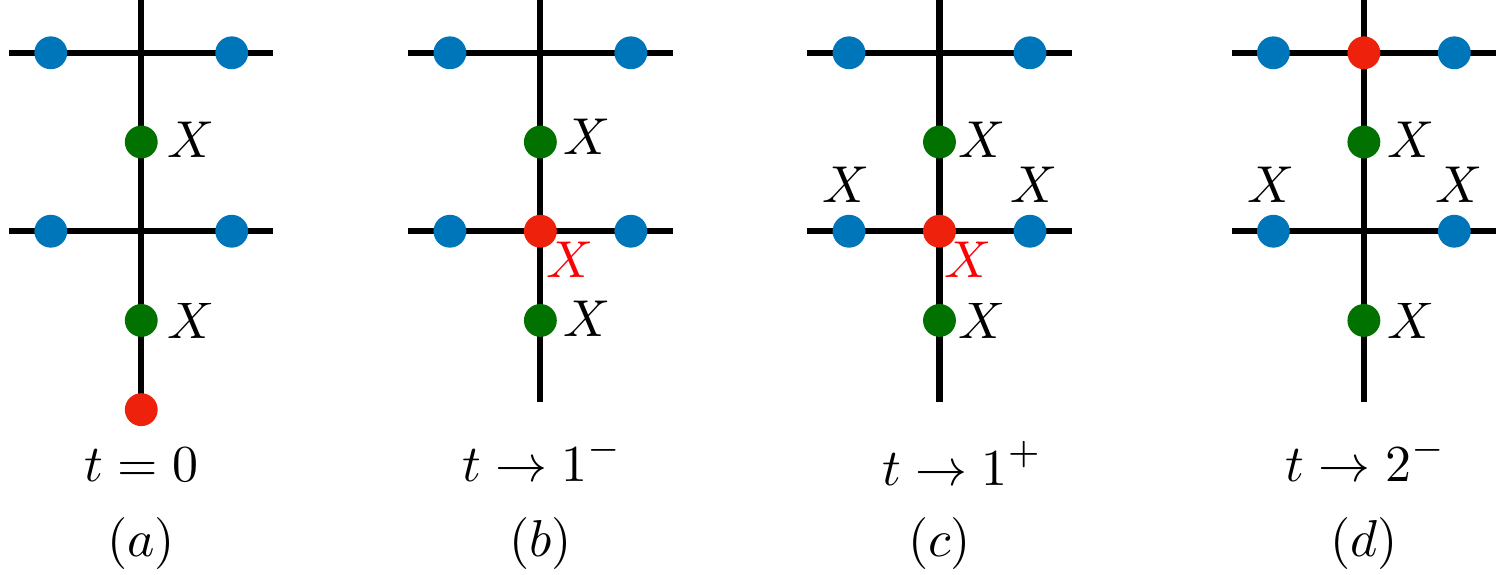}
\caption{Evolution of $XX$ stabilizers to the star operator $A_v =X^{\otimes 4}$ under the sequential circuit, at various time slices.}
\label{fig:star_evolution}
\end{figure}

As discussed in the main text, at $p_x = p_z = \frac{1}{2}$, the green and blue qubits are initialized at $ \ket{+}$ and $\ket{0}$, respectively. To see the emergence of the stabilizer $A_v=X^{\otimes 4}$ under the sequential unitary circuit, we notice that at $t=0$, since the green qubits are initialized at $\ket{+}$, the initial state has the product of two Pauli-Xs as a stabilizer acting on the two green qubits as shown in Fig.\ref{fig:star_evolution}(a). Next, the red qubit will entangle with the green qubit at $t=0.5$ via the controlled gate: $\ket{0}\bra{0}_a +   \ket{1}\bra{1}_a X_s$, where $a$ and $s$ refer to the ancilla (green) qubit and system (red) qubit, respectively. Under the conjugation of this gate, the red qubit will be attached by an $X$ operator, so the original $XX$ stabilizer at $t=0$ becomes an $XXX$ stabilizer, which will remain up to the time $t \to 1^-$, i.e., right before interacting with the blue qubits (Fig.\ref{fig:star_evolution}(b)). At the time $t=1$, the red qubit is entangled with the left and right neighboring blue qubits, both using the controlled-gate $\ket{+}\bra{+}_a +   \ket{-}\bra{-}_a Z_s$, implying the stabilizer will be further attached with two Pauli-Xs on those blue qubits (Fig.\ref{fig:star_evolution}(c)). After this, the red qubit will entangle with the green qubit at $t=1.5$ using the controlled gate $\ket{0}\bra{0}_a +   \ket{1}\bra{1}_a X_s$, which will annihilate the Pauli-X on the system red qubit (Fig.\ref{fig:star_evolution}(d)). Therefore, the original two-body XX stabilizer is mapped to the star operator $A_v =X^{\otimes 4}$ under the sequential circuit.

\begin{figure}[h]
\centering
\includegraphics[width=\columnwidth]{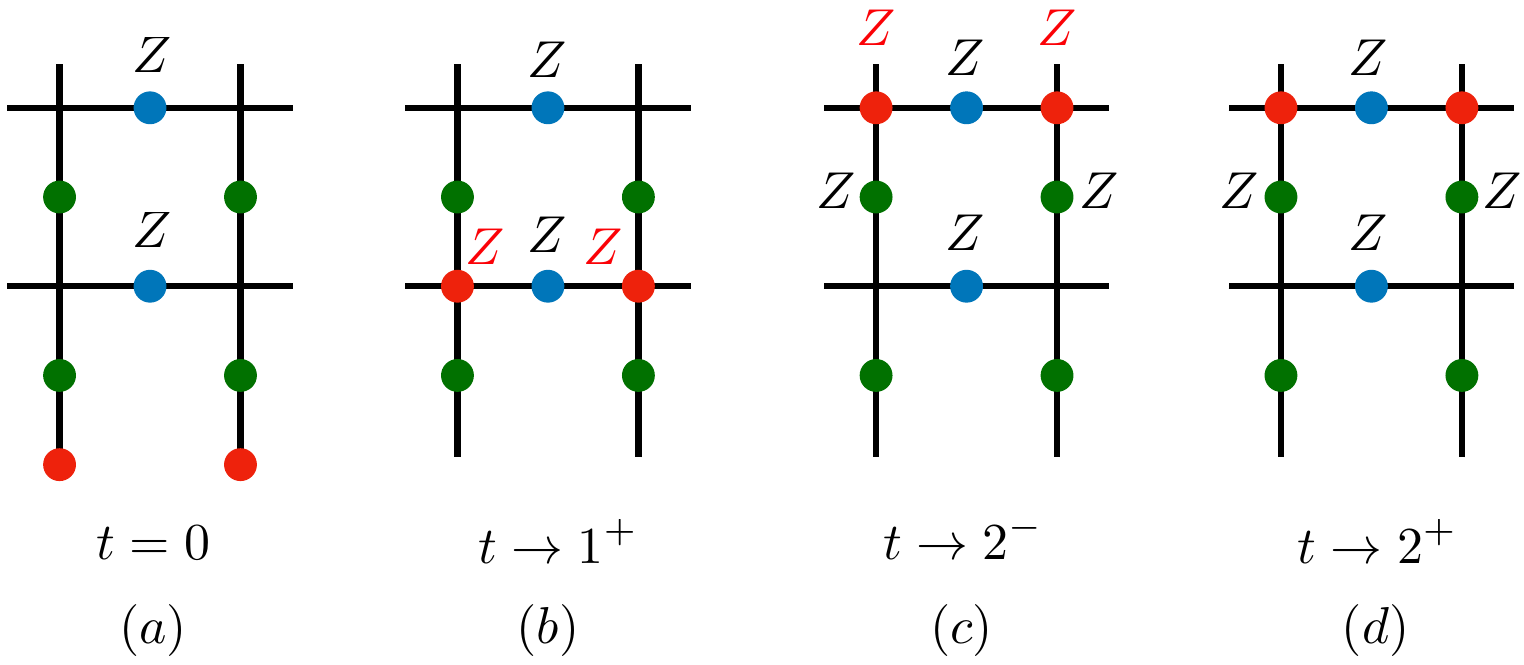}
\caption{Evolution of $ZZ$ stabilizers to the plaquette operator $B_p =Z^{\otimes 4}$ under the sequential circuit, at various time slices.}
\label{fig:plaquette_evolution}
\end{figure}

Now we follow a similar strategy to show the emergence of the plaquette stabilizer $B_p=Z^{\otimes 4}$ under the sequential circuit. At $t=0$, since all the blue qubits are in the state $\ket{0}$, one can define the $ZZ$ stabilizer on the two blue qubits shown in Fig.\ref{fig:plaquette_evolution}(a). At $t=1$, the two red qubits are entangled with the blue qubit in between, both using $\ket{+}\bra{+}_a +   \ket{-}\bra{-}_a Z_s$, after which the $ZZ$ stabilizer is decorated by two $Z$ operators on the red qubits (Fig.\ref{fig:plaquette_evolution}(b)). Then, those two red qubits will entangle with the two green qubits at $t=1.5$. This will generate Pauli-Zs on the green qubits, so one has a six-body stabilizer (Fig.\ref{fig:plaquette_evolution}(c)). Finally, at $t=2$, the red qubits interact with the blue qubit in between, leading to the emergence of the four-body plaquette stabilizer $B_p=Z^{\otimes 4}$  (Fig.\ref{fig:plaquette_evolution}(d)). 

Therefore, at $p_x = p_z = \frac{1}{2}$, the repeated quantum channel is exactly a sequential unitary circuit $U$ that prepares a toric code

\begin{equation}\label{eq:appendix_toric}
\ket{\text{TC}}  = U \left[  \ket{\psi_0}_s \prod_{a \in \text{x-link}} \ket{0}_a \prod_{a\in \text{y-link}  }\ket{+}_a  \right],  
\end{equation}
where $\ket{\psi_0}_s$ is the initial state of the system qubits, and the ancilla qubits in horizontal edges (x-links) and vertical edges (y-links) are initialized at $\ket{0}$ and $\ket{+}$, respectively. 

Away from $p_x=p_z =\frac{1}{2}$, the corresponding channel can be implemented by deforming the initial state (Eq.\ref{eq:z_deform} and Eq.\ref{eq:x_deform}):

\begin{equation}
    \ket{\psi_0}_s \prod_{a \in \text{x-link}} e^{g_z X_a}    \ket{+}_a \prod_{a\in \text{y-link}  }  e^{g_x Z_a}  \ket{0}_a 
\end{equation}
where the ancilla qubits are deformed with the imaginary time evolution $ e^{g_z X_a}$ or $ e^{g_x Z_a}$ with 

\begin{equation}
e^{2g_z} = \frac{1-p_z}{p_z}, ~~ e^{2g_x} = \frac{1-p_x}{p_x}.
\end{equation}  

Given the above deformed initial state, the sequential unitary circuit $U$ generates the following pure state $|\Psi^{2d}(g_x,g_z)\rangle$: 
\begin{equation}
\begin{split}
&|\Psi^{2d}(g_x,g_z)\rangle\\
&= U\left[      \ket{\psi_0}_s \prod_{a \in \text{x-link}} e^{g_z X_a}    \ket{+}_a \prod_{a\in \text{y-link}  }  e^{g_x Z_a}  \ket{0}_a 
 \right].  
 \end{split}
\end{equation}
The reduced density matrix obtained by tracing out all the ancilla qubits is the output of the repeated quantum channel.

Notably, in the sequential circuit $U$, the x-link ancilla qubits are entangled by the X-basis controlled-gate $\ket{+}\bra{+}_a  + \ket{-}\bra{-}_a Z_s$, so the deformation $ e^{g_z X_a}$ commutes with $U$. Similarly, the y-link ancilla qubits are entangled by the Z-basis controlled gates ($\ket{0} \bra{0}_a + \ket{1} \bra{1}_a X_s$), so the deformation $ e^{g_x Z_a}$ commutes with $U$ as well. Therefore, those imaginary time evolutions can be pushed to the left of $U$, implying (using Eq.\ref{eq:appendix_toric}): 
\begin{equation}\label{eq:deformed_appendix_2d}
|\Psi^{2d}(g_x,g_z)\rangle= \prod_{a \in \text{x-link}} e^{g_z X_a} \prod_{a\in \text{y-link}  }  e^{g_x Z_a} \ket{\text{TC}}, 
\end{equation}
which is a deformed toric-code state shown in Eq.\ref{eq:deformed_2d_toric}.

\section{Absence of finite-$g$ transition}\label{sec:transition}

Given the deformed toric code
\begin{equation}\label{}
|\Psi^{2d}(g_x,g_z)\rangle=\prod_{\ell \in x\,\text{link}}  e^{g_z X_{\ell }} \prod_{\ell \in y\,\text{link}}  e^{g_x Z_{\ell }} \ket{\text{TC}}, 
\end{equation}
here we show that its wavefunction overlap is analytic for any finite $g_x, g_z$, implying the absence of a transition. We also note that the analyticity of the wavefunction overlaps is commonly adopted to explore the stability of wavefunctions; see e.g. Ref.\cite{stability_tn_2021}. 

First, the wavefunction overlap is written as 

\begin{equation}
\begin{split}
&\braket{\Psi^{2d}(g_x,g_z) |\Psi^{2d}(g_x,g_z) }\\
&= \bra{\text{TC}}  \prod_{\ell \in x\,\text{link}}  e^{2g_z X_{\ell }} \prod_{\ell \in y\,\text{link}}  e^{2g_x Z_{\ell }}   \ket{\text{TC}}\\ 
& \propto \bra{\text{TC}}  \prod_{\ell \in x\,\text{link}} (1+  t_zX_\ell   )   \prod_{\ell \in y\,\text{link}}  (1+t_x  Z_\ell   )  \ket{\text{TC}}  
\end{split}
\end{equation}
with $t_x  \equiv   \tanh(2g_x ), t_z  \equiv   \tanh(2g_z )$.

The term  $\prod_{\ell \in y\,\text{link}}  (1+t_x  Z_\ell   ) $ can be expanded as a sum of all possible Pauli-Z insertions on y-links. Crucially, whenever those Zs do not form a closed loop, they will violate $X^{\otimes 4}$ and create $e$ charges on vertices, and hence, have vanishing contribution in the expectation value w.r.t. $\ket{\text{TC}}$. Only those Zs that extend from the top to the bottom boundaries may have a non-vanishing contribution. The weight of a given extended Z-string decays exponentially with the vertical lattice size $L_y$ (for any finite $g_x, g_z$), but the multiplicity is only polynomial in the horizontal lattice size $L_x$. Therefore, the contributions of these extended Z-strings will vanish in the thermodynamic limit (fixed $\frac{L_x}{L_y}$ and $L_y\to \infty$). Similar reasoning holds for $ \prod_{\ell \in x\,\text{link}} (1+  t_zX_\ell   )$. Therefore, the wavefunction overlap will follow 

\begin{equation}
 \braket{\Psi^{2d}(g_x,g_z) |\Psi^{2d}(g_x,g_z) }   =   1+...
\end{equation}
where all terms in $...$ will vanish in the thermodynamic limit, for finite $g_x, g_z$.

\section{MBQC perspective}\label{sec:mbqc}

The measurement-based quantum computation (MBQC) \cite{one_way_2001_mbqc,mbqc_rbh_2003,mbqc_2009} is a scheme to implement a spacetime quantum circuit acting on a system in $d$-spatial dimensions by measuring the bulk qubits of a resource state in one higher dimension. As discussed in Ref.\cite{Foliated_2016,foliation_anomaly_2024}, MBQC also admits a holographic correspondence between a bulk SPT order and the boundary topological order through the foliation construction. This correspondence is achieved via projective measurement, whereas our correspondence is achieved by a distinct operation, namely, a partial trace, and consequently, leading to a correspondence between boundary mixed-state orders and bulk topological orders. 

While the physical aspects of our framework differ from the holographic perspectives in MBQC, we will show below that our sequential-circuit holographic picture also admits an MBQC-like interpretation, although a post-selection is required away from the fixed-point limit. Specifically, by coupling a 1d input state to the bottom boundary of a 2d cluster-state SPT, we measure (and post-select) on a subset of qubits to exactly implement the noise channels discussed in the previous subsection (Eq.\ref{eq:noise}), thereby generating the same output mixed state on the top boundary of the 2d lattice.

To make contact with the MBQC, the essential step is to notice that the implementation of the X channel can be understood as an imperfect implementation of quantum teleportation, during which a Pauli-X is probabilistically applied. For simplicity,  we first consider $p_x=\frac{1}{2}$, so that a Pauli-X is applied with probability $\frac{1}{2}$, corresponding to the maximal decoherence in the X basis. The case for the general $p_x$ will be discussed later.

To implement the X-noise channel with $p_x = \frac{1}{2}$ on $\ket{\psi}_1 = \psi_+  \ket{+}_1 + \psi_- \ket{-}_1$ of the system qubit $1$, we can first introduce a new system qubit $2$ and one ancilla qubit (labeled by $a$), both initialized at $\ket{+}$. Namely, the initial state is written as 

\begin{equation}
\frac{1}{\sqrt{2}}\ket{\psi}_1 \ket{+}_2   ( \ket{0}_a + \ket{1}_a   ). 
\end{equation}   
We then apply the controlled gate $\text{CZ}_{1a} \text{CZ}_{2a}$, after which the state becomes 

\begin{equation}
\frac{1}{\sqrt{2}}  (    \ket{\psi}_1 \ket{+}_2   \ket{0}_a  + Z_1 \ket{\psi}_1 \ket{-}_2  \ket{1}_a ). 
\end{equation}
Then we project the qubit $1$ to $\ket{+}$, which can be implemented by measuring $X_1$ with +1 measurement outcome. The resulting state is 

\begin{equation}
\begin{split}
    & \psi_+  \ket{+}_2   \ket{0}_a  +     \psi_-  \ket{-}_2   \ket{1}_a \\
    \propto  &\ket{\psi}_2  \ket{+}_a  + X_2  \ket{\psi}_2  \ket{-}_a 
\end{split}
\end{equation}
where we have omitted a normalization. Tracing out the ancilla qubit then gives the desired X channel with $p_x = \frac{1}{2}$:   

\begin{equation}
\cE^x[\rho] =  \frac{1}{2} \rho   +  \frac{1}{2} X\rho X.
\end{equation}

We can immediately generalize the above result to the case where every qubit of the 1d system is subject to single-site $X$ noise: 

\begin{equation}
 \cE^x_{i}[\rho_0] =  (1-p_x) \rho_0 + p_x X_i \rho_0 X_i,     
\end{equation}
which can be schematically represented as

\begin{equation}\label{eq:x_figure}
\includegraphics[width=4cm]{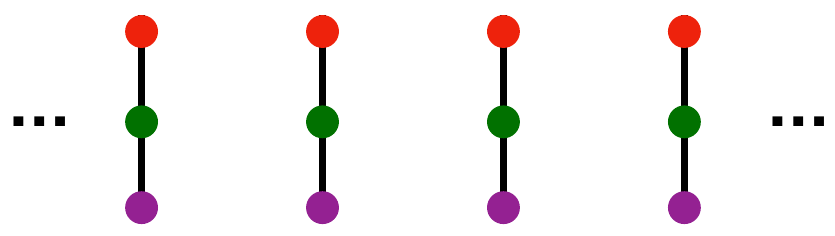}
\end{equation}
The above depicts the following process: starting with a 1d system of qubits (marked in purple), we introduce the ancilla green and red qubits initialized at $\ket{+}$, and extensively apply the 2-body CZ gates depicted by the black bonds. Then we project every purple qubit to $\ket{+}$. The output of the X noise channel is then given by the reduced density matrix on the red qubits (by tracing out the green qubits).

On the other hand, as we have discussed in the main text, to implement the ZZ channel with noise strength $p_z = \frac{1}{2}$ acting on the two system qubits $1$ and $2$, we introduce an ancilla qubit $a$ initialized at $\ket{+}$. By coupling the ancilla with the unitary gate $\text{CZ}_{1,a} \text{CZ}_{2,a}$ and tracing out $a$, the following ZZ channel acting on the two system qubits can be implemented:

	\begin{figure*}
		\centering
\includegraphics[width=0.85\textwidth]{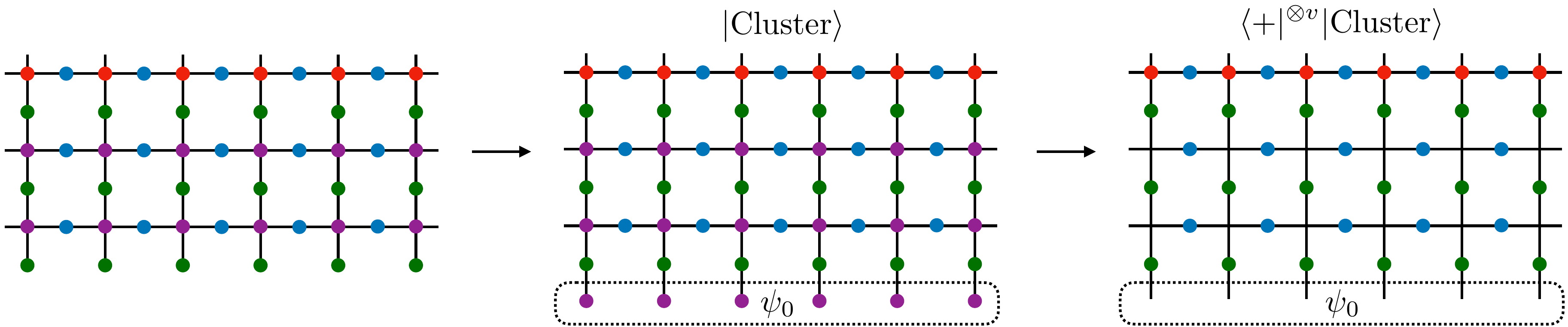}       
\caption{In the left panel, every qubit is initialized at $\ket{+}$ and then the CZ gate is extensively applied at the locations shown by the black lines. To utilize this 2d state as a resource within the MBQC framework, we couple it to a 1d system initialized at $\ket{\psi_0}$ to create a cluster state $\ket{\text{Cluster}}$ with a specific bottom boundary condition (middle panel). In the right panel, a subsequent projection on all vertex qubits to $\ket{+}$ gives rise to the 2d pure state generated by the sequential quantum circuit, and the reduced density matrix of the red qubits on the top boundary is the output of the repeated quantum channel.}
		\label{fig:mbqc_process}
	\end{figure*}

\begin{equation}
\cE^z[\rho] =  \frac{1}{2} \rho   +  \frac{1}{2} Z_1Z_2 \rho Z_1 Z_2.
\end{equation}

One can then extensively apply the ZZ channel acting on a 1d system, which can be schematically represented as
\begin{equation}\label{eq:zz_figure}
\includegraphics[width=4cm]{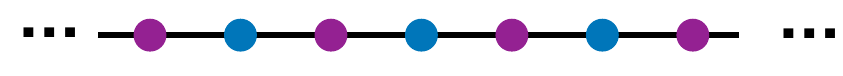}
\end{equation}
The above figure depicts the following process: starting with a 1d system of qubits (marked in purple), we introduce the ancilla blue qubits initialized at $\ket{+}$, and extensively apply the 2-body CZ gates depicted by the black bonds. The output of the ZZ noise channel is then given by the reduced density matrix on the purple qubits.

Based on the above understanding, the quantum channel ($\cE = \prod_i \cE^z_{i} \cE^x_{i}$) for $p_x=p_{z} = \frac{1}{2}$ can then be repeated applied by iterating the process shown in Eq.\ref{eq:x_figure} and Eq.\ref{eq:zz_figure}, in which one sequentially introduces rows of ancilla qubits in $\ket{+}$ followed by unitary entangling CZ gates and measurement/projection. In particular, the use of CZ gates and the projection allows us to reformulate the process in a way similar to MBQC; see Fig.\ref{fig:mbqc_process}. To see this, we first note that since all the CZ gates commute, we can first extensively apply CZ gates to create a 2d cluster state. Then we use a layer of CZ gates to entangle this cluster state to the 1d bottom boundary that specifies the initial state of the physical qubits. Denoting this cluster state with the specified bottom boundary by $\ket{\text{Cluster}}$. Projecting all vertex qubits into $\ket{+}$ state in $\ket{\text{Cluster}}$ gives rise to the 2d pure state  ($\langle +|^{\otimes v}  \ket{\text{Cluster}}$), which is exactly the one created by the sequential unitary circuit that implements the repeated quantum channel. It follows that the top boundary of this 2d pure state $\langle +|^{\otimes v}  \ket{\text{Cluster}}$ is the steady state of the quantum channel. In particular, one finds 

\begin{equation}
\langle +|^{\otimes v}  \ket{\text{Cluster}}  =  \ket{\overline{\text{TC}}}, 
\end{equation}
where $\ket{\overline{\text{TC}}}$ is exactly the 2d fixed-point toric-code wavefunction with the plaquette stabilizers $X^{\otimes 4}$ and the star stabilizers $Z^{\otimes 4}$. Note that $\ket{\overline{\text{TC}}}$ is related to the fixed-point toric code $\ket{\text{TC}}$ defined in Eq.\ref{eq:deformed_2d_toric} by a Hadamard rotation ($X\leftrightarrow Z$) acting on all edge qubits in the bulk, so they have the same 1d top boundary, which is the SW-SSB mixed state $  \propto (1+\prod_i X_i )$.

For general $p_x, p_z$, one can instead consider the deformed cluster state with imaginary-time evolution acting only on the edge qubits:
\begin{equation}\label{}
\prod_{\ell \in x\,\text{link}}  e^{g_z Z_{\ell }} \prod_{\ell \in y\,\text{link}}  e^{g_x X_{\ell }} \ket{\text{Cluster}}.
\end{equation} 

A subsequent projection on the vertex qubits commutes with the deformation on edges, giving rise to

\begin{equation}
  \prod_{\ell \in x\,\text{link}}  e^{g_z Z_{\ell }} \prod_{\ell \in y\,\text{link}}  e^{g_x X_{\ell }} \ket{ \overline{\text{TC}}},
\end{equation}
which is the same as the state in Eq.\ref{eq:deformed_2d_toric} (again, up to the Hadamard gates acting on the bulk qubits). To summarize, the repeated quantum channel can be implemented by preparing a deformed cluster state followed by a subsequent projection on vertex qubits.

\section{Counter example of holographic duality}\label{sec:pim} 
In this section, we present an explicit example of a quantum channel that preserves additional symmetry, and therefore, the dual isoTNS violates the $G$-injective condition defined in Sec. \ref{sec:main_g_injectivity}. Correspondingly, the holographic bulk exhibits GHZ-type or subsystem-symmetry breaking order, instead of topological order.

The example is the special limit of the quantum channel in Eq.~\ref{eq:noise} at $p_x=0$. Specifically, the 1d quantum channel is given by $ \cE =  \prod_i \cE^z_{i}$ with

\begin{equation}
   \cE^z_{i}[\rho_0] = \frac{1}{2}\rho_0 + \frac{1}{2} Z_i Z_{i+1}\rho_0 Z_i Z_{i+1}. 
\end{equation}

This channel only contains Kraus operators diagonal in the $Z$ basis, and therefore, any density matrix diagonal in the $Z_i$ basis is a steady state, and the steady-state manifolds have a dimension $2^{L_x}$. In particular, in addition to preserving the global $\mathbb{Z}_2$ symmetry generated by $\prod_{i=1}^{L_x }X_i$, this channel also has extensive symmetries generated by local $Z_i$ operator for $i\in \{1,2,..., L_x \}$. As such, this channel violates the $\mathbb{Z}_2$-injectivity condition. 

Due to the absence of the X-noise channel, we only need to introduce the ancilla qubits on the horizontal edges of the 2d lattice to implement the above ZZ noise channel. Following the same procedure described in Appendix.\ref{appendix:toric_code_derivation}, we find that the 2d wavefunction is characterized by the following stabilizers: 

\begin{equation}\label{stab1}
\includegraphics[width=5.5 cm]{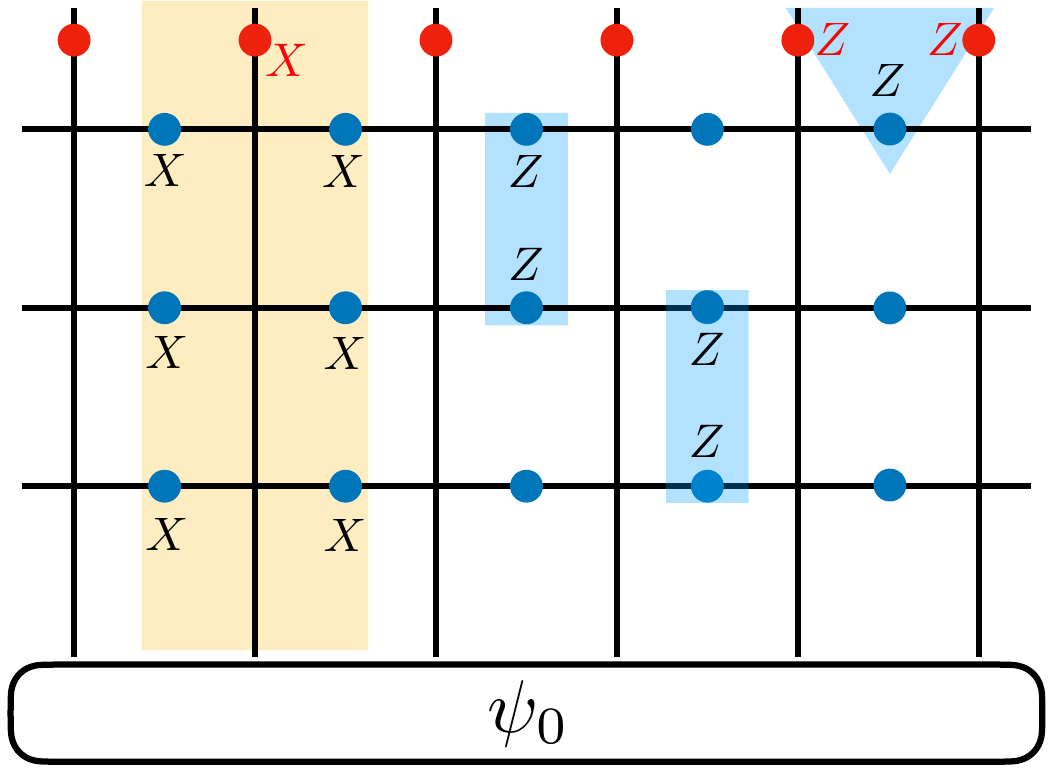}
\end{equation} 
Namely, the bulk is characterized by two-body ZZ stabilizers between the two qubits on consecutive rows at a fixed horizontal position. The boundary is characterized by the three-body $ZZZ$ stabilizers that couple the bulk qubits to the boundary qubits, as shown in the blue shaded regions. To see how these stabilizers emerge, one simple way is to consider the toric-code stabilizers shown in Fig.\ref{fig:2d_toric}, and then project all the vertical ancilla qubits to $Z=1$, which corresponds to the $p_x=0$ limit. Such a projection annihilates the XXXX stabilizers of the toric code and reduces the ZZZZ stabilizers to the two-body ZZ stabilizer.

By further assuming every qubit is initialized in $\ket{+}$, which is stabilized by on-site Pauli-X operators, a single Pauli X evolves into a nonlocal X-stabilizer supported on the yellow shaded region. The product of all such stabilizers cancels the bulk Pauli operators and leaves a boundary operator $\prod_{i=1}^{L_x}X_i$, which acts as a strong symmetry of the 1D mixed state. Correspondingly, the 1D mixed steady state is $\rho_s \propto (1+\prod_{i=1}^{L_x}  X_i )$ which exhibits SW-SSB order, even though the bulk does not exhibit a topological order; instead, the stabilizers shown in Fig.\ref{stab1} imply a GHZ-type symmetry-breaking order along each column in the bulk. On the other hand, if choosing the GHZ state $\ket{000\cdots} + \ket{111 \cdots}$ as an initial state, it is straightforward to see that the ZZ channel acts trivially, so it will become a steady state, which exhibits SN (strong-to-nothing)-SSB instead of SW-SSB \footnote{In fact, any GHZ-type state $\ket{\{\sigma_1, \sigma_2, ... , \sigma_{L_x}} +  \prod_{i=1}^{L_x} X_i\ket{\{\sigma_1, \sigma_2, ... , \sigma_{L_x}}$, with $\{\sigma_i\}$ being a computational Z-basis state, is a valid steady state that exhibit SN-SSB instead of SW-SSB.}. As such, in this fine-tuned channel, the steady state needs not exhibit SW-SSB, consistent with the fact that this channel preserves additional symmetries beyond the global $\mathbb{Z}_2$ symmetry generated by $\prod_i X_i$, and therefore violates the $G$-injectivity.

As a final remark, it is also useful to connect the extensive degeneracy of the ZZ channel to subsystem symmetry breaking of the 2d plaquette Ising model. In particular, in the 2d wavefunction shown in Fig.\ref{stab1}, in addition to the stabilizers shown in the figure, on each row of the ancilla bulk, there is a non-local stabilizer $\prod_{ i \in \text{x-link} } Z_i=1$ imposed as a constraint on the holographic wavefunction. This restricts individual rows of ancilla qubits to a symmetric sector, which allows us to perform a Kramers-Wannier duality transformation to each ancilla row: 

\begin{equation}\label{eq:kw}
\includegraphics[width=4.3 cm]{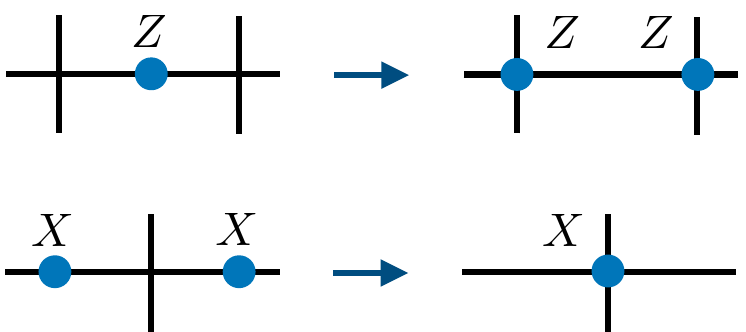}. 
\end{equation} 

Under such a non-local transformation, the bulk stabilizers take the form:
\begin{equation}\label{stab2}
\includegraphics[width=5 cm]{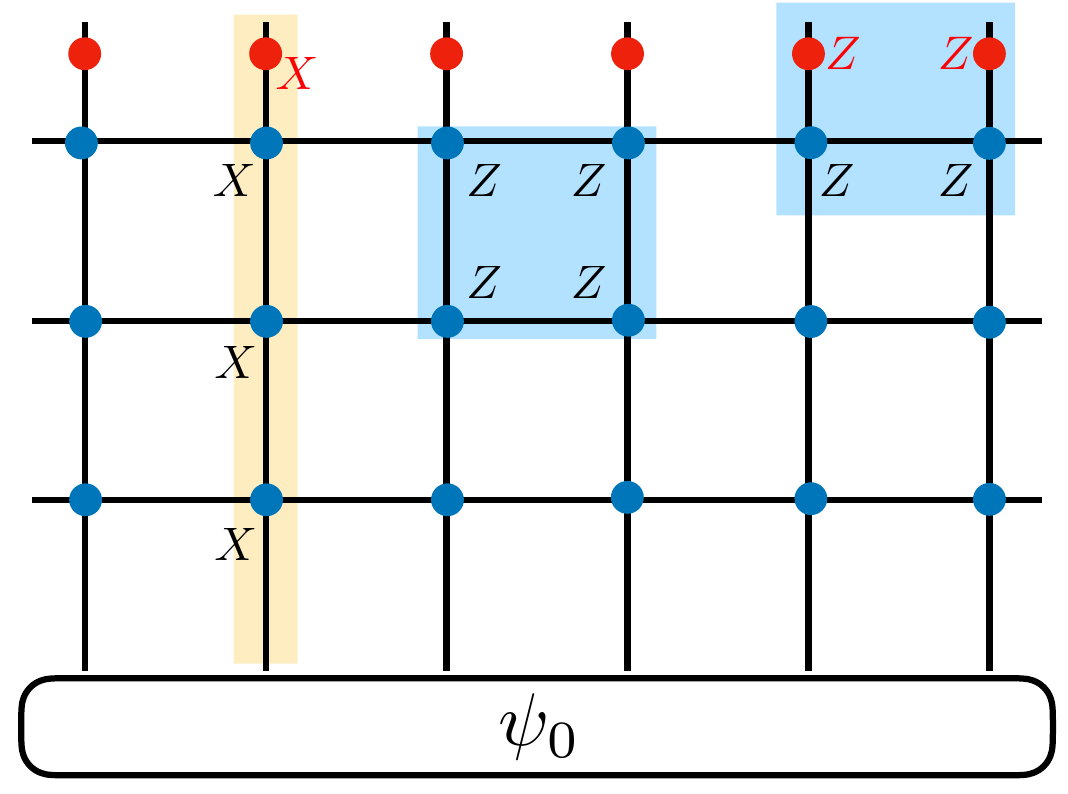}.
\end{equation} 

Notably, the resulting four-body interaction (blue shaded area in Eq.~\ref{stab2}) is precisely the plaquette Ising model (PIM) on the square lattice introduced in Ref.~\cite{xu2004strong,you2018subsystem}. A defining property of the PIM is its subsystem \(Z_2\) symmetry along any 1d rigid line of column or row: $\prod_{i \in \text{row}} X_i$ and $\prod_{i \in \text{column}} X_i$, which leads subsystem symmetry breaking and extensive ground-state degeneracy. We note that due to the X-type operator mapping rule in Eq.\ref{eq:kw}, the PIM is fixed at the symmetric sector $\prod_{i \in \text{row}} X_i =1$ along each row. The corresponding allowed degeneracy is therefore $2^{L_x}$ associated with the symmetry generator $\prod_{i \in \text{column}} X_i$ acting on each column.

To summarize, the present example shows that, for certain fine-tuned quantum channels, the steady state can be holographically dual to a 2d wavefunction with GHZ-type symmetry breaking order (Fig.\ref{stab1}) or subsystem symmetry breaking order (Fig.\ref{stab2}), rather than intrinsic topological order. This example is fine-tuned as the channel possesses additional local strong symmetries, which lead to an extensively degenerate steady-state manifold. In terms of isoTNS description, the isometric tensor possesses an additional virtual symmetry, thereby violating the G-injective condition. This is also consistent with the fact that the GHZ-type symmetry breaking order (Fig.\ref{stab1}) or subsystem symmetry breaking order is not robust under a generic perturbation. For instance, turning on \(p_x>0\) will immediately lift the extensive degeneracy, and lead to a topologically ordered bulk.

\section{Emergence of SW-SSB at finite depth}\label{appendix:finite_depth}
We here present a heuristic argument to show that for the non-maximal decoherence channel ($p_z\neq \frac{1}{2}$), the repeated quantum channel with finite depth is sufficient for the emergence of the SW-SSB in the output state.

\begin{figure}[t]
\centering
\includegraphics[width=0.9\columnwidth]{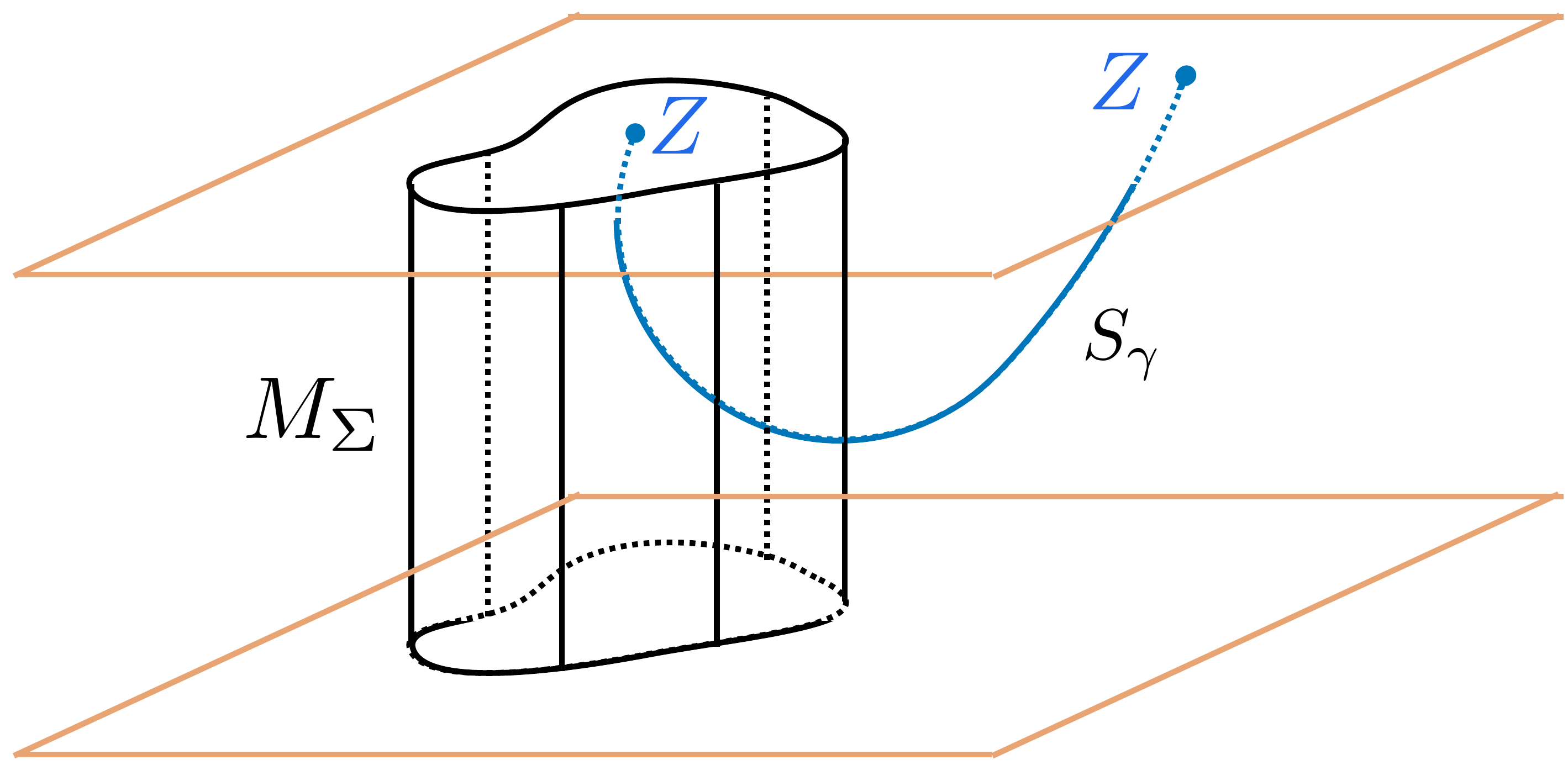}
\caption{The existence of the string symmetry operator $S_\gamma$ with $\gamma$ being any deformable path connecting two points on the top boundary indicates SW-SSB on the top boundary. For a constant but large $L_z$ (the linear size along the $z$ direction), the membrane $M_{\Sigma}$ will be suppressed, implying the emergence of $S_\gamma$ and therefore, SW-SSB on the top boundary.}
\label{fig:3d_brane}
\end{figure}

First, we notice that at the fixed-point limit $p_z=\frac{1}{2}$, the 3d holographic wavefunction is symmetric under the string operators $S_\gamma$ consisting of a product of Pauli-Zs, with two endpoints attached to the top boundary (see Fig.\ref{fig:3d_brane}). Such a string symmetry operator will be reduced to a weak symmetry $Z_i Z_j \rho  Z_i Z_j =\rho$ for the top-boundary reduced density matrix, which indicates a long-range order in the fidelity correlators and SW-SSB. 

At $p\neq \frac{1}{2}$, the fixed-point holographic toric code is deformed by $e^{g_z  \sum_{\ell \in x-y \text{  plane}}  X_\ell }$, and the question is: what is the required depth (namely, the linear size along the $z$ direction) so that $S_\gamma$ can be emergent? 

To estimate the depth, we first write  
\begin{equation}
e^{g_z  \sum_{\ell \in x-y \text{  plane}}  X_\ell } \propto \prod_{\ell \in x-y \text{  plane}} (1+ \tanh g_z X_\ell   )   
\end{equation} 
so the Pauli-X is applied with a probability amplitude $\tanh g_z$. Since the $S_\gamma$ string is deformable, in order to modify its expectation values, the application of Pauli-Xs must form a membrane $M_{\Sigma}$ extending along the $z$ direction from the top boundary to the bottom boundary, and the membrane $\Sigma$ must form a closed loop in the $x-y$ plane that encloses one endpoint of $S_\gamma$ on the top boundary. Such a membrane will necessarily intersect with $S_\gamma$ for any deformable string $\gamma$.

Now we estimate the probability of forming such a membrane $M_{\Sigma}$ with size $L_z$ along the $z$ direction and a circumference $L_{xy}$ on the $x-y$ plane. First, to form such a membrane of Pauli-Xs, the corresponding probability weight is $(\tanh g_z)^{ L_z L_{xy}}$. Meanwhile, the closed loop on the $x-y$ plane is deformable, so one needs to introduce a multiplicity factor $\sim 3^{L_{xy}}$. Therefore, the probability of forming a $M_{\gamma}$ to intersect with $S_\gamma$ is roughly

\begin{equation}  
(\tanh g_z)^{ L_z L_{xy}} 3^{ L_{xy}} =   (3(\tanh g_z)^{L_z}    )^{L_{xy}}
\end{equation}
As such, for a large but finite  $L_z$, the probability of forming $M_\Sigma$ decays exponentially with $L_{xy}$, meaning $S_{\gamma}$ can be emergent and gives rise to the desired SW-SSB on the top boundary. This argument is in spirit similar to Peierls' argument for the existence of an ordered phase in 2d Ising model \cite{peierls1936ising}.

\section{$G$-injective isoTNS and topological order}\label{sec:ginjective}

Here we provide detailed statements and proofs associated with the discussion in Sec.\ref{sec:main_g_injectivity}. We review the notions of \(G\)-injective iso-TNS and \(G\)-isometric tensor-network states. Specifically, by combining the corresponding tensor-network results with our holographic framework, we establish a general connection between generic strongly symmetric one-dimensional quantum channels and their dual two-dimensional topological wavefunctions.

The correspondence between strongly symmetric local quantum channels and topological order is generally expected, on physical grounds, to hold in the absence of fine-tuned channel structures, such as extra accidental symmetries. For a suitable subclass of quantum channels and their dual isoTNS, this expectation can be made rigorous, thereby excluding such fine-tuned cases. We identify this subclass by combining the channel-state duality with results from 2d tensor-network theory.

\textbf{Channel-isoTNS duality}: We first consider quantum-channel evolution in 1d. 
Given a translation-invariant, finite-depth local quantum channel \(\Lambda\) (possibly with a constant-size unit cell), we define its canonical dual isoTNS as follows: 
\begin{enumerate}
    \item Take each local quantum channel $\Lambda_a$ from $\Lambda$, find its Kraus decomposition with Kraus operators, such that
    \begin{equation}
        \Lambda_a[\rho] = \sum_{i = 1}^r T^i \rho \left(T^i\right)^\dag
    \end{equation}
    where $r$ denotes the minimal Kraus rank of the local quantum channel.
    \item Define the associated local tensor for the isoTNS as $T^i$, the entire isoTNS is then contracted from these local tensors (as described in Sec.~\ref{sec:iso}). 
\end{enumerate}
For a given quantum channel \(\Lambda\), its repeated application can be mapped to a canonical dual isoTNS with local physical dimension \(r\). Here \(r\) is the Kraus rank of the local channel and is therefore finite. Moreover, the canonical isoTNS is related to any other dual isoTNS by a local on-site isometry, reflecting the isometric freedom in the choice of Kraus operators. Such local isometries do not change the topological order of the isoTNS.

The central question is therefore: under what conditions does a strongly symmetric quantum channel give rise to a dual isoTNS that is provably topologically ordered? A strong \(G\) symmetry of the quantum channel is mapped, under the channel-state duality, to a virtual \(G\) symmetry of the dual isoTNS. Thus, the physical expectation that generic strongly symmetric channels exhibit SW-SSB steady states is translated, in the holographic picture, into the expectation that generic two-dimensional tensor-network states with a virtual \(G\) symmetry exhibit \(G\)-topological order.
In tensor-network theory, this notion of genericity can be made precise through the concept of \(G\)-injectivity, assuming that \(G\) acts as an on-site symmetry. We briefly review this concept here for completeness, and refer interested readers to Ref.~\cite{schuch2010peps} for a detailed discussion.

\textbf{G-injective tensor network}: Let $g\mapsto U_g$ be a semi-regular representation~\footnote{This means that it contains all the irreducible representations of $G$.} of $G$.
A translation-invariant 2d tensor-network state (TNS) is $G$-injective if it satisfies the following:
\begin{enumerate}
    \item The local tensors $T$ are $G$-symmetric under $U_g$ in the virtual space:
    \begin{equation}\label{eq:ginjective}
        \includegraphics[width = 0.5\linewidth]{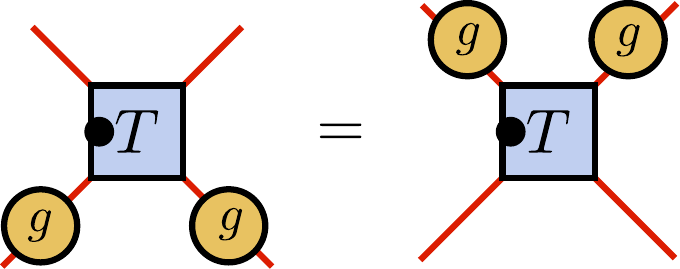},
    \end{equation}
    where we use the same convention as in Sec.~\ref{sec:iso}, and the left-handed black dot denotes the physical leg.
    \item Let $\mathcal{P}(T):(\mathbb{C}^D)^4\mapsto \mathbb{C}^d$ be the map from the virtual space to the physical space. There exists a left inverse $\mathcal{P}(T)^{-1}$ such that $\mathcal{P}(T)^{-1}\mathcal{P}(T) = I|_{U}$, where $I|_{U}$ denotes the orthogonal projector onto the $U_g$-invariant subspace, omitting the normalization:
    \begin{equation}
        \includegraphics[width = 0.65\linewidth]{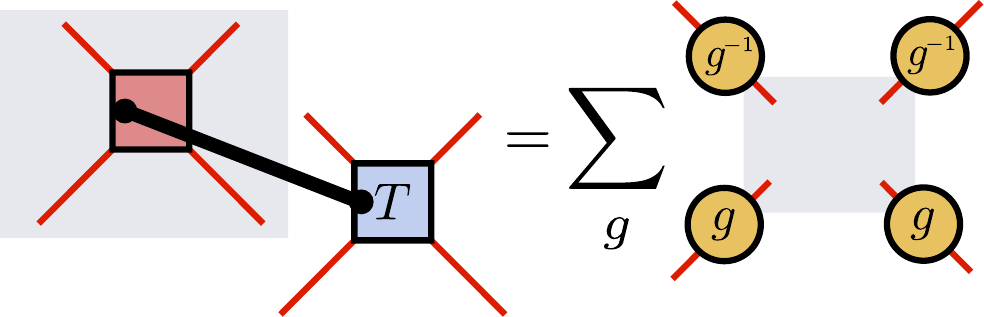}
    \end{equation}
    where on the left-hand side, contracting the physical legs of the red tensor with those of the $T$-tensor implements the left inverse $\mathbb{P}(T)^{-1}$.
\end{enumerate}
In other words, the map \(\mathcal{P}(T)\) is injective within the \(U_g\)-invariant subspace. Note that, in some cases, one needs to block several local tensors into a new local tensor for injectivity to hold. Upon blocking more sites, the dimension of the range of \(\mathcal{P}(T)\) grows faster than the dimension of the domain of \(\mathcal{P}(T)\). This means that, for a generic non-fine-tuned \(T\) with virtual \(G\) symmetry, we expect the blocked map to eventually become injective after blocking \(O(1)\) local tensors. Imposing \(G\)-injectivity therefore explicitly rules out fine-tuned cases. In particular, it implies that there are no accidental virtual symmetries other than \(G\).

$G$-injective TNS has the following important property:

\begin{theorem}[Theorem 5.9 from Ref.\cite{schuch2010peps}]
    For $G$-injective TNS, there exists a translation-invariant, local frustration-free parent Hamiltonian on a torus with a $K$-fold degenerate ground-state subspace, where $K$ is the number of particle types in the quantum double model of $G$. 
\end{theorem}
Although this result alone does not prove that a \(G\)-injective TNS is topologically ordered, the exact ground-state degeneracy strongly suggests that the ground state is either symmetry-breaking or topologically ordered. It is worth emphasizing that this statement concerns \(G\)-injectivity for a \textit{general TNS}, without imposing the additional isoTNS condition that the transfer matrix defines a CPTP map. In what follows, we show that imposing both the isoTNS structure and \(G\)-injectivity rules out the symmetry-breaking interpretation, thereby establishing that the state realizes topological order.

In fact, for the subclass of $G$-injective TNS known as $G$-isometric TNS-where $g\mapsto U_g$ is the left regular representation of $G$ (up to a basis change)~\footnote{The left regular representation is defined on a vector space of dimension $|G|$ and satisfies $U_g\ket{h} = \ket{gh}$ for $g,h\in G$.} and the left inverse of the map $\mathcal{P}(T)$ satisfies $\mathcal{P}(T)^{-1} = \mathcal{P}(T^\dagger)$-one can rigorously establish topological order. These \textit{\(G\)-isometric TNS}, which should not be confused with \textit{\(G\)-injective isoTNS}, are mathematically more tractable because they correspond to fixed points of real-space renormalization-group flow and therefore exhibit zero correlation length.

\begin{theorem}[From Ref.\cite{schuch2010peps}]
    $G$-isometric TNS are fixed-point wavefunctions with $G$-topologically order. 
\end{theorem}
In Ref.~\cite{schuch2010peps}, the topological order of $G$- isometric TNS is established by verifying a set of defining properties, including the existence of a commuting parent Hamiltonian with topological ground-state degeneracy, the local indistinguishability of the ground states, a nonzero topological entanglement entropy, and the structure of anyonic excitations.

In particular, we quote the result on the nontrivial topological entanglement entropy for $G$-isometric TNS: 
\begin{theorem}[Theorem 6.9 from Ref.\cite{schuch2010peps}]
Consider a $G$-isometric TNS, the reduced density operator $\rho_D$ of any topologically trivial block $D$ has rank $|G|^{|\partial D|-1}$, its von
Neumann entropy $S$ and all Rényi entropies are 
\begin{equation}
    S(\rho_D) = a(|\partial D|) - \log |G|.
\end{equation}

\end{theorem}
$\log |G|$ is the constant correction to the area-law scaling of the entropy, and is known as topological entanglement entropy \cite{Levin-2006,kitaev-preskill}.

We can now translate these results into statements about 1D strongly symmetric quantum channels and their dual 2D isoTNS.

\begin{defn}[$G$-injective quantum channel]
    A 1d translation-invariant (up to a constant-size unit cell) local quantum channel $\Lambda$ is $G$-injective if its canonical dual isoTNS is $G$-injective. 
\end{defn}
Here, the \(G\)-injectivity condition for quantum channels provides a mathematically sufficient criterion for characterizing `generic local quantum channels' with no additional strong symmetries beyond \(G\). In the isoTNS language, this condition rules out any additional virtual symmetries on the virtual bonds beyond the virtual \(G\)-symmetry. Consequently, the degenerate dominant eigenvectors of the transfer matrix arise solely from the distinct \(G\)-charge sectors.
In what follows, we demonstrate that the holographic dual of such a channel, described by a \(G\)-injective isoTNS, has a parent Hamiltonian with nontrivial ground-state degeneracy and is therefore expected to realize a topologically ordered phase.

To make this connection explicit, we now introduce an analogue of \(G\)-isometric TNS within the class of \(G\)-injective isoTNS. We first clarify the terminology. The term `isoTNS' refers to a local isometric structure that ensures the tensor network defines a trace-preserving completely positive quantum channel. By contrast, a `\(G\)-isometric TNS' refers to a special fixed-point subclass of \(G\)-injective TNS with additional algebraic structure. Thus, despite the shared use of the word `isometric,' the notion of a \(G\)-isometric TNS is distinct from that of an isoTNS. 

Although \(G\)-isometric TNS and \(G\)-injective isoTNS are distinct notions, we demonstrate that they belong to the same phase of matter by establishing the following result:
\begin{prop}\label{prop}
    $G$-isometric TNS form a subclass of isoTNS; they correspond to $G$-injective (fixed-point) isoTNS. 
\end{prop}
\begin{proof}
    By definition, $G$-isometric TNS satisfy
    \begin{equation}
        \includegraphics[width = 0.65\linewidth]{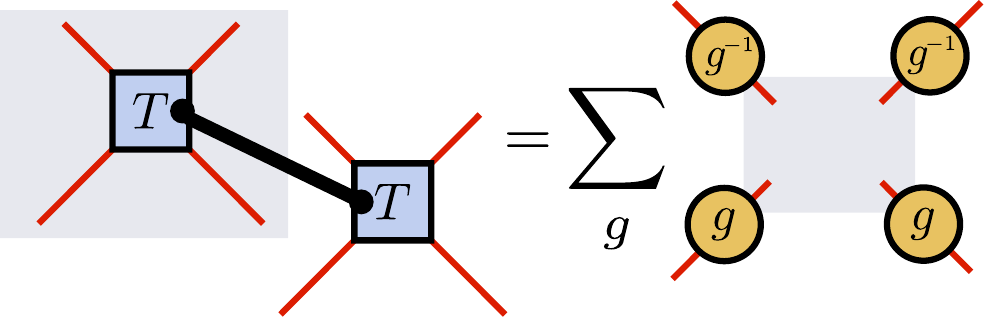},\label{sm:eq:iso}
    \end{equation}
    where the tensor with right-handed black dot denotes $T^\dag$. To be isoTNS, we need to show that the identity operator contracting from the top virtual space of Eq.~\eqref{sm:eq:iso} is a fixed-point eigenvector (up to an overall normalization of the $T$ tensor). Contracting the identity operator from the top leads to
    \begin{equation}
        F = \sum_g \Tr[U_{g^{-1}}]^2U_g\otimes U_g.
    \end{equation}
    In the basis $\{|h\rangle:h\in G\}$, the operator $U_g$ permutes basis vectors by
$|h\rangle\mapsto |gh\rangle$.
The trace of a permutation matrix equals the number of fixed basis vectors. Thus
$\operatorname{Tr}(U_g)
=
\#\{h\in G:gh=h\}$. Therefore
\begin{equation}
\Tr[U_g]
=
\begin{cases}
|G|, & g=e,\\
0, & g\neq e.
\end{cases}
\end{equation}
So we have $F = |G|^2U_e\otimes U_e  = |G|^2I$.
\end{proof}

This proposition shows that $G$-isometric TNS form a subclass of isoTNS. Consequently, although a generic $G$-injective TNS need not itself be an isoTNS, every $G$-isometric TNS can be regarded as a fixed-point representative within the broader class of $G$-injective isoTNS. From the dynamical perspective, $G$- injective isoTNS describe $G$-symmetric quantum channel dynamics without additional structure beyond the symmetry itself. The corresponding $G$-isometric TNS then play the role of fixed-point dynamics. Since no extra fine-tuned structure is present, we expect a generic $G$-injective isoTNS to be smoothly deformable to its fixed-point $G$-isometric representative without encountering any singularity or phase transition. 
This leads to the following conjecture:
\begin{center}\label{coro}
\textit{A $G$-injective isoTNS state exhibits $G$ topological order.}
\end{center}
The key physical intuition is that $G$-injective isoTNS and their corresponding $G$-isometric fixed points belong to the same quantum phase. They should therefore share the same universal long-distance properties, including topological order, anyonic structure, and long-range entanglement.

\begin{defn}[$G$-isometric quantum channel]
    A 1d translation-invariant (up to a constant-size unit cell) finite-depth local quantum channel $\Lambda$ is $G$-isometric if its canonical dual isoTNS is $G$-isometric.
\end{defn}
Such \(G\)-isometric quantum channels represent fixed-point symmetric dynamics:
after a finite number of iterations, the channel projects any initial state onto
its steady-state manifold, because the associated isoTNS has zero correlation
length. By the result on \(G\)-isometric TNS, they provably map to \(G\)-topologically
ordered isoTNS. The SW-SSB order in the steady state then follows from the
mutual anomaly inherited from the 1-form symmetries in the bulk of the
\(G\)-topologically ordered isoTNS, as explained in the main text.

Assuming that the quantum channel is \(G\)-isometric, each individual local
quantum channel \(\Lambda_i\) acts on
\(\mathcal{B}(\mathbb{C}^{|G|}\otimes \mathbb{C}^{|G|})\) and is isomorphic to
the following explicit form:
\begin{equation}
    \Lambda_i[\rho_i] = \frac{1}{|G|^2}\sum_g \Tr[(U_g\otimes U_g)\rho_i]U_{g^{-1}}\otimes U_{g^{-1}}.
\end{equation}
This channel has the following properties:
\begin{enumerate}
\item $\Lambda_i$ is strongly $G$-symmetric.
    \item $\Lambda_i$ is unital, i.e. $\Lambda_i[I] = I$, where $I$ is the identity operator on the local support $i$.
    \item $\Lambda_i$ generates fixed-point dynamics, i.e. $\Lambda_i^2 = \Lambda_i$. In fact, $\Lambda_i$ is the orthogonal projection onto
$\text{span}\{U_g\otimes U_g:g\in G\}$.
\end{enumerate}
They therefore provide a class of \(G\)-isometric quantum channels with 1d
SW-SSB steady states and dual 2d topologically ordered bulks.

Finally, we remark that \(G\)-injectivity pertains to on-site symmetries.
More generally, symmetries need not be on-site and can instead be represented
by matrix product operators (MPOs) \cite{schuch2010peps,williamson:2016,csahinouglu2021,bultinck2017anyons}. In this broader setting, the notion of \(G\)-injectivity can be extended to MPO-injectivity with respect to an MPO algebra. The resulting class of topological phases generalizes
quantum double models of \(G\) to the broader family of string-net models.

\section{Holographic bulk of the 2d 0-form SW-SSB}\label{appendix:2d_0-form_SW-SSB}
In Sec.\ref{sec:2d}, we consider the quantum channel $\cE = \prod_i \cE^z_{i} \cE^x_{i}$ (Eq.\ref{eq:2d_noise}), defined as
\begin{equation}\label{appendix:0_form_channel}
\begin{split} 
   &\cE^z_{ij}[\rho_0] = (1-p_z) \rho_0 + p_z  Z_i Z_{j}\rho_0 Z_i Z_{j}. \\
   &  \cE^x_{i}[\rho_0] =  (1-p_x) \rho_0 + p_x X_i \rho_0 X_i.
\end{split}
\end{equation}
The steady state of the channel is the SW-SSB density matrix $\rho \propto 1+ \prod_i X_i$. Here, we present the detailed operator mapping to show that the bulk exhibits a 3d toric-code topological order. We will focus on the fixed-point limit ($p_x =p_z =\frac{1}{2}$), and the non-fixed point limit can be easily obtained by applying an imaginary-time deformation, as in Eq.\ref{eq:deformed_appendix_2d}.

\begin{figure}[h]
\centering
\includegraphics[width=0.6\columnwidth]{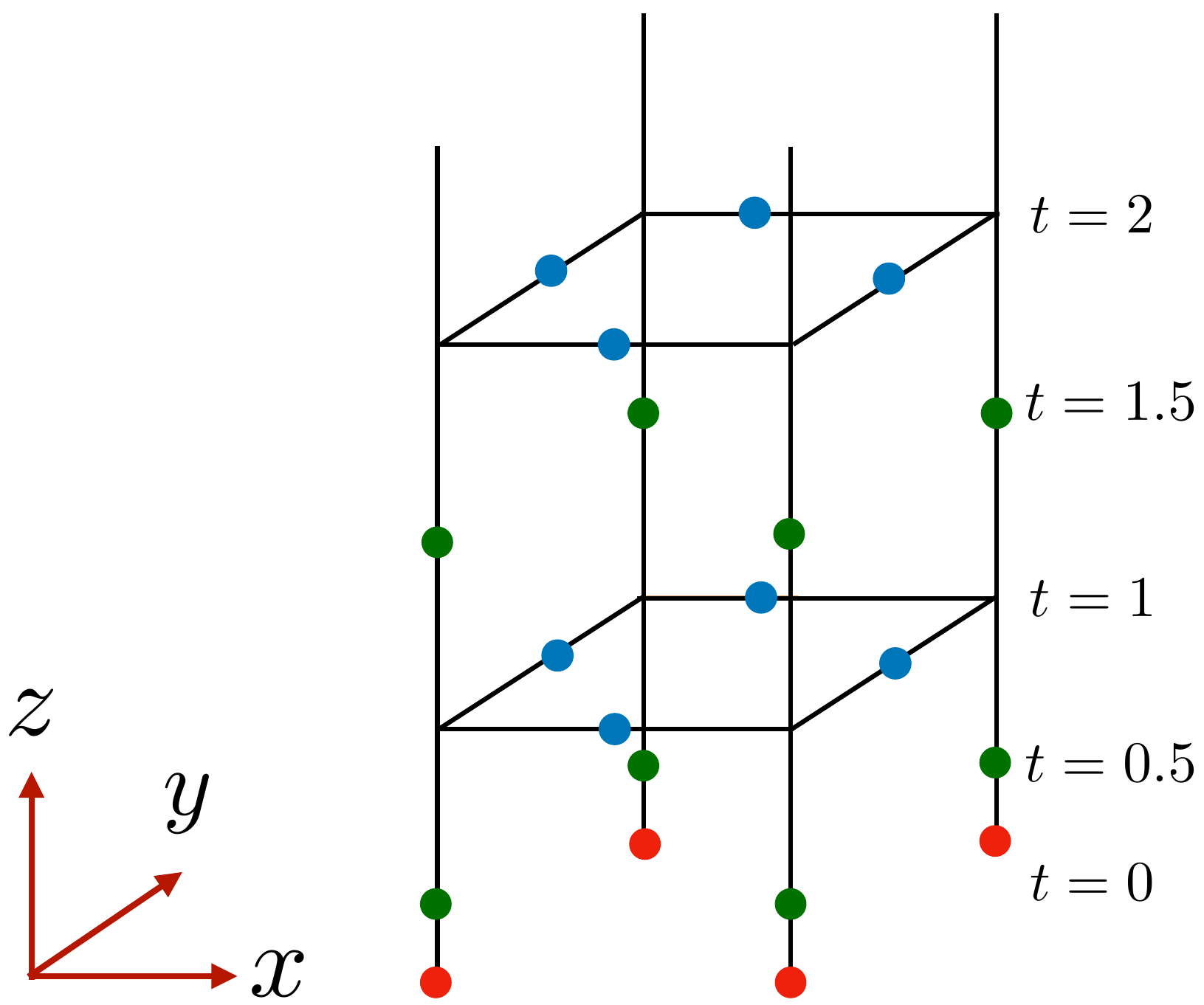}
\caption{The input of the repeated quantum channel (Eq.\ref{appendix:0_form_channel}) is denoted by the red qubits. They constantly travel upward (along the $z$ direction) to unitarily entangle with the green and blue ancilla qubits. Tracing out the ancilla qubits leads to the output of the quantum channel.}
\label{fig:3d_toric_evolution}
\end{figure}

To repeatedly implement the channel, the input red system qubits are constantly moving up to sequentially interact with the green and blue ancilla qubits, which are responsible for the ZZ noise and X noise, as shown in Fig. \ref{fig:3d_toric_evolution}. In particular, the green and blue ancilla qubits are initialized at $\ket{+}$ and $\ket{0}$, respectively. The unitary gate that entangles the ancilla green qubits and the system red qubits is given by  
\begin{equation}
u_x= \ket{0} \bra{0}_a + \ket{1} \bra{1}_a X_s.
\end{equation}

The unitary gate that entangles the ancilla blue qubits and the system red qubits is given by  
\begin{equation}
u_z= \ket{+} \bra{+}_a + \ket{-} \bra{-}_a Z_s, 
\end{equation}

\begin{figure}[h]
\centering
\includegraphics[width=\columnwidth]{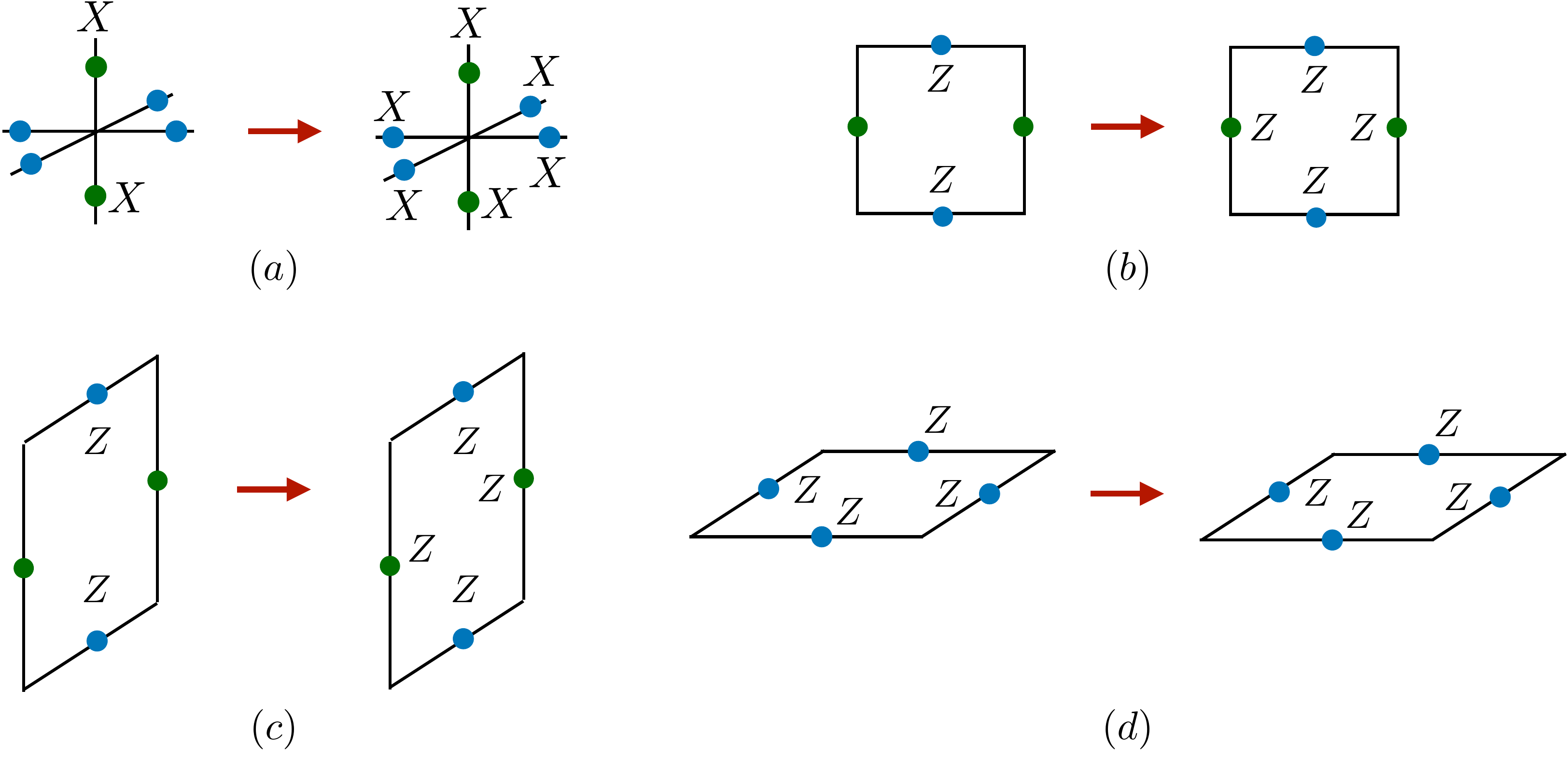}
\caption{Operator mapping under the sequential unitary circuit $U$ that implements the repeated application of the channel defined in Eq.\ref{appendix:0_form_channel}.}
\label{fig:3d_toric_stabilizer_dynamics}
\end{figure}

Since the green qubits are initialized at $\ket{+}$, the initial state before applying the sequential unitary circuit $U$ has the XX stabilizer on the two neighboring qubits on the $z$ links (Fig.\ref{fig:3d_toric_stabilizer_dynamics}(a)). One can check that after the conjugation of $U$, it is mapped to $X^{\otimes 6}$ on the 6 links around a vertex, i.e., the vertex stabilizers of the 3d toric code. On the other hand, since all the ancilla (blue) qubits on the $x-y$ plane are initialized in $\ket{0}$, the initial state has the $Z^{\otimes 2}$ stabilizer on the two links of an $x-z$ plane plaquette (Fig.\ref{fig:3d_toric_stabilizer_dynamics}(b)). Under the conjugation of $U$, it is mapped to the plaquette operator $Z^{\otimes 4}$ in the $x-z$ plane. Similarly, the $Z^{\otimes 2}$ stabilizer on the two links of a $y-z$ plane plaquette (Fig.\ref{fig:3d_toric_stabilizer_dynamics}(c)) is mapped to the plaquette operator $Z^{\otimes 4}$ in the $y-z$ plane. Finally, the initial state of the ancilla qubits also has the plaquette operator $Z^{\otimes 4}$ on the $x-y$ plane (Fig.\ref{fig:3d_toric_stabilizer_dynamics}(d)), and it is invariant under the sequential unitary circuit $U$. The existence of these star and plaquette stabilizers therefore indicates a 3d toric-code topological order.

\section{Holographic bulk of the 2d 1-form SW-SSB}\label{appendix:1-form_swssb}

In Sec.\ref{sec:sw-SSB_1-form}, we consider the following channel $\cE = \prod_v \cE_v^z \prod_\ell \cE_\ell^x$ defined as
\begin{align}\label{appendix:2dhigher}
   \cE^x_{\ell}[\rho] =  (1-p_x) \rho+ p_x X_\ell\rho X_\ell . \nonumber\\
     \cE^z_{v}[\rho] = (1- p_z)  \rho   + p_z     A_v \rho A_v,    
\end{align}
where $A_v $ is a $Z^{\otimes 4}$ acting on the four links around a vertex. Given any input state with the strong 1-form symmetry generated by $B_p= \prod_{\ell \in p} X_{\ell}$, the output state is $\rho \propto \prod_p  (1+B_p)$. Here we show that the holographic wavefunction is a 3d toric-code topological order with $m$-loop condensed boundary.

Specifically, we will focus on the fixed-point limit $p_x=p_z  = \frac{1}{2}$, and show the emergence of the fixed-point toric-code stabilizers.

\begin{figure}[th]
\centering
\includegraphics[width=0.75\columnwidth]{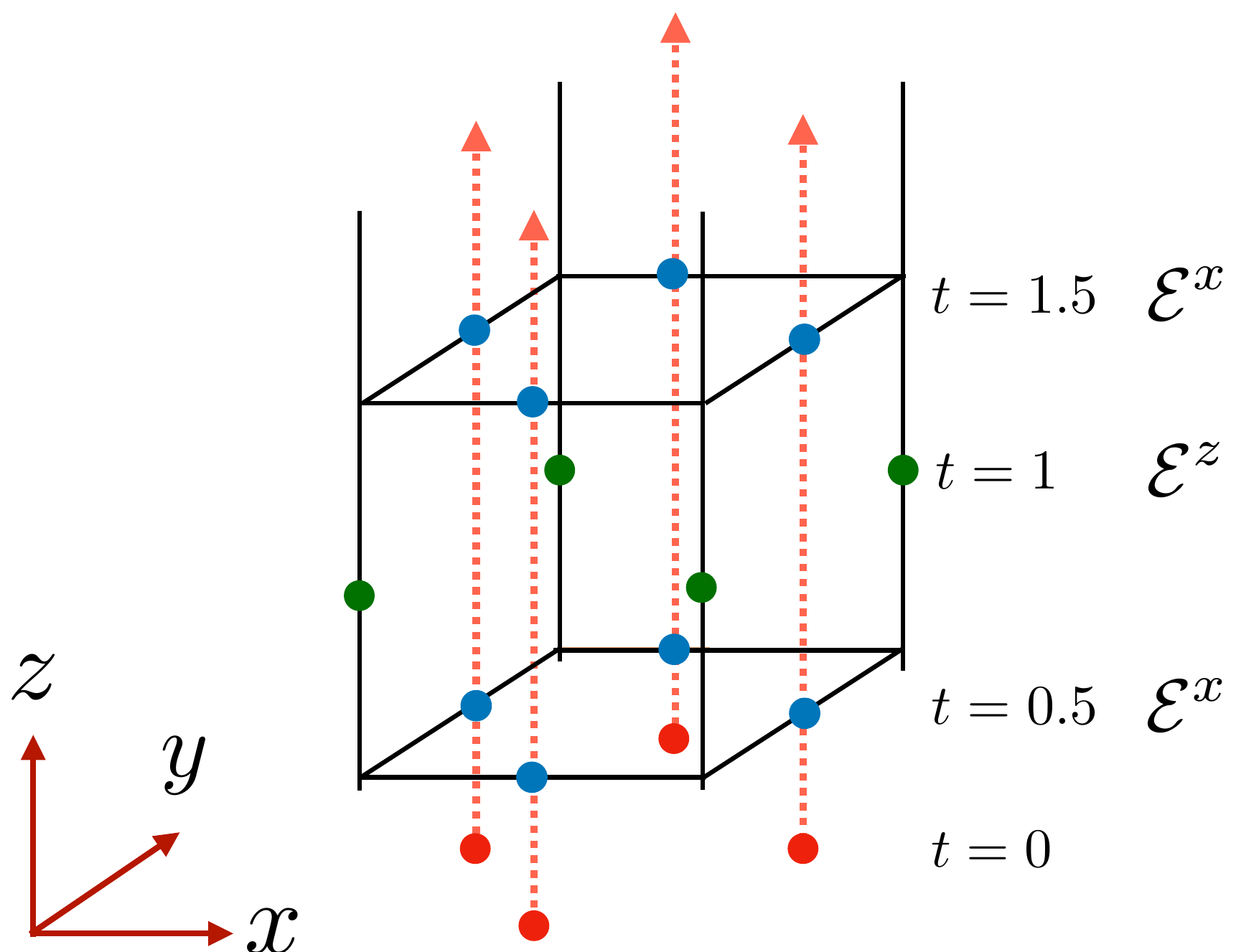}
\caption{The input of the repeated quantum channel (Eq.\ref{appendix:2dhigher}) is denoted by the red qubits. They constantly travel upward (along the $z$ direction) to unitarily entangle with the green and blue ancilla qubits. Tracing out the ancilla qubits leads to the output of the quantum channel.}
\label{fig:3d_1_form_toric_dynamics}
\end{figure}

To repeatedly implement the channel, the input red system qubits are constantly moving up to sequentially interact with the green and blue ancilla qubits, which are responsible for the Z-type noise and X-type noise, as shown in Fig. \ref{fig:3d_1_form_toric_dynamics}.

The blue ancilla qubits live on the links in the $x-y$ plane. They are initialized at $\ket{+ }$. The X-type noise can be implemented by entangling a system (red) qubit and a blue qubit via the gate
\begin{equation}
u_x= \ket{0} \bra{0}_a + \ket{1} \bra{1}_a X_s.
\end{equation}

This process can be explained more explicitly through Fig.\ref{fig:3d_1_form_toric_dynamics}: in the time slice when $t=0.5$, all the red qubits travel to the links on the x-y plane. A single red qubit will entangle with another blue qubit at the same location using the $u_x$ gate.

The green ancilla qubits live on the links oriented along the $z$ direction. They are initialized at $\ket{0 }$. The Z-type noise can be implemented by entangling a system (red) qubit and a green qubit via the gate
\begin{equation}
u_z= \ket{+} \bra{+}_a + \ket{-} \bra{-}_a Z_s.
\end{equation}
More explicitly, in the time slice when $t=1$ (Fig.\ref{fig:3d_1_form_toric_dynamics}), all the red qubits travel to the centers of the x-z plane plaquettes and the centers of the y-z plane plaquettes; a single red qubit will entangle with its two neighboring green qubits, both using the $u_z$ gate. Equivalently, a single green qubit will entangle with four green qubits on the four neighboring plaquettes, all using the $u_z$ gate.

\begin{figure}[h]
\centering
\includegraphics[width=\columnwidth]{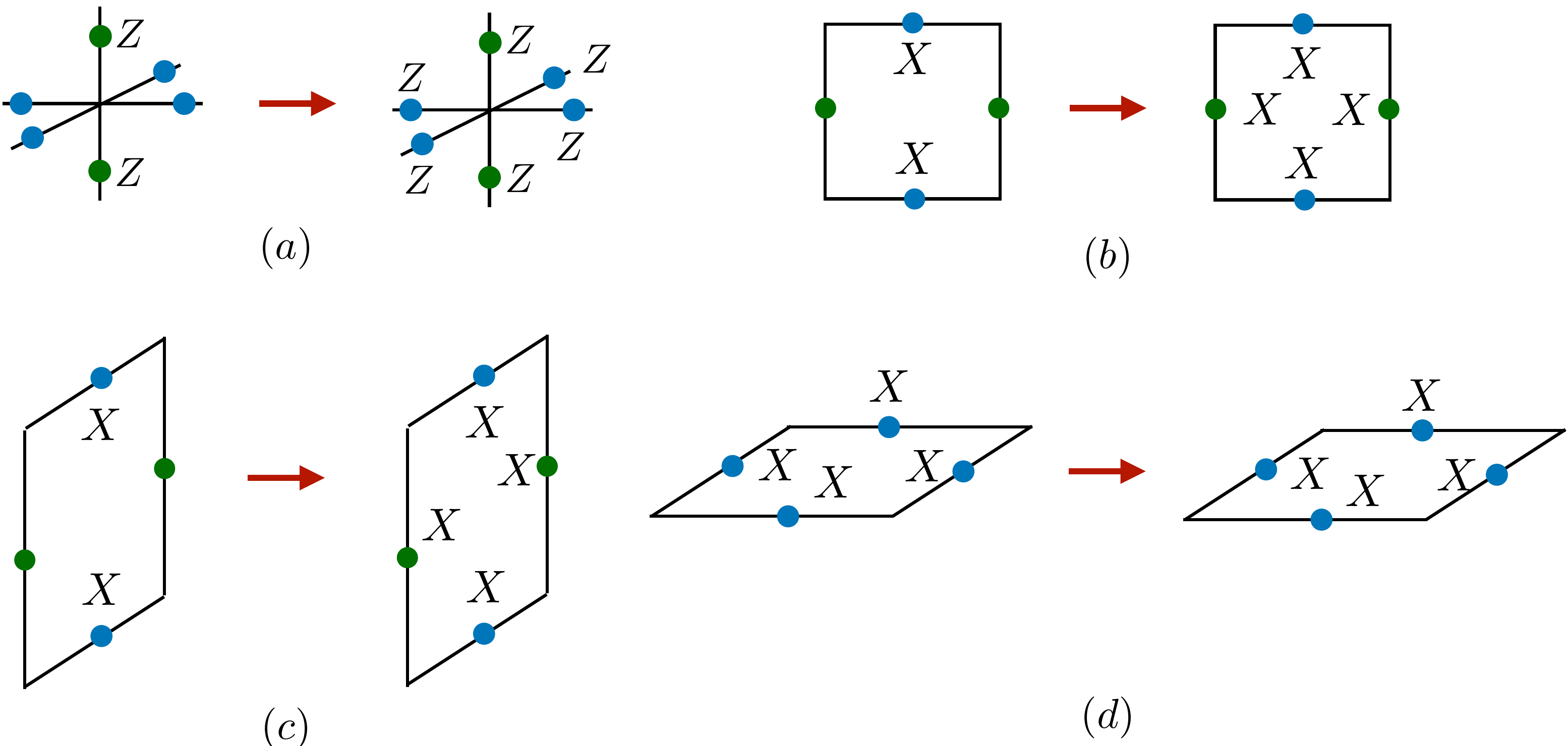}
\caption{Operator mapping under the sequential unitary circuit $U$ that implements the repeated application of the channel defined in Eq.\ref{appendix:2dhigher}.}
\label{fig:3d_1_form_toric_operator_mapping}
\end{figure}

With the application of the sequential unitary circuit, one can derive the operator mapping rules under the circuit conjugation (Fig. \ref{fig:3d_1_form_toric_operator_mapping}). The emergence of the star and plaquette stabilizers indicates the 3d toric-code topological order. We note that the operator mapping in (d) requires the red qubits to have the strong 1-form symmetry generated by $B_p =\prod_{\ell \in p }X_\ell$. This is because when the red qubits interact with the blue qubits via the onsite $u_x$ gate, the $X^{\otimes 4}$ stabilizer on the blue qubits is, in fact, mapped to a product of $X^{\otimes 4}$ (on the blue qubits) and $B_p$ (on the red qubits). Only when $B_p=1$,  $X^{\otimes 4}$ of the blue qubits can be invariant under the circuit dynamics.

\section{Holographic bulk of the 2d fermionic 1-form SW-SSB}\label{appendix:1-form_swssb_fermion}

In Sec. \ref{sec:fermion_1_form}, we consider the channel $ \cE = \prod_v \cE_v  \prod_\ell  \cE_\ell $ with

\begin{equation}\label{appendix:fermion_channel}
\begin{split}
   &\cE_{\ell}[\rho] = p_\ell \rho +  (1-p_\ell) X_\ell Z_{\ell +\frac{\hat{x}}{2}-\frac{\hat{y}}{2}}\rho    X_\ell Z_{\ell +\frac{\hat{x}}{2}-\frac{\hat{y}}{2}}\\
   & \cE_v[\rho] = p_v \rho + (1-p_v) A_v \rho A_v 
    \end{split}
\end{equation}\label{appendix:2dhigherfer}

Here we discuss details regarding the emergence of the 3d holographic wavefunction. We will focus on the fixed-point limit $p_\ell=p_v = \frac{1}{2}$, and show the emergence of the fixed-point 3d fermionic toric-code stabilizers.

\begin{figure}[t]
\centering
\includegraphics[width=0.7\columnwidth]{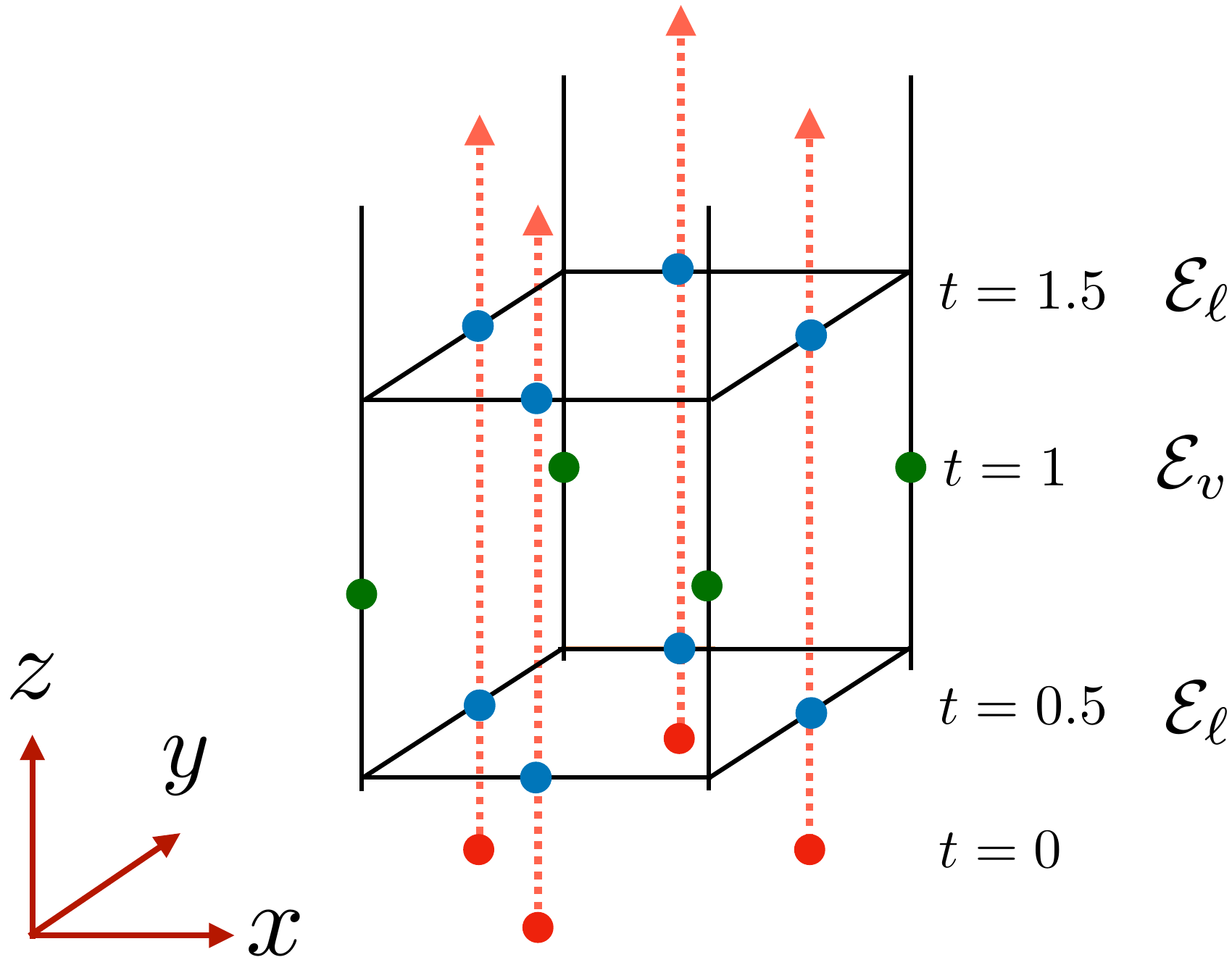}
\caption{The input of the repeated quantum channel (Eq.\ref{appendix:fermion_channel}) is denoted by the red qubits. They constantly travel upward (along the $z$ direction) to unitarily entangle with the green and blue ancilla qubits. Tracing out the ancilla qubits leads to the output of the quantum channel.}

\label{fig:3d_fermion_channel}
\end{figure}

To repeatedly implement the channel, the input red system qubits are constantly moving up to sequentially interact with the blue and green ancilla qubits, which are responsible for the $\cE_\ell$ channel and $\cE_v$, as shown in Fig. \ref{fig:3d_fermion_channel}.

	\begin{figure*}
		\centering
\includegraphics[width=0.7\textwidth]{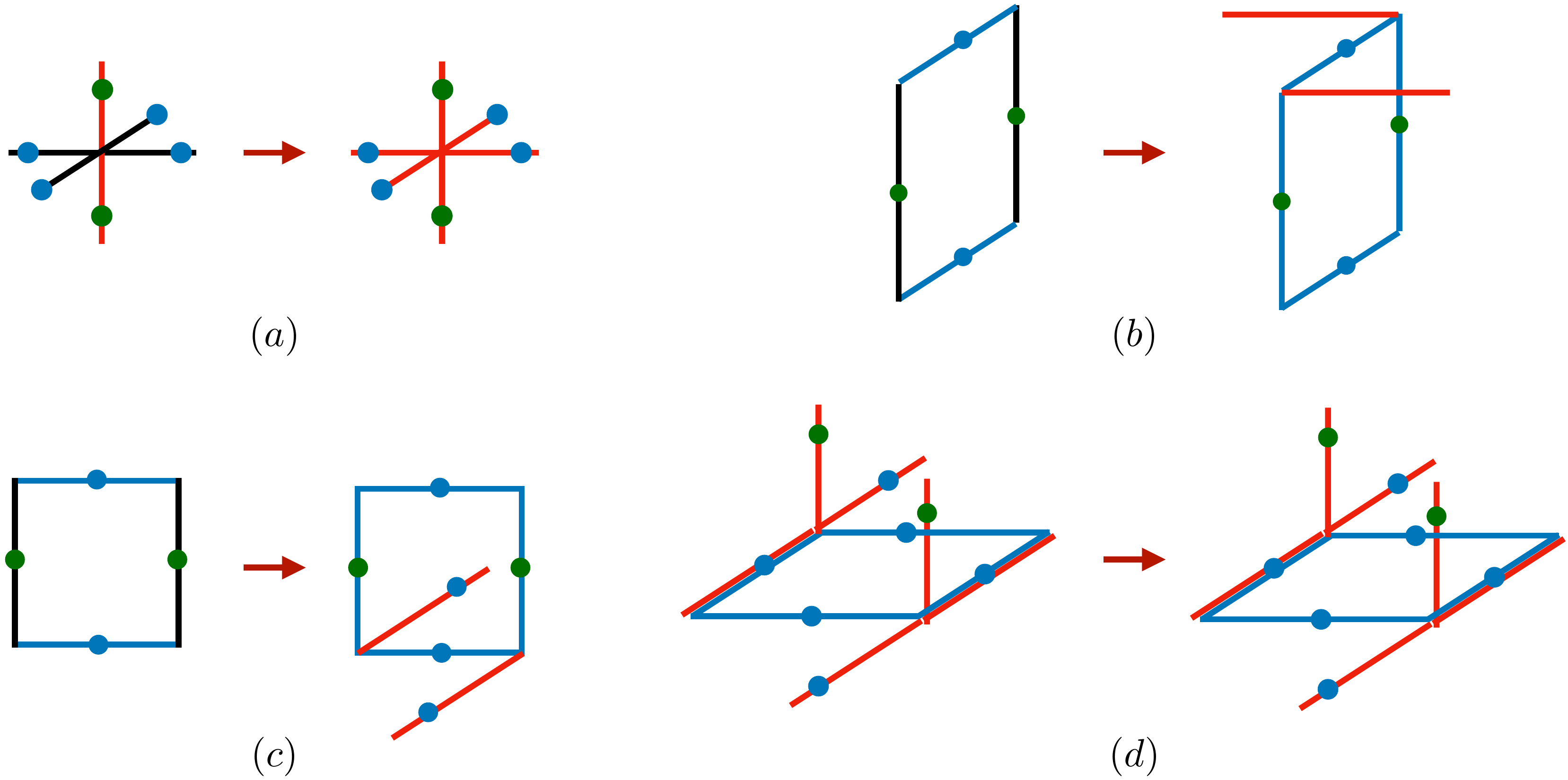}       
\caption{Operator mapping under the sequential unitary circuit that implements the repeated application of the channel defined in Eq.\ref{appendix:fermion_channel}. The red and blue links denote the Pauli-X and Pauli-Z, respectively; no Pauli operators are supported on black links.}
		\label{fig:3d_fermion_mapping}
	\end{figure*}

The blue ancilla qubits live on the links in the $x-y$ plane. They are initialized at $\ket{+}$. $\cE_\ell$ can be implemented by entangling a system (red) qubit and a blue qubit via the gate
\begin{equation}
u_\ell = \ket{0} \bra{0}_{a,\ell} + \ket{1} \bra{1}_{a,\ell}   X_{\ell} Z_{\ell + \frac{x}{2} - \frac{y}{2}}.
\end{equation}
where $\ket{0}_{a,\ell}, \ket{1}_{a,\ell} $ and the state of the blue ancilla qubit at the link $\ell$, and $ X_{\ell} Z_{\ell + \frac{x}{2} - \frac{y}{2}}$ is the operator acting on the two system qubits.

This process can be explained more explicitly through Fig.\ref{fig:3d_fermion_channel}: in the time slice when $t=0.5$, all the red qubits travel to the links on the x-y plane and they interact with the blue qubits with $\prod_\ell  u_{\ell}$. Notice that neighboring $u_{\ell}$ may not commute, so we would need to specify the time order for the application of the $u_\ell$ gate. Our choice is shown as follows:

\begin{equation*}
\includegraphics[width=4cm]{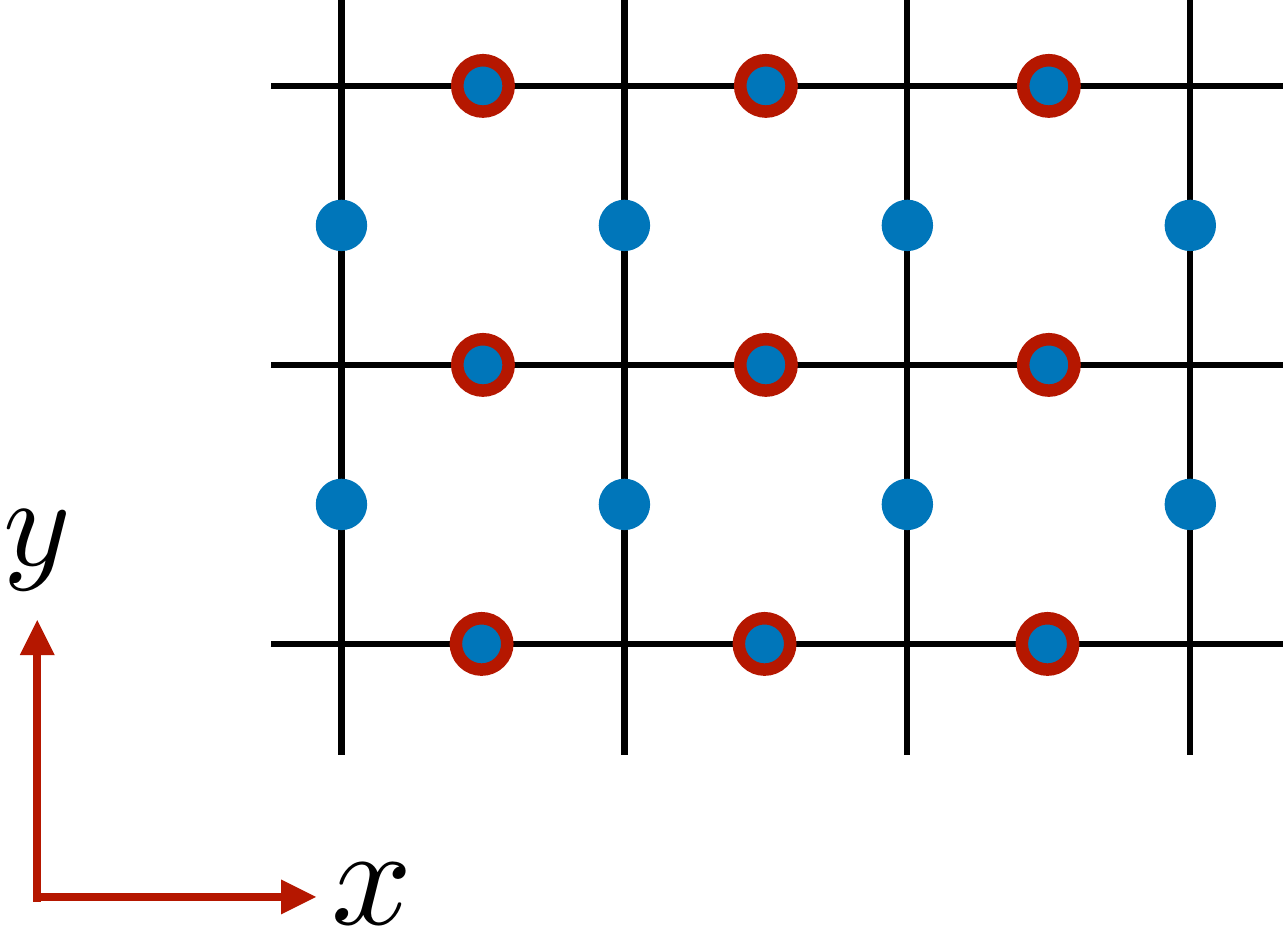}
\end{equation*}

We first simultaneously apply $u_\ell$ on all the x-links, and then simultaneously apply $u_\ell$ on all the y-links. While the choice of the order of the gate applied affects the detail of the 3d bulk stabilizers, it does not affect the quantum channel on the system red qubits. 

The green ancilla qubits live on the links oriented along the $z$ direction. They are initialized at $\ket{0 }$. The Z-type noise can be implemented by entangling a system (red) qubit and a green qubit via the gate
\begin{equation}
u_z= \ket{+} \bra{+}_a + \ket{-} \bra{-}_a A_v.
\end{equation}
More explicitly, in the time slice when $t=1$ (Fig.\ref{fig:3d_1_form_toric_dynamics}), all the red qubits travel to the centers of the x-z plane plaquettes and the centers of the y-z plane plaquettes; a single red qubit will entangle with its two neighboring green qubits, both using the $u_z$ gate. Equivalently, a single green qubit will entangle with four green qubits on the four neighboring plaquettes, all using the $u_z$ gate.

With the application of the sequential unitary circuit, one can then derive the operator mapping rules under the circuit conjugation, shown in Fig. \ref{fig:3d_fermion_mapping}.  

\section{Non-CPTP channel for deformed toric code} \label{sec:non_cptp_appendix}

Below we consider the deformed toric code with homogeneous string tension introduced in Sec.\ref{sec:beyond_channel}:   
\begin{equation}\label{eq:dftc}
\ket{\psi_\beta} \propto e^{\beta \sum_i Z_i } \ket{\text{TC}}.
\end{equation}
By tuning $\beta$, this wavefunction exhibits a phase transition between a topologically ordered phase and a charge-confined phase \cite{Fradkin_deformed_2007, Chamon_state_2008}.
Here, we show that this wavefunction can be prepared by a sequential nonunitary circuit involving measurements and post-selection. Therefore, the corresponding quantum channel acting on the 1d boundary system defines a non-CPTP map.

As a warm-up, we first review the isoTNS construction of the fixed-point toric code $\ket{\text{TC}}$, in the plumbing construction (Eq.\ref{eq:plumbing}). Specifically, the isoTNS takes the form

\begin{equation}\label{toric_fixed_point:channel}
\includegraphics[width=5.5 cm]{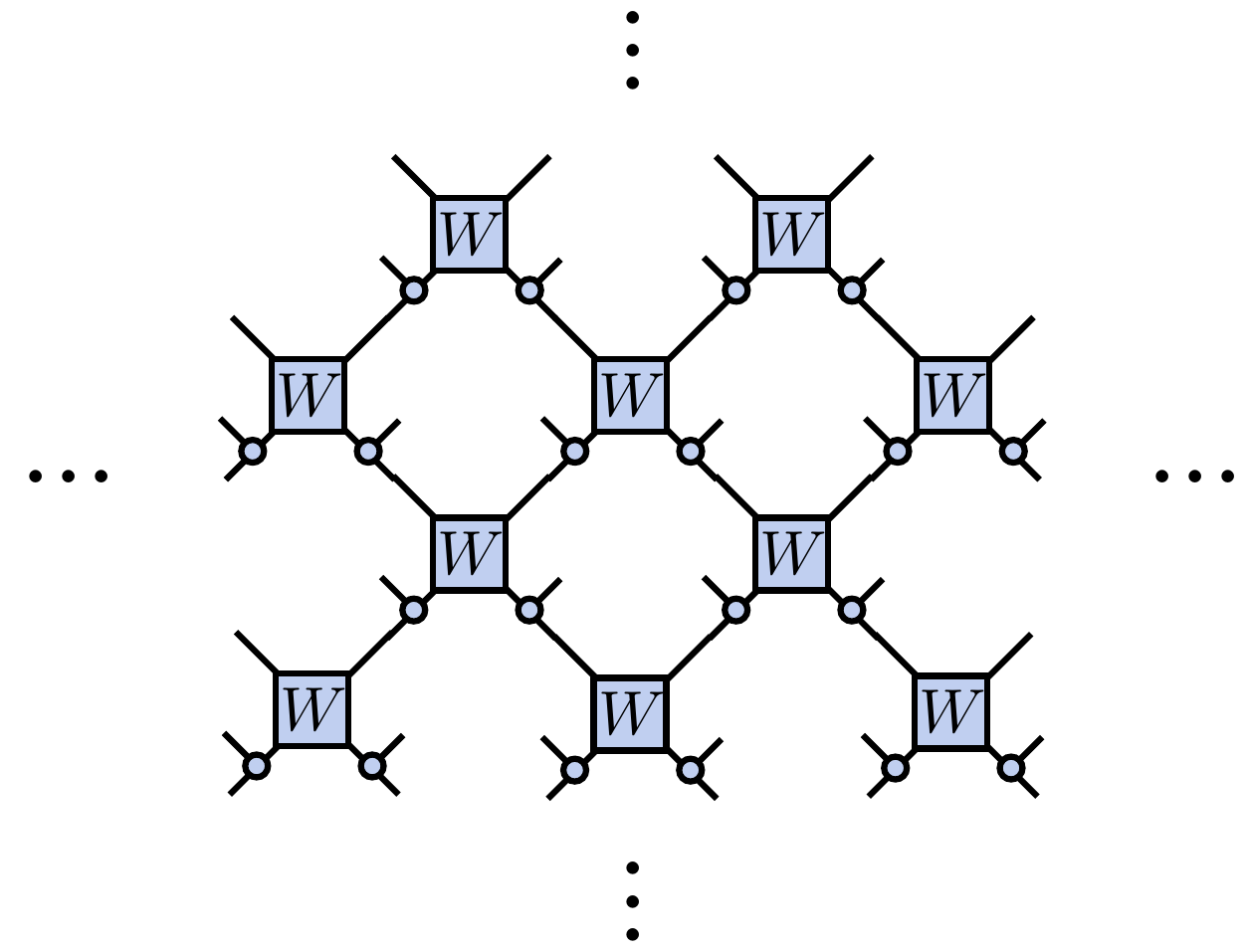}.
\end{equation} 
Each local tensor involves four virtual legs and two physical legs. Importantly, it is uniquely specified by the following stabilizer symmetries: 

\begin{equation}\label{toric_code_symmetry}
\includegraphics[width=5.5 cm]{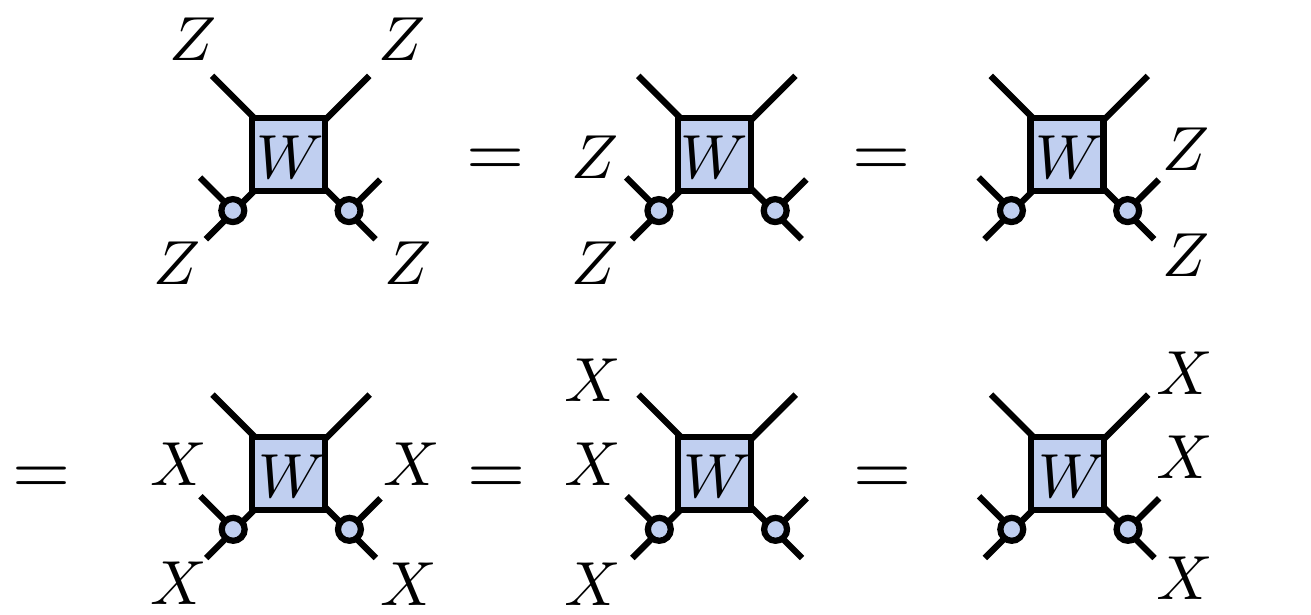}.
\end{equation} 
These conditions imply that the isoTNS describes the fixed-point toric code, with $Z^{\otimes 4}$ on the four physical legs around a vertex and $X^{\otimes 4}$ on the four physical legs around a plaquette. In particular, the local tensor defines an isometry $V: (\mathbb{C}^{2})^{\otimes 2}  \to (\mathbb{C}^{2})^{\otimes 4}$. The bottom two virtual legs are the input. The top two virtual legs and the two physical legs are the output:

\begin{equation}\label{}
\includegraphics[width=1.5cm]{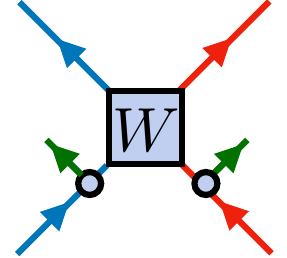}
\end{equation} 

Such an isometry can then be enlarged as a four-qubit unitary gate $U$ that entangles the two virtual qubits (whose worldlines are colored in blue and red, respectively) and the two ancilla qubits initialized at $\ket{00}$. As such, the local tensor defines a two-qubit quantum channel when tracing out the ancilla qubits:

\begin{equation}
 \cE[\rho_s] = \text{Tr}_{a} \left[  U \rho_s \otimes  (\ket{00} \bra{00})_a U^{\dagger} \right], 
\end{equation}  
where $s$ and $a$ denote the virtual-leg qubits and the physical-leg qubits, respectively. 

The action of $U$ can also be specified by the stabilizer symmetries shown in Eq.\ref{toric_code_symmetry}, which can be viewed as an operator mapping from the input to the output. Therefore, the isoTNS of the toric code in Eq.\ref{toric_fixed_point:channel} corresponds to a process where a 1d system of qubits unitarily interacts with rows of ancilla qubits (which are the physical legs of the isoTNS).

Equipped with the understanding of the fixed-point toric code tensor, we now discuss the case where every physical leg of the tensor is subject to additional imaginary-time deformation $e^{\beta  Z}$. In this case, the local tensor no longer defines an isometry. This can be seen as follows: due to the delta tensor used in the plumbing construction, applying the imaginary-time deformation to a physical leg is equivalent to applying the same action on the virtual leg. Specifically, the corresponding local tensor becomes 

\begin{equation}\label{}
\includegraphics[width=3cm]{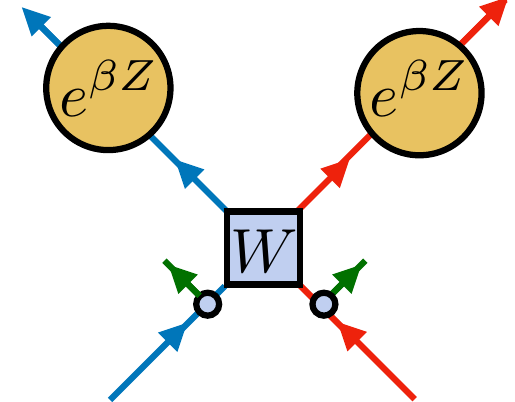}. 
\end{equation} 

In other words, after applying the four-qubit entangling gate $U$, the two virtual qubits are further subject to the non-unitary evolution $e^{\beta  Z}$ individually. As such, the overall quantum channel acting on the two virtual qubits is not CPTP. In particular, the channel can be implemented with post-selection via the following equality:

\begin{equation} \label{eq:imaginary_time}
\bra{i}_a e^{-i t Z_a  \otimes Z}\ket{+}_a  \propto e^{\beta Z}, ~~\tan t= \tanh \beta
\end{equation}
The subscript $a$ denotes an additional ancilla qubit, $\ket{+}_a = \frac{1}{\sqrt{2}} \left( \ket{0 }_a + \ket{1} \right)_a$ is the Pauli-X +1 eigenstate,  $\ket{i}_a = \frac{1}{\sqrt{2}}  \left(  \ket{0}_a + i \ket{1}_a \right)$ is the Pauli-Y +1 eigenstate, and $Z_a$ is the Pauli-Z acting on the ancilla qubit. Eq.\ref{eq:imaginary_time} implies that the non-unitary evolution $ e^{\beta Z}$ can be implemented by the following process involving post-selection: (i) prepare an ancilla qubit in $\ket{+}_a$, (ii) perform the two-qubit unitary evolution $ e^{-i t Z_a  \otimes Z}$ with the evolution time $t=  \tan^{-1}(  \tanh \beta )$, and (iii) post-select on the ancilla qubit in the $+1$ Pauli-Y eigenstate $\ket{i}_a$. With this ingredient, we see that the non-CPTP channel associated with the deformed toric code can be implemented with unitary gates and post-selection.

\bibliography{ref.bib}
 \end{document}